\pdfoutput=1 

\documentclass[hyper,letterpaper,11pt]{JHEP3}
\usepackage{cite}
\usepackage{amsmath,amssymb,multirow,array}
\usepackage{graphicx}

\newcommand{\N}{\mathcal{N}}
\newcommand{\jt}{\langle J^0 \rangle}
\newcommand{\beq}{\begin{equation}}
\newcommand{\eeq}{\end{equation}}

\title{Moduli Spaces of Cold Holographic Matter}
\author{Martin Ammon$^1$\footnotemark[1]\,, Kristan Jensen$^2$\footnotemark[2]\,, Keun-Young Kim$^3$\footnotemark[3]\,, Jo\~ao N. Laia$^4$\footnotemark[4]\,, and Andy O'Bannon$^4$\footnotemark[4]\, \\
$^1$ Department of Physics and Astronomy, University of California, \\
\hspace{.3cm}Los Angeles, CA 90095, United States \\
$^2$ Department of Physics and Astronomy, University of Victoria, \\ 
\hspace{.3cm}Victoria, BC V8W 3P6, Canada \\
$^3$ Institute for Theoretical Physics, University of Amsterdam, Science Park 904,\\
\hspace{.3cm}Postbus 94485, 1090 GL Amsterdam, The Netherlands \\
$^4$ Department of Applied Mathematics and Theoretical Physics, \\
\hspace{.3cm}University of Cambridge, Cambridge CB3 0WA, United Kingdom
}

\footnotetext[1]{E-mail address: \email{ammon@physics.ucla.edu}}
\footnotetext[2]{E-mail address: \email{kristanj@uvic.ca}}
\footnotetext[3]{E-mail address: \email{K.Y.Kim@uva.nl}}
\footnotetext[4]{E-mail addresses: \email{J.Laia,A.OBannon@damtp.cam.ac.uk}}

\abstract{We use holography to study $(3+1)$-dimensional $\N=4$ supersymmetric Yang-Mills theory with gauge group $SU(N_c)$, in the large-$N_c$ and large-coupling limits, coupled to a single massless $(n+1)$-dimensional hypermultiplet in the fundamental representation of $SU(N_c)$, with $n=3,2,1$. In particular, we study zero-temperature states with a nonzero baryon number charge density, which we call holographic matter. We demonstrate that a moduli space of such states exists in these theories, specifically a Higgs branch parameterized by the expectation values of scalar operators bilinear in the hypermultiplet scalars. At a generic point on the Higgs branch, the R-symmetry and gauge group are spontaneously broken to subgroups. Our holographic calculation consists of introducing a single probe D$p$-brane into $AdS_5 \times \mathbb{S}^5$, with $p=2n+1=7,5,3$, introducing an electric flux of the D$p$-brane worldvolume $U(1)$ gauge field, and then obtaining explicit solutions for the worldvolume fields dual to the scalar operators that parameterize the Higgs branch. In all three cases, we can express these solutions as non-singular self-dual $U(1)$ instantons in a four-dimensional space with a metric determined by the electric flux. We speculate on the possibility that the existence of Higgs branches may point the way to a counting of the microstates producing a nonzero entropy in holographic matter. Additionally, we speculate on the possible classification of zero-temperature, nonzero-density states described holographically by probe D-branes with worldvolume electric flux.}

\preprint{DAMTP-2012-53\\CCTP-2012-22}

\begin{document}

\section{Introduction}
\label{S:intro}

Consider a system with a global $U(1)$ symmetry. A compressible state of that system is a state with a nonzero $U(1)$ charge density that varies smoothly as a function of the chemical potential $\mu$. The best-understood examples of compressible states are superfluids, in which the $U(1)$ is spontaneously broken, solids, in which translational symmetry is spontaneously broken to a discrete subgroup, and Fermi liquids, in which neither the $U(1)$ nor translational symmetries are broken. In a superfluid or a solid, the light degrees of freedom include the Goldstone boson(s) of the spontaneous symmetry breaking, while in a Fermi liquid the light degrees of freedom include Landau quasi-particles.

Some real materials are compressible but are not superfluids, solids, or Fermi liquids, a prime example being the ``strange metal'' phase of many materials, including the normal (non-superconducting) state of high-$T_c$ cuprates. The degrees of freedom in strange metals remain mysterious, primarily because strong electron-electron correlations preclude a quasi-particle description. The existence of strange metals, and other exotic compressible states, raises a general question: can we classify compressible states of matter?

The Anti- de Sitter/Conformal Field Theory (AdS/CFT) correspondence \cite{Maldacena:1997re,Gubser:1998bc,Witten:1998qj}, and its generalizations, collectively called gauge-gravity duality or holography, may help answer that question \cite{2011arXiv1108.1197S,Huijse:2011hp}. AdS/CFT is the conjectured equivalence between certain strongly-coupled CFT's and certain weakly-coupled theories of gravity in an AdS spacetime of one higher dimension. In some sense the CFT ``lives'' on the co-dimension-one AdS boundary, hence the name ``holography.'' The AdS/CFT dictionary~\cite{Witten:1998qj,Gubser:1998bc} equates the on-shell bulk action with the generating functional of CFT correlation functions. The conserved current of a global $U(1)$ symmetry of the CFT is dual to a $U(1)$ gauge field in AdS, and nonzero-density states are dual to spacetimes with nonzero electric flux at the AdS boundary. AdS/CFT provides many examples of compressible states involving strongly-interacting degrees of freedom, including states that are not superfluids, solids, or Fermi liquids, the principal example being the AdS-Reissner-Nordstr\"om charged black brane. Holography has the potential to reveal universal properties of compressible states, including perhaps some guiding principle(s) to classify them.

We will use holography to study compressible states in three systems, namely the field theories arising from the $(n+1)$-dimensional intersection of $N_c$ D3-branes with $N_f$ D$p$-branes~\cite{Karch:2002sh,Karch:2000gx,Constable:2002xt}, with $p=2n+1 = 7,5,3$, so that $n=3,2,1$ respectively. We will call these D-brane intersections the D3/D$p$ systems and the corresponding field theories the D3/D$p$ theories. These theories are $(3+1)$-dimensional $\N=4$ supersymmetric Yang-Mills (SYM) theory, with gauge group $SU(N_c)$ and Yang-Mills coupling $g_{YM}$, coupled to $N_f$ $(n+1)$-dimensional hypermultiplets \cite{Karch:2002sh,Karch:2000gx,DeWolfe:2001pq,Constable:2002xt,Erdmenger:2002ex} in the fundamental representation of $SU(N_c)$, \textit{i.e.} flavor fields. Recall that a hypermultiplet contains both fermions and scalars, which we will call quarks and squarks respectively. We will consider only massless flavor fields, unless stated otherwise. The resulting theories preserve eight Poincar\'e supercharges. 

When $N_f>1$ each of these theories possesses a moduli space of supersymmetric (SUSY) vacua parameterized by the expectation values of gauge-invariant scalar operators bilinear in the squarks. This moduli space is called the ``Higgs branch'' of the theory. For all points on the Higgs branch, the eight Poincar\'e supercharges are preserved, while at a generic point on the Higgs branch both the R-symmetry and the gauge group are spontaneously broken to subgroups. When $N_f=1$, the Higgs branch is absent in the $n=3$ case, but remains in the $n=2,1$ cases.

We will work in the Maldacena limits~\cite{Maldacena:1997re}, meaning first we take the 't Hooft large-$N_c$ limit, $N_c \to \infty$ with $g_{YM}^2 \to 0$ with the 't Hooft coupling $\lambda \equiv g_{YM}^2 N_c$ fixed, followed by the limit $\lambda \to \infty$. We will keep $N_f$ finite as $N_c \to \infty$, so that $N_f \ll N_c$, and work in the probe limit, meaning we expand all observables in the small parameter $N_f/N_c$ and only retain terms of order $N_c^2$ and of order $N_fN_c$.

In these limits, $\N=4$ SYM is holographically dual to type IIB supergravity in the near-horizon geometry of the D3-branes, $AdS_5 \times \mathbb{S}^5$, with $N_c$ units of Ramond-Ramond (RR) five-form flux on the $\mathbb{S}^5$. The probe flavors are dual to probe D$p$-branes extended along an asymptotically $AdS_{n+2} \times \mathbb{S}^n$ submanifold inside $AdS_5 \times \mathbb{S}^5$. A $U(N_f)$ worldvolume gauge field and $(9-p)$ scalars propagate on the worldvolume of the D$p$-branes, with an action given by a Dirac-Born-Infeld (DBI) term plus Wess-Zumino (WZ) terms describing the coupling to RR fields, including in particular the RR five-form.

The scalar operators that parameterize the Higgs branch are dual to fields on the D$p$-brane that, upon Kaluza-Klein reduction to the $AdS_{n+2}$ submanifold, are scalars. Points on the Higgs branch appears in the bulk as solutions for these fields that are static, normalizable, and do not affect the value of the on-shell D$p$-brane action. For example, in the $n=3$ case with $N_f>1$, a point on the Higgs branch appears in the bulk as an instanton of the D7-brane worldvolume non-Abelian gauge field~\cite{Guralnik:2004ve,Guralnik:2004wq,Guralnik:2005jg,Erdmenger:2005bj,Arean:2007nh} in the $\mathbb{S}^3$ and $AdS_5$ radial directions. The size and orientation moduli of these instantons are isomorphic to the moduli of the Higgs branch, and thus encode the pattern of the spontaneous breaking of the R-symmetry. For example, the size modulus maps to the squark expectation value.\footnote{A squark expectation value is not gauge invariant, and thus strictly speaking is unobservable, so the identification of the size modulus with the squark exepectation value is really just a mnemonic device. To our knowledge, the precise dictionary between the moduli of bulk instantons and the expectation values of \textit{gauge-invariant} field theory operators has not been determined.} Moreover, these instanton solutions endow the D7-branes with D3-brane charge, indicating the spontaneous breaking of $SU(N_c)$ to a subgroup.\footnote{In order to remain in the probe limit, the D7-branes must carry a D3-brane charge much less than $N_c$.} The $N_f=1$ case is special, since the D3/D7 theory then has no Higgs branch. In the holographic description, the corresponding statement is that when $N_f=1$ the D7-brane worldvolume gauge field is Abelian, in which case an instanton solution has no size modulus, and furthermore is singular at its core. Statements analogous to the above also apply for the probe D5- and D3-branes~\cite{Arean:2006vg,Arean:2007nh,Constable:2002xt}, with the exception that when $N_f=1$, non-singular, normalizable solutions holographically dual to points on the Higgs branch exist.

Each of the D3/D$p$ theories, with massless flavor fields, enjoys a global $U(N_f)$ flavor symmetry. We will call the overall diagonal $U(1)$ subgroup of this $U(N_f)$ baryon number. We will produce compressible states by introducing a nonzero baryon number charge density. In the Maldacena and probe limits, the components of the conserved baryon number current $J^{\mu}$, with $\mu = 0,1,\ldots, n$, are dual to the components of the $U(1)$ worldvolume D$p$-brane gauge field in the same directions. States with nonzero baryon density correspond to $U(1)$ gauge field solutions with nonzero electric flux through the boundary of the $AdS_{n+2}$. We call such compressible states of the D3/D$p$ systems ``holographic matter.'' We will work at zero temperature, hence our holographic matter will be ``cold.''

These compressible states of the D3/D$p$ theories have been extensively studied, using holography, in refs.~\cite{Nakamura:2006xk,Kobayashi:2006sb,Karch:2007br,Karch:2008fa,Kulaxizi:2008kv,Myers:2008me, Nickel:2010pr,Ammon:2011hz,Davison:2011ek,Ammon:2012je}. These calculations have shown that this holographic matter breaks neither the baryon number $U(1)$ nor any continuous translational symmetry. Moreover, to date no evidence of a Fermi surface has been found in holographic matter. In other words, these states are not superfluids, solids, or Fermi liquids. Indeed, these states have various unusual properties, such as an extensive ground state degeneracy~\cite{Karch:2008fa} and a spectrum of low-energy fluctuations that appears to be controlled by a $(0+1)$-dimensional CFT~\cite{Jensen:2010ga,Nickel:2010pr,Ammon:2011hz}.

We will work with $N_f=1$. In that case, for $n=2$ (the probe D5-brane), the results of ref.~\cite{UW:WIP} showed that the zero-density Higgs branch survives the introduction of the nonzero charge density, despite the fact that the charge density breaks all SUSY. We will show that in \textit{all three} of our cases, $n=3,2,1$, at nonzero charge density a Higgs branch exists. For $n=2,1$ we thus demonstrate that the zero-density Higgs branch survives the introduction of nonzero density. For the $n=3$ case, which has no Higgs branch at zero density, the existence of a Higgs branch at nonzero density is especially surprising: apparently, in this case the nonzero density \textit{creates} a moduli space where one did not previously exist.

Our calculation consists of introducing electric flux on the worldvolume of a single probe D$p$-brane and then obtaining normalizable solutions for the D$p$-brane worldvolume fields dual to the scalar operators that parameterize the Higgs branch. The essential ingredient for obtaining these solutions is something that we call the ``effective metric''\cite{Chen:2009kx}, which is a metric on the $\mathbb{R}^{n+1}$ spanned by the $\mathbb{S}^n$ and the $AdS_{n+2}$ radial direction. The effective metric is conformally equivalent to the flat metric, with a conformal factor determined by the electric flux.\footnote{Crucially, the effective metric is \textit{not} the open string metric \cite{Seiberg:1999vs}.} Importantly, this conformal factor has a zero that effectively ``cuts out'' a ball around the origin of $\mathbb{R}^{n+1}$ whose radius is proportional to the chemical potential. 

In all three of our cases, we can write our solutions as $U(1)$ field strengths in $\mathbb{R}^4$ self-dual with respect to the effective metric, although doing so when $n=2,1$ requires adding fictitious spatial directions. These self-dual $U(1)$ field strengths solve the equations of motion derived from the DBI-plus-WZ action, are localized in $\mathbb{R}^4$, and have finite action. We therefore call them $U(1)$ instantons. Our $U(1)$ instantons are singular, characterized by field strengths that blow up near their core. We have two ways to dealing with these singularities. For $n=2,1$, a singularity simply indicates that the probe D$p$-brane, which appears as a defect in $AdS_5$, bends and stretches to spatial infinity in the field theory directions transverse to the defect. Such singularities are hidden at the point at infinity, and thus are physically acceptable. For all of $n=3,2,1$ we have a second option, however: we may hide the singularity in the ball excised by the worldvolume electric flux. The resulting instanton solution is nonsingular everywhere in the physical region outside the ball. In some sense, the effective metric ``de-singularizes'' the instantons simply by excising the region of space where the singularities would otherwise be found. Notice that an \textit{electric flux} de-singularizes our $U(1)$ instantons, in contrast to another well-known mechanism for de-singularizing $U(1)$ instantons, namely spatial non-commutativity~\cite{Nekrasov:1998ss}. For the $n=2,1$ cases, solutions with singularities inside the ball plus those with singularities outside the ball together describe all points on the nonzero-density Higgs branch. For $n=3$ only the solutions with singularities inside the ball describe the nonzero-density Higgs branch.

Our $U(1)$ instantons have no size modulus, however in our solutions free parameters appear that are dual holographically to the expectation values of the gauge-invariant scalar operators that parameterize the nonzero-density Higgs branches. Expressing our solutions as instantons enables us to derive various properties of the solutions easily, including the fact that they do not affect the value of the on-shell D$p$-brane action. Translating to the field theory we learn that, starting from the compressible states studied in refs.~\cite{Karch:2007br,Karch:2008fa}, moving onto the Higgs branch does not change the value of the free energy.

The existence of Higgs branches adds to the growing list of unusual properties of holographic matter, and raises a number of questions. Are these Higgs branches artifacts of the large-$N_c$ and/or large-$\lambda$ limits? Can the moduli possibly survive finite-$N_c$ and/or finite-$\lambda$ corrections, having no obvious symmetry to protect them? What about other types of probe D-branes, in other holographic spacetimes, besides D3/D$p$? How generic is the appearance of a Higgs branch in compressible states of such systems? Can we predict when a Higgs branch will appear? We will not answer these questions, but we will indulge in some speculation about them, and about other questions. For example, for the last question, the crucial role of the WZ terms in our, and many similar, calculations suggests that a classification of compressible states described holographically by probe D-branes with electric flux may be possible, similar to the K-theory classification of D-brane systems describing topological insulators~\cite{Ryu:2010hc,Ryu:2010fe}, which are incompressible states. In addition, we speculate on whether and how the existence of nonzero-density Higgs branches may point the way to a counting of the microstates responsible for the extensive ground state degeneracy in holographic matter.

This paper is organized as follows. In section~\ref{S:4ND} we review cold holographic matter. In sections~\ref{S:d3d7}, \ref{S:d3d5}, and \ref{S:d3d3} we present our instanton solutions and discuss their properties in the D3/D7, D3/D5, and D3/D3 systems, respectively. In section~\ref{S:class} we discuss similar results for systems besides D3/D$p$ and we speculate about a classification of compressible states. We end in section~\ref{S:discuss} with a summary of our results, and speculations about various possible extensions of our work.

\section{Review: Cold Holographic Matter}
\label{S:4ND}

As explained in the introduction, in type IIB string theory we will study the $(n+1)$-dimensional intersection of $N_c$ D3-branes with $N_f$ D$p$-branes, with $p=2n+1=7,5,3$. We call these the D3/D$p$ systems \cite{Karch:2002sh,Karch:2000gx,DeWolfe:2001pq,Erdmenger:2002ex,Constable:2002xt}. We summarize these three systems collectively in the following array:
\begin{center}
\begin{tabular}{c|cccccccccc}
& $x^0$ & $x^1$ & $x^2$ & $x^3$ & $x^4$ & $x^5$ & $x^6$ & $x^7$ & $x^8$ & $x^9$ \\
\hline
D3 & X & X & X & X & & & & & & \\
D7 & X & X & X & X & X & X & X & X & & \\
D5 & X & X & X & & X & X & X &  & & \\
D$3'$ & X & X & & & X & X & & & &
\end{tabular}
\end{center}
For the D3/D3 case, we denote the D$p$-branes as D$3'$-branes to distinguish them from the $N_c$ D3-branes. Strings with both ends on the D3-branes give rise at low energies to $(3+1)$-dimensional $\N=4$ SYM with gauge group $SU(N_c)$ and Yang-Mills coupling squared $g_{YM}^2 = 4 \pi g_s$, with $g_s$ the string coupling. The bosonic symmetry of this theory is $SO(4,2) \times SO(6)$. The $SO(4,2)$ spacetime symmetry is the conformal group in (3+1) dimensions, \textit{i.e.} $\N=4$ SYM is a CFT. In particular, the theory is scale-invariant, hence the beta function of $g_{YM}$ vanishes. $SO(6)$ is the R-symmetry of the theory, corresponding to the $SO(6)$ rotational symmetry in the $(x^4, \ldots, x^9)$ directions.

Each of our D3/D$p$ intersections is a so-called four Neumann-Dirichlet (4ND) intersection, meaning that the open strings between the D3-branes and D$p$-branes have mixed Neumann-Dirichlet boundary conditions in four directions. (Equivalently, each intersection has four directions in which only one type of D-brane is extended.) These intersections preserve half the SUSY that D3-branes alone preserve. In particular, the open strings between the D3-branes and D$p$-branes give rise to $N_f$ hypermultiplets, in the $N_c$ representation of $SU(N_c)$, localized at the $(n+1)$-dimensional intersection. For $n=2,1$ these hypermultiplets thus propagate along a defect in the D3-brane worldvolume theory, for example, for $n=2$ the hypermultiplets propagate along the $\mathbb{R}^{2,1}$ sitting at a fixed value of $x^3$, which we take to be $x^3=0$. For $n=1$, we take the hypermultiplets to propagate along the $\mathbb{R}^{1,1}$ defined by $x^3=0$ and $x^2=0$. Recall that a hypermultiplet includes fermions and their scalar superpartners, which we will call quarks and squarks, respectively.

For $n=3$, the $SO(4,2)$ spacetime symmetry of $\N=4$ SYM is broken to $SO(3,1)$, and in particular scale invariance is broken. For $n=2$, the $SO(4,2)$ is broken to $SO(3,2)$, the subgroup of conformal transformations that leaves the subspace $x^3=0$ invariant. In this case, scale invariance is preserved: the beta function of $g_{YM}$ remains zero. For $n=1$, the conjecture of ref.~\cite{Constable:2002xt} is that $SO(4,2)$ is broken to $SO(2,2)$, and again scale invariance is preserved. The couplings of the hypermultiplets break the $SO(6)$ R-symmetry of $\N=4$ SYM down to $SO(n+1) \times SO(5-n)$, corresponding to rotations in the $(x^4,\ldots,x^{4+n})$ and $(x^{5+n},\ldots,x^9)$ subspaces, respectively. A subgroup of the $SO(n+1) \times SO(5-n)$ is the R-symmetry of the remaining SUSY.

Separating the D$p$-branes from the D3-branes in a mutually transverse direction will give the open strings stretched between them a nonzero length, and hence will give the hypermultiplets a nonzero mass. Such a deformation will break part of the $SO(5-n)$ symmetry, as is obvious from the array above. In what follows, we will choose not to perform any such mass deformation: we work only with massless hypermultiplets until section~\ref{S:discuss}, where we discuss how to include a nonzero hypermultiplet mass in our analysis.

In addition to the global symmetries discussed above, $\N=4$ SYM theory with $N_f$ massless hypermultiplets enjoys a global $U(N_f)$ flavor symmetry. We will call the overall diagonal $U(1)$ subgroup baryon number, and denote the associated conserved current as $J^{\mu}$ with $\mu = 0,\ldots, n$. We will study states with a nonzero baryon number charge density, that is, states with nonzero $\jt$. Such a nonzero $\jt$ is produced by a nonzero density of strings stretched between the D3-branes and the D$p$-brane. Notice that both the quarks and the squarks are charged under baryon number.

To describe these theories holographically, we will take the Maldacena and probe limits. The Maldacena limits are the 't Hooft large-$N_c$ limit, $N_c \to \infty$ and $g_{YM}^2 \to 0$ with the 't Hooft coupling $\lambda \equiv g_{YM}^2 N_c$ fixed, followed by taking $\lambda \gg 1$. The probe limit consists of keeping $N_f$ fixed as $N_c \to \infty$, expanding all observables in the small parameter $N_f/N_c$, and only retaining terms up to order $N_f N_c$. Indeed, starting now we will take $N_f=1$. Roughly speaking, the probe limit consists of neglecting quantum effects due to the flavor fields. For example, in the $n=3$ case, in the probe limit we neglect the flavor contribution to the beta function of $\lambda$. In these limits, all three of the D3/D$p$ theories are scale-invariant, so the only scales in any problem are those that we introduce by hand. In what follows, the only scale we will introduce in the field theory is $\jt$, or equivalently the chemical potential $\mu$.

$\N=4$ SYM in the Maldacena limits is dual to type IIB supergravity in the near-horizon geometry of the D3-branes, $AdS_5 \times \mathbb{S}^5$ with $N_c$ units of RR five-form flux on the  $\mathbb{S}^5$. We write 
the $AdS_5 \times \mathbb{S}^5$ metric and the RR five-form $F_5$ as
\begin{subequations}
\beq
\label{eq:metric}
ds^2 = Z^{-1/2}(r) \eta_{\mu\nu} dx^{\mu} dx^{\nu} + Z^{1/2}(r) \left( dr^2 + r^2 ds_{\mathbb{S}^5}^2 \right), \qquad Z(r) \equiv R^4/r^4,
\eeq
\beq
\label{eq:F5}
F_{5} = \frac{4}{R} \left( \textrm{vol}_{AdS_5} + \textrm{vol}_{\mathbb{S}^5} \right),
\eeq
\end{subequations}
where in eq.~\eqref{eq:metric} $\mu,\nu \in \{ 0,\dots,3 \}$, $\left[\eta_{\mu\nu}\right] = \textrm{diag}(-1,1,1,1)$, $r$ is the $AdS_5$ radial coordinate, with the boundary at $r \to \infty$ and the Poincar\'e horizon at $r\to 0$, and $ds^2_{\mathbb{S}^5}$ is the metric of a round $\mathbb{S}^5$. The radius of curvature of $AdS_5$ and $\mathbb{S}^5$ is $R$, where $R^4 = 4 \pi g_s N_c \alpha'^2$, with $\alpha'$ the string length squared. In eq.~\eqref{eq:F5}, $\textrm{vol}_{AdS_5}$ and $\textrm{vol}_{\mathbb{S}^5}$ denote the volume forms of $AdS_5$ and $\mathbb{S}^5,$ respectively. For later use, we will also define a RR four-form potential $C_4$ via $F_5 = d C_4$. We will choose a gauge such that $\left(C_4\right)_{0123} = Z(r)^{-1}$, which is the only component of $C_4$ that we will need. Starting now, we use units in which $R \equiv 1$.

The probe flavor degrees of freedom appear in the bulk as probe D$p$-branes, with $p=2n+1$, extended along $AdS_{n+2} \times \mathbb{S}^n$. The D$p$-brane action, $S_p$, is the sum of a DBI term, $S_{DBI}$, and WZ terms, $S_{WZ}$, which in our cases take the form
\begin{subequations}
\begin{eqnarray}\label{TotS}
S_{p} &=& S_{DBI} + S_{WZ},\\
S_{DBI} &=& - T_{p} \int d^{p+1} \xi \sqrt{- \det\left( P[G]_{ab} + F_{ab} \right)}, \\
S_{WZ} &=& T_{p} \int P[C_4] \wedge e^F,
\end{eqnarray}
\end{subequations}
where $T_p$ is the D$p$-brane tension, $T_p = g_s^{-1} (2\pi\sqrt{\alpha'})^{-(p+1)}$, $\xi^a$ denote the worldvolume coordinates, with $a=1,\dots,p+1$, $P[G]_{ab}$ and $P[C_4]$ are the pullbacks of the metric $G$ and of $C_4$ to the D$p$-brane worldvolume, and $F=dA$ is the field strength of the $U(1)$ worldvolume gauge field. Notice that compared to the usual convention (for example that of ref.~\cite{Polchinski:1998rq}), we have absorbed a factor of $(2\pi\alpha')$ into $F$.

To specify the embedding of each D$p$-brane into $AdS_5 \times \mathbb{S}^5$ we must specify its position in directions transverse to its worldvolume. The transverse directions appear in the D$p$-brane worldvolume theory as scalar fields, via the pullbacks $P[G]_{ab}$ and $P[C_4]$. To make the $SO(n+1)\times SO(5-n)$ isometries preserved by the D$p$-brane explicit, we re-write the part of the metric transverse to the D$3$-branes as
\beq
dr^2 + r^2 ds_{\mathbb{S}^5}^2 = d\rho^2 + \rho^2 ds^2_{\mathbb{S}^n} + \sum\limits_{M=1}^{5-n} (dy^M)^2, \qquad r^2 = \rho^2 + \sum\limits_{M=1}^{5-n} (y^M)^2.
\eeq
In these coordinates, each D$p$-brane is extended along the $\mathbb{S}^n$, the radial direction $\rho$, the time direction, and $n$ spatial directions. The radial coordinate $\rho$ and the $\mathbb{S}^n$ together span $\mathbb{R}^{n+1}$; this space will play a crucial role in what follows. The worldvolume scalars include the $y^M$ in the internal space as well as, for $p=5$, the transverse direction $x^3$ in $AdS_5$, and for $p=3$, the transverse directions $x^3$ and $x^2$.

We now need an ansatz for the worldvolume fields. The baryon number charge density operator $J^0$ is dual to $A_0$, so to describe nonzero-density states in our system we must introduce a nonzero $A_0$. We will impose a number of symmetries on these nonzero-density states, which will constrain our ansatz. We will demand time-translation symmetry. In the $(x^1,\dots,x^n)$ directions, we will demand translational, rotational, and parity invariance. For the $n=2,1$ cases we also demand reflection symmetry about the defect in the $(x^{n+1},..,x^{3})$ directions. We then cannot allow any worldvolume fields to depend on the directions of $\mathbb{R}^{n,1}$, nor can we introduce a constant magnetic field in the $\mathbb{R}^{n,1}$ directions. Furthermore, we will demand that the $SO(n+1) \times SO(5-n)$ symmetry remain unbroken. The $SO(n+1)$ symmetry forbids any worldvolume field from depending on the $\mathbb{S}^n$ coordinates, and forces the components of the worldvolume gauge field on the $\mathbb{S}^n$ to vanish. The $SO(5-n)$ symmetry forces $y^M=0$ for all $M$. Notice that $y^M=0$ describes massless hypermultiplets: in the original D3/D$p$ intersection, nonzero $y^M$ imply a nonzero separation between the D-branes, while in holographic terms, the worldvolume scalar $\sqrt{\sum_M^{5-n} (y^M)^2}$ is dual to the hypermultiplet mass operator. These symmetries allow only $A_0$ to be nonvanishing, with dependence only on $\rho$.

With our ansatz, in the action $S_p$ the integration over the directions $x^0,x^1,\ldots,x^n$ trivially produces a factor of $\textrm{vol}(\mathbb{R}^{n,1})$, the volume of $\mathbb{R}^{n,1}$, so for convenience we will define an action density $s_p \equiv S_p/ \textrm{vol}(\mathbb{R}^{n,1})$.\footnote{To be explicit, we choose static gauge, $\xi^1 = x^0$, and identify all remaining worldvolume coordinates with those of the $AdS_{n+2} \times \mathbb{S}^n$ submanifold of the background geometry. Furthermore, we work in a gauge with $A_{\rho}=0$, so that $F_{\rho 0} = \partial_{\rho} A_0$.} Inserting our ansatz into $s_p$, we find
\beq
\label{smallS}
s_p = - T_p \text{vol}(\mathbb{S}^n)\int d\rho \,\rho^n \sqrt{1  - A_0'(\rho)^2},
\eeq
where $\text{vol}(\mathbb{S}^n)$ is the volume of a unit-radius $S^n$ and the prime denotes $\partial_{\rho}$. Since $s_p$ depends only on the derivative $A_0'(\rho)$, we find a conserved quantity $d$,
\beq
\label{eq:density}
d \equiv \frac{\delta s_p}{\delta A_0'(\rho)}.
\eeq
The conserved quantity $d$ is related to the baryon number density $\langle J^0 \rangle$ as $\langle J^0 \rangle = (2\pi\alpha')  d$ \cite{Kobayashi:2006sb,Karch:2007br}. We can easily solve eq.~\eqref{eq:density} for $A_0'(\rho)$ in terms of $d$. Defining
\beq
\label{eq:rho0}
\rho_0^{2n} \equiv \frac{d^2}{T_p^2 \, \textrm{vol}(\mathbb{S}^n)^2},
\eeq
the solution for $A_0'(\rho)$ can be expressed as
\beq
\label{eq:solA0}
A_0'(\rho) = \frac{1}{\sqrt{1 + \rho^{2n}/\rho_0^{2n}}}.
\eeq
Eq.~\eqref{eq:solA0} can be integrated, with a boundary condition $A_0(\rho=0)=0$, giving a solution for $A_0(\rho)$ in terms of a hypergeometric function, but in what follows we will only need $A_0'(\rho)$.

In solving for $A_0'(\rho)$, we have essentially solved an electrostatics problem, with a DBI action, with no explicit source charges. By Gauss's law, however, we know that any nonzero electric flux must be produced by some source charges. Indeed, we can easily locate the source charges in our case: for our solution of $A_0(\rho)$, we impose $A_0(\rho=0)=0$, but from eq.~\eqref{eq:solA0} we can see that the derivative $A_0'(\rho=0)=1$ is nonzero, indicating a kink singularity in the solution which corresponds to a delta-function source at $\rho=0$. Physically, that source is a density of strings.  If we imagine that the $N_c$ D3-branes producing the background geometry and RR five-form flux are ``hiding'' behind the Poincar\'e horizon, that is, that they are sitting at the ``bottom'' of $AdS_5$, $r=0$, then a nonzero density of strings with one end on the $N_c$ D3-branes and one end on the D$p$-brane can similarly ``hide'' behind the Poincar\'e horizon, and their endpoints on the D$p$-brane will act as a source of electric flux on the D$p$-brane worldvolume. In our case, these strings are uniformly distributed on $\mathbb{R}^{n,1}$. In section~\ref{S:d3d5} we will argue that we should in fact excise the Poincar\'e horizon from the D$p$-brane worldvolume. Recall that the Poincar\'e horizon corresponds to the ``point at infinity'' in the field theory, so excising the Poincar\'e horizon from the D$p$-brane, and hence any sources sitting at the point at infinity, is natural. Upon excising the Poincar\'e horizon, the density of strings is truly ``hidden'': the string endpoints which would source the electric flux are absent from the spacetime, and only their flux remains. As a result, we regard the kink singularity in $A_0(\rho)$ as a boundary condition that we impose at $\rho=0$.

Our solution for $A_0(\rho)$ teaches us a general lesson about sources for worldvolume fields: if a solution for the worldvolume fields has a kink singularity corresponding to a source at the point at infinity, we may excise the source and replace it with boundary conditions. We will present a rigorous argument for this in section~\ref{S:d3d5}. Notice that we can reach the point at infinity in two ways. The first is to fix values of $(x^1,x^2,x^3)$ and then take $\rho \to 0$, in which case we approach the Poincar\'e horizon. The second way is to fix $\rho$ and take any of $(x^1,x^2,x^3)$ to infinity. In sections~\ref{S:d3d7},~\ref{S:d3d5}, and~\ref{S:d3d3}, we will encounter various types of sources sitting at the point at infinity, approached in both ways. We will always excise such sources.

Plugging the solution for $A_0'(\rho)$ in eq.~\eqref{eq:solA0} back into the action eq.~\eqref{smallS} and performing the $\rho$ integration (suitably regulating divergences \cite{Karch:2005ms,Karch:2007br}), we obtain the on-shell D$p$-brane action. Via the AdS/CFT dictionary, the on-shell D$p$-brane action is equivalent to \textit{minus} the order $N_f N_c$ contribution to the field theory free energy in the grand canonical ensemble, \textit{i.e.} the grand potential. A Legendre transform then gives us the free energy in the canonical ensemble, \textit{i.e.} the Helmholtz free energy. Crucially, notice that, because of a minus sign, \textit{maximizing} the D$p$-brane action corresponds to \textit{minimizing} the Helmholtz free energy.

As shown in ref.~\cite{Karch:2007br}, the nonzero-density states of the D3/D$p$ systems described by the solution for $A_0'(\rho)$ above are indeed compressible. These states have been studied intensively in refs.~\cite{Nakamura:2006xk,Kobayashi:2006sb,Karch:2007br,Karch:2008fa,Kulaxizi:2008kv,Myers:2008me, Nickel:2010pr,Ammon:2011hz,Davison:2011ek,Ammon:2012je}, and exhibit many unusual properties. To date, no Fermi surface has been detected in these states, so they do not appear to be Landau Fermi liquids. Moreover, to date no evidence has been found to indicate that the baryon number $U(1)$ is spontaneously broken in these states, so they do not appear to be Bose liquids either.

These compressible states actually have a nonzero extensive ground state degeneracy~\cite{Karch:2008fa}, \textit{i.e.} a nonzero thermodynamic entropy density proportional to (up to purely numerical prefactors) $d \propto \jt/\sqrt{\lambda}$. Such a degeneracy suggests instability, since generically we expect any perturbation to break the degeneracy and drive the system to a new, presumably non-degenerate, ground state. These states are known to be stable against thermodynamic fluctuations, however, in the sense that the Hessian of the free energy in the space of temperature $T$ and chemical potential $\mu$ has non-negative eigenvalues~\cite{Benincasa:2009be}. We hasten to add that both the nonzero entropy and the thermodynamic stability may be artifacts of the large-$N_c$ and/or large-$\lambda$ limits. For example, corrections in $N_c$ and/or $\lambda$ may lift the extensive degeneracy of states.

For the D3/D7 system, this compressible state is also stable against dynamical (\textit{i.e.} finite-frequency and finite-momentum) fluctuations: a holographic calculation revealed that the spectrum of excitations about this state is tachyon-free \cite{Ammon:2011hz}. Most remarkably, the spectrum appears to be controlled by a mysterious (0+1)-dimensional CFT~\cite{Jensen:2010ga,Nickel:2010pr,Ammon:2011hz}. The spectrum includes a sound mode with a dispersion relation nearly identical in form to that of Landau's zero sound mode in a Landau Fermi liquid, although given the many differences with a Landau Fermi liquid, the similarity is almost certainly superficial.

Crucially for us, in the D3/D7 system the spectrum of fluctuations about this compressible state also includes purely imaginary modes that suggest the existence of a moduli space. More precisely, in these compressible states of the D3/D7 theory, holographic calculations reveal poles in the retarded two-point functions of certain scalar operators dual to certain Kaluza-Klein modes of the D7-brane gauge field components on the $\mathbb{S}^3$ wrapped by the D7-brane. We call these operators $\mathcal{O}_l^-$ in section~\ref{S:d3d7}. The poles are identical in form to the diffusive pole in the retarded two-point function of a conserved current in standard hydrodynamics: for frequency $\omega$ and momentum $k$, the poles occur when $\omega = - i D k^2$, with some ``diffusion constant'' $D$. In ref.~\cite{Ammon:2011hz} these poles were dubbed ``R-spin diffusion'' modes, and for a few of them the value of $D$ was computed numerically. Notice that $D \propto 1/\mu$ follows simply from dimensional analysis. The existence of these gapless modes suggests that if we deform the theory in these directions in field space, that is, if we give these operators nonzero expectation values, then so long as those expectation values are constant in space ($k=0$) such deformations will cost zero energy ($\omega=0$). In other words, the existence of these gapless modes suggests the existence of a moduli space.\footnote{We give many thanks to Dam Son for pointing this out to us.} We will prove in the next section that such a nonzero-density moduli space indeed exists, by finding explicit solutions for the worldvolume fields dual to the $\mathcal{O}_l^-$ describing points on that moduli space. For the D3/D5 and D3/D3 systems we will also find nonzero-density moduli spaces, which implies the existence of ``R-spin modes'' in those systems as well.

\section{D3/D7 and Self-dual Instantons}
\label{S:d3d7}

In the D3/D7 system~\cite{Karch:2002sh},
\begin{center}
\begin{tabular}{c|cccccccccc}
& $x^0$ & $x^1$ & $x^2$ & $x^3$ & $x^4$ & $x^5$ & $x^6$ & $x^7$ & $x^8$ & $x^9$ \\
\hline
D3 & X & X & X & X & & & & & & \\
D7 & X & X & X & X & X & X & X & X & & \\
\end{tabular}
\end{center}
the flavor fields break the $SO(6)$ R-symmetry down to $SO(4) \times SO(2)$. An $SU(2) \subset SO(4)$ combines with the $SO(2)$ to form the remaining R-symmetry of the theory. The $SO(4) \times SO(2)$ symmetry corresponds to rotations in the $(x^4,\dots,x^7)$ and $(x^8,x^9)$ directions, respectively. We will re-label the coordinates $(x^4,\dots,x^7)$, along the D7-brane but transverse to the D3-branes, as $z^i$ with $i=1,\dots,4$. 

 In the Maldacena and probe limits, we obtain a probe D7-brane extended along $AdS_5 \times \mathbb{S}^3$ inside $AdS_5 \times \mathbb{S}^5$. The D7-brane action takes the form
\beq
\label{D7action0}
S_{7}=-T_{7}\int d^{8}\xi \sqrt{\det(-P [G]_{ab}+ F_{ab})} + \frac{1}{2}T_{7} \int P[C_{4}]\wedge F\wedge F.
\eeq
Notice in particular the form of the WZ term in this case, which involves $F \wedge F$. For the D7-brane worldvolume fields we will consider an ansatz more general than that of section~\ref{S:4ND}: we will demand all the same symmetries as in section~\ref{S:4ND}, except we will \textit{not} impose the $SO(4)$ symmetry. The most general ansatz we can then write is
\beq
\label{D7fields}
A(\xi) = A_0(z) dx^0 + A_i(z) dz^i,
\eeq
with all other worldvolume fields vanishing. Given that the ansatz in eq.~\eqref{D7fields} is more general than that of section \ref{S:4ND}, solutions of the form in eq.~\eqref{D7fields} have the potential to describe many different kinds of field theory states, not just compressible states. We will specialize to compressible states later; for now, we will keep our analysis as general as possible. Upon inserting eq.~\eqref{D7fields} into eq.~\eqref{D7action0}, we can write the D7-brane action density as
\beq
\label{D7action1}
s_{7} =  - T_{7} \int d^4z\left[\sqrt{\det(g_{ij} +Z^{-1/2} f_{ij})} - \frac{1}{8}Z^{-1}\tilde{\epsilon}^{ijkl} f_{ij} f_{kl}\right],
\eeq
where $\tilde{\epsilon}^{ijkl}$ is the Levi-Civita symbol (a tensor density), with $\tilde{\epsilon}^{1234} \equiv +1$, and we have defined the effective metric\footnote{Our effective metric may formally be written as  $g_{ij} \propto P[G]_{ij} - F_{i\mu}P[G]^{\mu\nu}F_{\nu j}$, where $\mu,\nu={0,1,2,3}$, which is different from the open string metric, $P[G]_{ij} - F_{ia}P[G]^{ab}F_{bj}$ \cite{Seiberg:1999vs}, since in the latter $a,b$ run over all eight worldvolume directions of the D7-brane.} and field strength in the $\mathbb{R}^4$ spanned by $\rho$ and the $\mathbb{S}^3$:
\beq
\label{gij}
g_{ij} \equiv \delta_{ij}- \partial_i A_0 \partial_j A_0, \qquad f_{ij} \equiv \partial_i A_j - \partial_j A_i.
\eeq
The factor $Z$ appearing in eq.~\eqref{D7action1} is the warp factor appearing in the background metric, eq.~\eqref{eq:metric}, evaluated on the D7-brane worldvolume: $Z=1/\rho^4$.

Our task is to find solutions for $A_0(z)$ and $A_i(z)$, a problem formally similar to that of static four-dimensional DBI electromagnetism in a curved background geometry. We can simplify our task greatly by exploiting two useful results of ref.~\cite{Gibbons:2000mx}. In fact, the arguments of ref.~\cite{Gibbons:2000mx} are very general, so let us state them in general terms, and then apply them to our system. Consider a four-dimensional Riemannian manifold with metric $\mathcal{G}_{ij}$, where $i,j=1,\dots,4$, and an antisymmetric matrix $\mathcal{F}_{ij}$. We begin with Minkowski's inequality,
\beq
\label{mink1}
\left[\det \left(\mathcal{G}_{ij} + \mathcal{F}_{ik} \mathcal{F}^k_{~j}\right)\right]^{1/4} \geq \left[\det \mathcal{G}_{ij}\right]^{1/4} + \left[\det \mathcal{F}_{ik} \mathcal{F}^k_ {~j}\right]^{1/4}.
\eeq
Next we bring the left-hand-side of eq.~\eqref{mink1} into a form useful to us. Using the (anti)symmetry properties of $\mathcal{G}_{ij}$ and $\mathcal{F}_{ij}$, we can easily show that
\beq
\det \left(\mathcal{G}_{ij} + \mathcal{F}_{ij}\right) = \det \left(\mathcal{G}_{ij} -\mathcal{F}_{ij} \right),
\eeq
which then implies
\beq
\label{detsquare}
\left[\det \left(\mathcal{G}_{ij} + \mathcal{F}_{ij}\right)\right]^2 = \det \left(\mathcal{G}_{ij} + \mathcal{F}_{ij}\right) \det \left(\mathcal{G}_{ij} -\mathcal{F}_{ij} \right) = \det \left(\mathcal{G}_{ij} + \mathcal{F}_{ik} \mathcal{F}^k_{~j}\right) \det \mathcal{G}_{ij}.
\eeq
Using the result of eq.~\eqref{detsquare} to re-write the left-hand-side of eq.~\eqref{mink1}, we find
\beq
\sqrt{\det\left( \mathcal{G}_{ij} + \mathcal{F}_{ij} \right)} \geq \sqrt{\det \mathcal{G}_{ij}} + \sqrt{\det \mathcal{F}_{ij}}.
\eeq
Next we observe that
\beq
\sqrt{\det \mathcal{F}_{ij}} = \frac{1}{8} \left|\tilde{\epsilon}^{ijkl}\mathcal{F}_{ij}\mathcal{F}_{kl}\right|,
\eeq
which is a topological invariant. We thus arrive at the first useful result,
\beq
\label{mink2}
\sqrt{\det\left( \mathcal{G}_{ij} + \mathcal{F}_{ij} \right)} \geq \sqrt{\det \mathcal{G}_{ij}} +  \frac{1}{8} \left|\tilde{\epsilon}^{ijkl}\mathcal{F}_{ij}\mathcal{F}_{kl}\right|.
\eeq
Now for the second useful result: Minkowski's inequality is saturated when $\mathcal{F}_{ik} \mathcal{F}^k_{~j} \propto \mathcal{G}_{ij}$, which implies that $\mathcal{F}_{ij}$ is (anti-)self-dual with respect to $\mathcal{G}_{ij}$,
\beq
\mathcal{F}_{ij} = \pm \frac{1}{2} \epsilon_{ijkl} \mathcal{F}^{kl}, \qquad \epsilon^{ijkl} \equiv \tilde{\epsilon}^{ijkl}/\sqrt{\det \mathcal{G}_{ij}}.
\eeq
(Notice that $\epsilon^{ijkl}$ is the Levi-Civita tensor, not the tensor density.) The bottom line is that, in a given topological sector, (anti-)self-dual $\mathcal{F}_{ij}$'s saturate the inequality in eq.~\eqref{mink2}. Following ref.~\cite{Gibbons:2000mx}, we will refer to the inequality in eq.~\eqref{mink2} as a ``topological bound.''

To apply these results to our system, we simply take $\mathcal{G}_{ij} \to g_{ij}$ and $\mathcal{F}_{ij} \to Z^{-1/2} f_{ij}$. We thus find that the D7-brane action density obeys a bound,
\beq
\label{d7bound}
s_{7} \leq -T_{7}\int d^4z \left[ \sqrt{\det g_{ij}}+\frac{1}{8} Z^{-1}\left(|\tilde{\epsilon}^{ijkl}f_{ij}f_{kl}|-\tilde{\epsilon}^{ijkl}f_{ij}f_{kl} \right)\right].
\eeq
Due to the sign of the WZ term, the bound in eq.~\eqref{d7bound} is only saturated for $f_{ij}$ that are \textit{self-dual} with respect to $g_{ij}$,
\beq
\label{selfdual}
 f_{ij}=+\frac{1}{2}\epsilon_{ijkl} f^{kl}, \qquad \epsilon^{ijkl} \equiv \tilde{\epsilon}^{ijkl}/\sqrt{\det g_{ij}}.
\eeq
For such self-dual $f_{ij}$ the D7-brane action density reduces to
\beq
\label{E:simpleS1}
s_{7}= -T_{7}\int d^4z \sqrt{\det g_{ij}},
\eeq
which is independent of $A_i(z)$, due to the cancellation between the DBI and WZ terms involving $f_{ij}$ in eq.~\eqref{d7bound}\footnote{This cancellation was previously observed for the non-Abelian D7-brane action, expanded to second order in the field strength in refs.~\cite{Guralnik:2004ve,Guralnik:2004wq,Guralnik:2005jg}, as well to all orders in the field strength with the symmetrized trace prescription in ref.~\cite{Chen:2009kx}.}.

Self-dual $f_{ij}$ extremize the action, and thus also solve the equations of motion for the $A_i(z)$. That simplifies our task greatly: we do not need to solve the full, non-linear, equations for $A_i(z)$, but only the self-duality condition in eq.~\eqref{selfdual}, which is linear. The task of solving for $A_0(z)$ is also greatly simplified. Given that self-dual $f_{ij}$ extremize the action, and that the result, eq.~\eqref{E:simpleS1}, is independent of the $A_i(z)$, the only non-trivial equations of motion are those that follow from eq.~\eqref{E:simpleS1}. In other words, the equation of motion for $A_0(z)$ derived from eq.~\eqref{D7action1}, when evaluated on self-dual $f_{ij}$, reduces to the equation of motion for $A_0(z)$ derived from eq.~\eqref{E:simpleS1}. In effect, then, when solving for $A_0(z)$ we can ignore $f_{ij}$.

We can thus write a simple two-step recipe. The first step is to solve the equation of motion for $A_0(z)$ derived from eq.~\eqref{E:simpleS1}, which is simply a problem in DBI electrostatics. That solution then determines the effective metric in eq.~\eqref{gij}, which in turn determines the self-duality condition for $f_{ij}$ in eq.~\eqref{selfdual}. The second step is to solve this self-duality condition. The self-duality condition is linear, so if we obtain multiple solutions to it then we may obtain new solutions simply by linear superposition. Moreover, if the effective metric has some isometries, then we may obtain new solutions by acting on known solutions with the isometries. For any solutions $A_0(z)$ and $A_i(z)$ obtained by this recipe, the on-shell action will be sensitive only to $A_0(z)$.

Not only do self-dual $f_{ij}$ contribute nothing to the on-shell action, but also they contribute nothing to the D7-brane's stress-energy tensor, as shown in ref.~\cite{Gibbons:2000mx}. More generally, however, we expect variational derivatives of the action, when evaluated on a solution, to depend on both $A_0(z)$ and $A_i(z)$, \textit{i.e.} the action itself and the stress-energy tensor are special cases.

Self-dual $f_{ij}$ have a simple interpretation in terms of D-brane physics. The instanton number density on the D7-brane worldvolume, which is proportional to $F \wedge F$, acts as a source for $C_4$, hence any solution with nonzero $F \wedge F$ endows the D7-brane with some D3-brane charge density, and generically represents D3-branes dissolved into the D7-brane. Self-dual $f_{ij}$ have nonzero, positive instanton number density. Indeed, in our conventions, self-dual $f_{ij}$ correspond to dissolved D3-branes while anti-self-dual $f_{ij}$ correspond to dissolved anti-D3-branes. By charge conservation, any dissolved D3-branes must come from the $N_c$ D3-branes producing the background geometry and RR flux.

Let us illustrate our recipe first using the trivial solution $A_0(z)=0$. In this case the dual field theory has no SUSY Higgs branch, so our goal is in fact to demonstrate the \textit{absence} of non-trivial, regular, self-dual $f_{ij}$. Taking $A_0(z)=0$, the effective metric in eq.~\eqref{gij} reduces to the flat metric on $\mathbb{R}^4$. Because the self-duality equation is invariant under the $SO(4)$ isometries of $\mathbb{R}^4$, we may decompose the gauge field into vector spherical harmonics. These fall into two types, distinguished by how they transform under $SO(4)\approx SU(2)\times SU(2)$. The first type, the ``genuine'' vector harmonics $\mathcal{Y}^{l,\pm}_{\alpha}$, transform under the $\left(\frac{l\mp 1}{2},\frac{l\pm 1}{2}\right)$-representation, where the integer $l$ satisfies $l\geq 1$ and where $\alpha=1,2,3$ labels the angles of the $\mathbb{S}^3$. The $\mathcal{Y}^{l,\pm}_{\alpha}$ satisfy (among other things)
\beq
\label{eqy}
\epsilon^{\alpha\beta\gamma}\partial_{\beta}\mathcal{Y}_{\gamma}^{l,\pm} = \pm (l+1)\mathcal{Y}^{l,\pm, \alpha}\,,
\eeq
where $\epsilon^{\alpha\beta\gamma}$ is the Levi-Civita tensor of $\mathbb{S}^3$, and the Greek indices are raised and lowered with the unit $\mathbb{S}^3$ metric. The second type of vector spherical harmonics are those built by taking derivatives of the scalar spherical harmonics. We will not need the explicit form for this second type of vector spherical harmonic, because the only non-trivial solutions to the self-duality condition eq.~\eqref{selfdual} are those built from the genuine vector spherical harmonics. For a given $\mathcal{Y}^{l,\pm}_{\alpha}$, we straightforwardly find a solution to eq.~\eqref{selfdual} $A_{\alpha}(z) \propto \rho^{\pm (l+1)} \mathcal{Y}^{l,\pm}_{\alpha}$, with $A_{\rho}(z)=0$. The solutions $\propto \rho^{l+1}$ blow up at large $\rho$, so these solutions are non-normalizable near the $AdS_5$ boundary. In field theory terms, these solutions can describe external sources for the scalar operators dual to the $A_{\alpha}$. The solutions $\propto \rho^{-(l+1)}$ approach zero as $\rho \to \infty$ and thus are normalizable; these solutions can describe states in which the dual scalar operators have nonzero expectation values. Notice, however, that the solutions $\propto \rho^{-(l+1)}$ diverge as $\rho\to 0$. By linear superposition of these solutions, we can construct a more general solution,
\beq
\label{E:selfdualSoln}
A_{\alpha}(z) = \sum_{l=1}^{\infty}\left( b_l\, \rho^{l+1}\mathcal{Y}^{l,+}_{\alpha}+\frac{c_l}{\rho^{l+1}}\mathcal{Y}^{l,-}_{\alpha}\right), \qquad A_{\rho}(z)=0\,,
\eeq
where the $b_l$ and $c_l$ are arbitrary constants. The self-duality condition is invariant under the isometries of the flat metric, so we can also obtain new solutions by acting on known solutions with those isometries. Indeed, by applying isometries to the solution in eq.~\eqref{E:selfdualSoln}, we can construct \textit{the most general self-dual solutions} to the $A_i(z)$ equations of motion. 

A generic self-dual gauge field of the form in eq.~\eqref{E:selfdualSoln} will be localized around a set of singularities in $\mathbb{R}^4$. Since this gauge field configuration is localized and has finite action eq.~\eqref{D7action0}, we call it a $U(1)$ instanton. The presence of singularities in the gauge field implies two things: first, the $U(1)$ instanton has no size modulus (it has shrunk to zero size), and second the instanton number density $F\wedge F$, which is the local D3-brane charge density on the D7-brane, diverges near the singularity. The singularities are physical: not only the gauge field but also the field strength, and its derivatives, diverge at the singularities, so these singularities cannot be removed by any gauge transformation. As a result the $U(1)$ instantons we have found are inadmissible within our approximations (the Maldacena and probe limits), so we must discard them. The only potential subtlety arises for $U(1)$ instantons which are singular at the point at infinity, meaning either at $\rho=0$ or at nonzero $\rho$ but with any of $(x^1,x^2,x^3)$ going to infinity. As we will argue in section~\ref{S:d3d5}, we should excise this point at infinity from the D7-brane worldvolume. In that case, such a $U(1)$ instanton, while non-singular on the worldvolume, is inconsistent with the boundary condition of regularity at the point at infinity, and so again is inadmissible. We have thus demonstrated the \textit{absence} of any non-trivial, regular, self-dual $f_{ij}$, consistent with the absence of a zero-density Higgs branch in the D3/D7 theory.

Let us now specialize to solutions for $A_0(z)$ and $A_i(z)$ that describe compressible states. Following our recipe, we first need a solution for $A_0(z)$ that describes a compressible state. We have such a solution, namely the one in section~\ref{S:4ND}, eq.~\eqref{eq:solA0} with $n=3$,
\beq
\label{eq:solA02}
A_0'(\rho) = \frac{1}{\sqrt{1+\rho^6/\rho_0^6}}\,, \qquad \qquad \rho_0^6 \equiv \frac{d^2}{T_7^2 \, \textrm{vol}(\mathbb{S}^3)^2}\,.
\eeq
Inserting the solution for $A_0'(\rho)$ from eq.~\eqref{eq:solA02} into the $g_{ij}$ in eq.~\eqref{gij}, we find the effective metric
\beq
\label{gij1}
g_{ij} \, dz^i dz^j = \frac{\rho^6}{\rho^6 + \rho_0^6} \, d\rho^2 + \rho^2 ds^2_{\mathbb{S}^3}\,.
\eeq
If we change to a different radial coordinate,\footnote{In anticipation of the D3/D5 and D3/D3 systems, let us record the appropriate change of radial coordinate and the resulting conformal factor for the D$p$-brane, with $p=2n+1$,
\beq
\label{rhobar}
\bar{\rho} \equiv \rho \left(\frac{1+\sqrt{1+\rho_0^{2n}/\rho^{2n}}}{2}\right)^{1/n}, \qquad \Omega(\bar{\rho}) = \left(1-\frac{{\rho}_0^{2n}}{4\bar{\rho}^{2n}}\right)^{1/n}\,.
\eeq
Additionally, the Ricci scalar of the effective metric is $n(n+1)\rho_0^{2n}/\rho^{2n+2}$.}
\beq
\label{E:d3d7rhobar}
\bar{\rho} \equiv \rho\left(  \frac{1+\sqrt{1+{\rho}_0^6/\rho^6}}{2}\right)^{1/3}\,,
\eeq
then the effective metric in eq.~\eqref{gij1} becomes conformally equivalent to the flat metric,
\beq
\label{gij2}
g_{ij} \, dz^i dz^j = \Omega(\bar{\rho})^2 \left( d\bar{\rho}^2 + \bar{\rho}^2 ds^2_{\mathbb{S}^3}\right)\,, \qquad \Omega(\bar{\rho}) = \left(1-\frac{{\rho}_0^6}{4\bar{\rho}^6}\right)^{1/3}\,.
\eeq
The radial coordinate $\rho$ is valued on the positive real line, $\rho\in \mathbb{R}^+$, while from eq.~\eqref{E:d3d7rhobar} we see that $\bar{\rho}\in[2^{-1/3}\rho_0,\infty)$. The conformal factor $\Omega(\bar{\rho})$ vanishes at the lower endpoint $\bar{\rho}=2^{-1/3}\rho_0$, so the effective metric is actually conformally equivalent to $\mathbb{R}^4 \backslash \mathbb{B}^4$, that is, $\mathbb{R}^4$ with all points inside a four-ball $\mathbb{B}^4$ of radius $2^{-1/3}\bar{\rho}_0$ excised. The existence of this $\mathbb{B}^4$ is a uniquely \textit{stringy} effect, corresponding to the fact that the D7-brane's effective tension goes to zero deep in the bulk. The effective metric is actually singular at $\rho=0$: its Ricci scalar is $+12\rho_0^6/\rho^8$. In what follows, this singularity will not produce any singularities in any physical quantity that we will study, either in the bulk or in the field theory. The same statements will apply for the D3/D5 and D3/D3 systems in sections~\ref{S:d3d5} and \ref{S:d3d3}. To be clear, we do discuss various singular solutions, however the singularities in those solutions do not arise from the curvature singularity of the effective metric. Whether the curvature singularity of the effective metric has any physical meaning we leave as an open question.

\begin{figure}[t]
  \centering
  \includegraphics[width=0.7\textwidth]{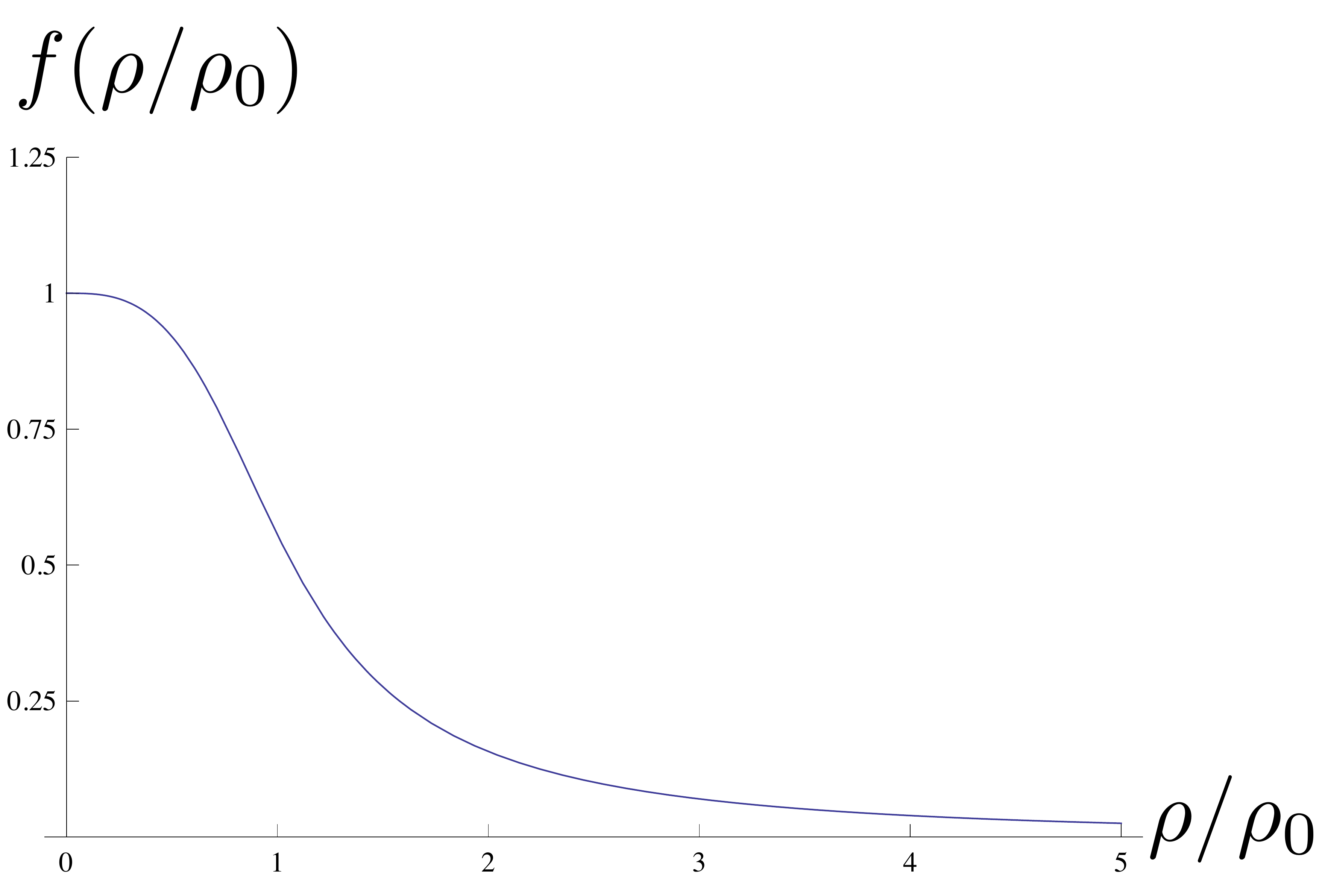}
    \caption{The function $f(\rho/\rho_0) = \left[\frac{ \sqrt{1+{\rho}_0^6/\rho^6} - 1}{\sqrt{1+{\rho}_0^6/\rho^6} + 1}  \right]^{1/3}$ appearing in the solution for $A_{\alpha}(z)$ in eq.~\protect\eqref{examplesol2}, plotted as a function of $\rho/\rho_0$. The Poincar\'e horizon is at $\rho/\rho_0 \to 0$, while the $AdS_5$ boundary is at $\rho/\rho_0 \to \infty$. This function and all of its derivatives are finite for all $\rho \in [0,\infty)$.}
    \label{fig:solfig1}
\end{figure}

The second step in our recipe is to solve the self-duality condition for $f_{ij}$ in eq.~\eqref{selfdual}. The key observation here is that the self-duality condition is conformally invariant. We can thus ignore the conformal factor in the effective metric in eq.~\eqref{gij2}, and solve the self-duality condition using the flat metric $d\bar{\rho}^2+\bar{\rho}^2ds^2_{\mathbb{S}^3}$, which is trivial to do. The general solution is the same as that in eq.~\eqref{E:selfdualSoln} (plus solutions obtained from it by acting with the isometries of flat space), but with $\rho\to \bar{\rho}$. In the field theory, we do not want any external sources besides the chemical potential $\mu$, so we will ignore all non-normalizable solutions. To begin studying the normalizable solutions in detail, let us focus first on a simple example, namely a solution of the form $\bar{\rho}^{-(l+1)} \mathcal{Y}^{l,-}_{\alpha}$, with $l=1$,
\beq
\label{examplesol}
A_{\alpha}(z) = c \, \frac{1}{\bar{\rho}^2} \, \mathcal{Y}^{1,-}_{\alpha}, \qquad A_{\bar{\rho}}(z)=0\,,
\eeq
with $c$ a finite constant. In terms of the radial coordinate $\rho$, this solution is
\beq
\label{examplesol2}
A_{\alpha}(z) = c \, \frac{2^{2/3}}{{\rho}_0^2} \left[\frac{\sqrt{1+{\rho}_0^6/\rho^6} - 1}{\sqrt{1+{\rho}_0^6/\rho^6} + 1}  \right]^{1/3} \, \mathcal{Y}^{1,-}_{\alpha}, \qquad A_{\rho}(z)=0\,.
\eeq
The function of $\rho$ in eq.~\eqref{examplesol2} and all of its derivatives are non-singular for all $\rho \in \mathbb{R}^+$. In particular, deep in the bulk, $\rho \to 0$, the solution approaches a nonzero constant, $A_{\alpha}(z) \to c \, \frac{2^{2/3}}{{\rho}_0^2}$. In fig.~\ref{fig:solfig1} we plot the $\rho$-dependence of the solution for $A_{\alpha}(z)$ in eq.~\eqref{examplesol2}. The solution in eq.~\eqref{examplesol2} nevertheless appears to be singular: if we take derivatives of the $A_{\alpha}(z)$ in eq.~\eqref{examplesol2} in $\mathbb{S}^3$ directions and then take $\rho \to 0$, the solution always approaches a constant, but the value of that constant depends on the direction in $\mathbb{R}^4$ along which we approach $\rho \to 0$. The solution thus appears to have a kink singularity, and must be supported by some dipole-like ($l=1$) source at $\rho=0$. That source would in principle fix the value of $c$. As we mentioned above, and as we will argue in detail in section~\ref{S:d3d5}, we should excise the point $\rho=0$ from the D7-brane worldvolume. In that case $c$ is fixed by boundary conditions near $\rho=0$. Notice that with $\rho=0$ excised, the solution is regular for any finite value of $c$.

What is the physical meaning of $c$? Consider first the bulk interpretation of $c$. The instanton number density $F \wedge F$ of the solution in eq.~\eqref{examplesol} will be proportional to $c^2$. We may thus think of $c^2$ as measuring a number of D3-branes that have dissolved into the D7-brane. We expect $c^2$ to obey a quantization condition, since the number of D3-branes should be quantized, although that quantization is not visible in the supergravity approximation to the full string theory. In fact, we can interpret the dipole-like source that we excised at $\rho=0$ as a dipole-like distribution of D3-branes that have dissolved into the D7-brane.

In the field theory, $c$ sets the expectation value of the dual scalar operator. The squarks $q$ and $\tilde{q}$ of the hypermultiplet form a doublet of the $SU(2)$ R-symmetry, which we denote as $Q = (q,\tilde{q}^{\dagger})^T$. As shown in ref.~\cite{Kruczenski:2003be}, the $l=1$ mode of $A_{\alpha}(z)$ is dual to a dimension-two Lorentz scalar operator bilinear in $Q$, transforming in the $(1,0)$ representation of the $SO(4)$ global symmetry, and neutral under the $SO(2)$. In other words, the $l=1$ mode of $A_{\alpha}(z)$ is dual to a scalar operator transforming as a vector of the $SU(2)$ R-symmetry. To be explicit, the operator dual to the $l=1$ mode of $A_{\alpha}(z)$ is
\beq
\label{oidef}
\mathcal{O}^I = Q^{\dagger} \sigma^I Q\,,
\eeq
with $\sigma^I$ the Pauli matrices of the $SU(2)$ R-symmetry, with $I = 1,2,3$. Near the $AdS_5$ boundary, $\rho \to \infty$, the solution in eq.~\eqref{examplesol} behaves at leading order as $A_{\alpha}(z) \propto c \, \rho^{-2}$, the expected scaling with $\rho$ for a normalizable field dual to a scalar operator of dimension two. A straightforward exercise (in holographic renormalization) then shows that $\langle \mathcal{O}^I \rangle \propto c$.

We can now state our main result precisely: if we begin with the solution for $A_0(z)$ in eq.~\eqref{eq:solA02} and introduce nonzero $c$, then the on-shell D7-brane action does not depend on $c$. In field theory terms, if we begin in the compressible states described in section~\ref{S:4ND}, then introducing a nonzero value of $\langle \mathcal{O}^I \rangle$ does not change the Helmholtz free energy, nor the grand potential in the grand canonical ensemble. We also know that nonzero $c$ does not change the value of the energy, defined as the expectation value of the ``$x^0x^0$'' component of the field theory stress-energy tensor, at least to order $N_fN_c$. To see why, recall that the components of the D7-brane stress-energy tensor with indices in the $(x^0,\dots,x^3)$ directions (when integrated over $\rho$ and the $\mathbb{S}^3$) are precisely equal to the hypermultiplets' order $N_f N_c$ contribution to the expectation value of the field theory stress-energy tensor, as shown in ref.~\cite{Karch:2008uy}. Given that any self-dual solution for $A_i(z)$ does not affect the D7-brane stress-energy tensor, we immediately conclude that the energy of the field theory is independent of $c$.

We have thus found a moduli space parameterized by $\langle \mathcal{O}^I \rangle$. We can trivially extend the moduli space to be infinite-dimensional by taking a linear superposition of the $l=1$ mode of $A_{\alpha}(z)$ with all of the $l>1$ modes, producing the solution
\beq
\label{E:SDinstanton}
A_{\alpha}(z) = \sum_{l=1}^{\infty} c_l \, \frac{1}{\bar{\rho}^{l+1}} \, \mathcal{Y}_{\alpha}^{l,-}\,, \qquad A_{\bar{\rho}}(z) = 0\,,
\eeq
with $c_l$ arbitrary constants. We expect the $c_l$'s to obey a quantization condition: roughly speaking, we expect the instanton number $\int F \wedge F$ to look like a sum of the $c_l^2$, and that sum should represent the number of D3-branes dissolved into the D7-brane. In fact, each $l$ mode of $A_{\alpha}(z)$ of eq.~\eqref{E:SDinstanton} requires a source at $\rho=0$, with the same value of $l$. As in the $l=1$ case, we excise the point $\rho=0$ from the D7-brane, however intuitively we can think of these sources as higher multipole distributions of D3-branes that have dissolved into the D7-brane.

Each $l$ mode of $A_{\alpha}(z)$ is dual to a scalar operator bilinear in $Q$, which we will call $\mathcal{O}_l^-$. The operator $\mathcal{O}_l^-$ has dimension $\Delta = l+1$, transforms in the $(\frac{l+1}{2},\frac{l-1}{2})$ representation of $SO(4)$, and is neutral under the $SO(2)$. Schematically, the $\mathcal{O}_l^-$ look like the $\mathcal{O}^I$ in eq.~\eqref{oidef}, but with adjoint scalars, with traceless, symmetrized $SO(4)$ indices, sandwiched between $Q^{\dagger}$ and $Q$ \cite{Kruczenski:2003be}. We ultimately expect the $c_l$ to be isomorphic to the expectation values of the $\mathcal{O}_l^-$'s. Determining the exact isomorphism requires careful holographic renormalization. The moduli space parameterized by these scalar expectation values is our Higgs branch. Since the $\mathcal{O}_l^-$ are charged under $SO(4)$, at a generic point on the Higgs branch we expect $SO(4)$ to be broken to a subgroup. Furthermore, if $F\wedge F$ is nonzero and hence the D7-brane carries D3-brane charge, then in the field theory we expect $SU(N_c)$ to be broken to a subgroup.

We can generate the most general normalizable self-dual $f_{ij}$ by acting on the solution in eq.~\eqref{E:SDinstanton} with isometries of the flat metric $d\bar{\rho}^2+\bar{\rho}^2ds^2_{\mathbb{S}^3}$. Some of these solutions will have singularities at nonzero values of $\rho$. These singularities represent pointlike instantons, similar to the ones we found above in our discussion of the zero-density SUSY Higgs branch. Here again, such singular solutions are inadmissable within the supergravity approximation to string theory.

\begin{figure}[t]
\begin{center}
\includegraphics[width=5in]{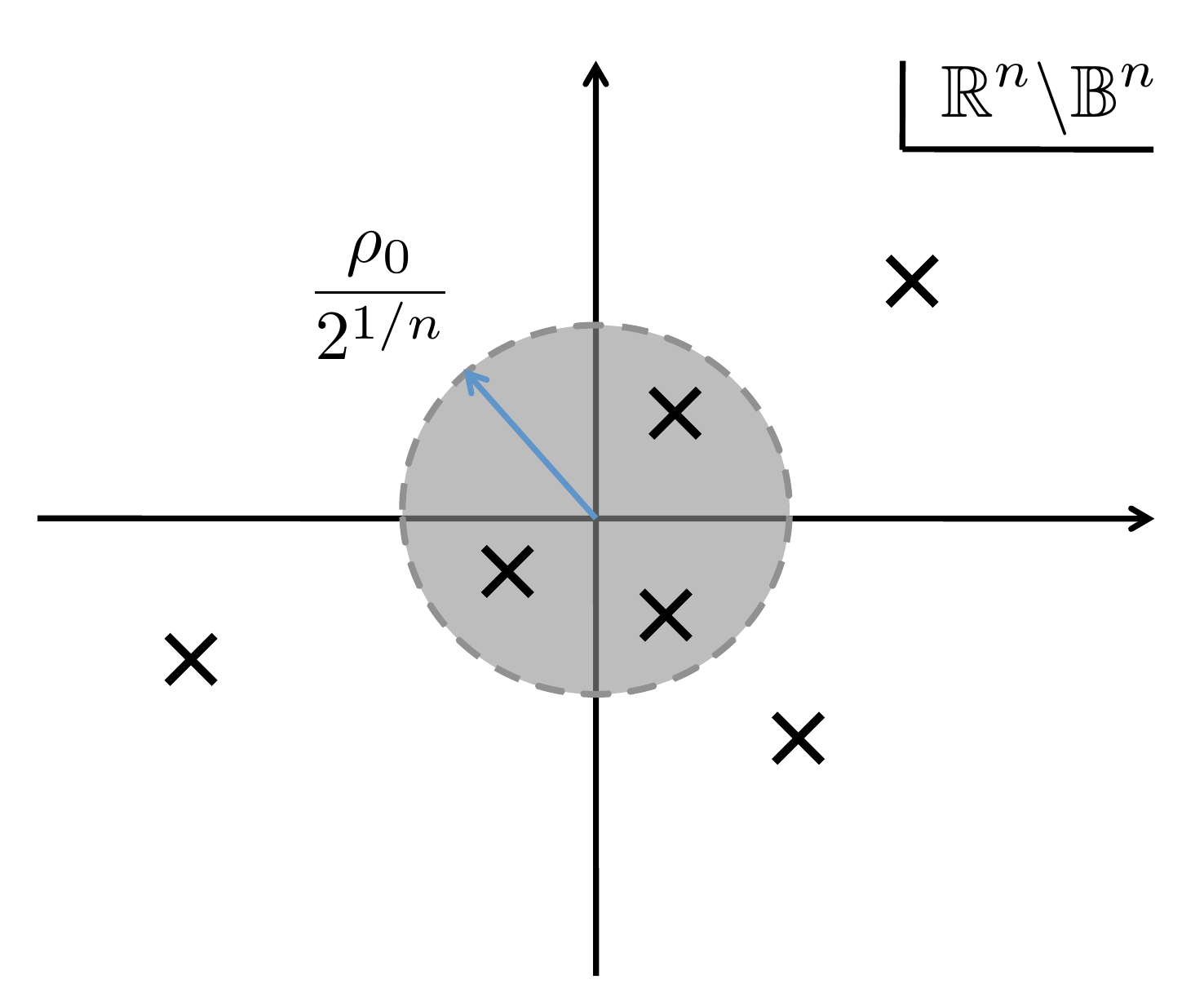}
\caption{\label{F:d3d7_higgs}The space $\mathbb{R}^{n+1}\backslash\mathbb{B}^{n+1}$, with $n=3,2,1$ and with $n-1$ directions suppressed for clarity. The shaded region centered at the origin represents a ball $\mathbb{B}^{n+1}$ of radius $\rho_0/2^{1/n}$. All points inside this $\mathbb{B}^{n+1}$ are excised from the space. In bulk terms, the surface of this $\mathbb{B}^{n+1}$, indicated by the dashed line, is the Poincar\'e horizon. An `$\mathbf{\times}$' denotes the position of a singularity (and center) of a self-dual $U(1)$ instanton. For $n=3$, the instanton solution falls off as a power law $A_{\alpha}\propto 1/\bar{\rho}^{l+1}$ near the singularity, with integer $l\geq1$. Instantons can be localized outside the ball or ``inside the ball,'' although since the points inside the ball have been excised, the latter is simply a mnemonic device.  Instantons centered outside the ball are singular solutions, which we discard when $n=3$ but which have physical meaning when $n=2,1$. Instantons centered inside the ball are regular everywhere in the physical region (outside the ball). For $n=3$ only solutions with singularities inside the ball describe points on the nonzero-density Higgs branch. For $n=2,1$, solutions with singularities outside the ball plus those with singularities inside the ball together describe all points on the nonzero-density Higgs branch.}
\end{center}
\end{figure}

Broadly speaking, then, our system has two classes of $U(1)$ instantons, those which have singularities at nonzero $\rho$, outside the excised $\mathbb{B}^4$, and those with singularities ``inside the ball,'' although since the $\mathbb{B}^4$ is excised this is really just a mnemonic device. We present an illustration of instanton solutions with singularities inside and outside the $\mathbb{B}^4$ in fig.~\ref{F:d3d7_higgs}. The instantons with singularities outside the ball are genuinely singular field configurations which we discard. The instantons with singularities inside the ball, our solutions in eq.~\eqref{E:SDinstanton}, are completely non-singular on the D7-brane worldvolume. Intuitively, these solutions represent multipole distributions of D3-branes dissolved into the D7-brane at $\rho=0$. These non-singular, normalizable solutions describe points on a Higgs branch in the dual field theory. As we recalled above, at zero density the D3/D7 theory has no Higgs branch. The fact that a Higgs branch emerges at nonzero density is therefore remarkable.

The interior of the ball is a useful mnemonic device because varying the strengths, values of $l$, and positions of the singularities inside the ball corresponds to moving around in the space of the $c_l$. To see why, recall that each singularity is supported by some D3-brane source, as explained above. Suppose we introduce a single such D3-brane source, with some value of $l$, sitting at $\bar{\rho}=0$, at the center of the ball but outside the physical region $\bar{\rho} \in[2^{-1/3}\rho_0,\infty)$. The resulting solution will be that in eq.~\eqref{E:SDinstanton} with only a single one of the $c_l$'s nonzero, and all other $c_l$'s zero. Changing the strength of the source changes the value of the single nonzero $c_l$. In other words, a source at the origin has two degrees of freedom, a strength and a value of $l$, that allow us to pick a direction (value of $l$) and to move in that direction (strength of the source) in the space of $c_l$'s. We have a third degree of freedom as well, the positions of the sources inside the ball. For example, suppose we begin with a single source at the center of the ball and then displace it to a nonzero $\bar{\rho}$ still inside the ball. Clearly such a solution will preserve less of the $SO(4)$ symmetry than the solution supported by a source at the origin, and so will generically involve $c_l$'s with many different $l$ values. This third degree of freedom thus corresponds to redistributing the relative weights of the $c_l$'s in the solution of eq.~\eqref{E:SDinstanton}. Notice that we expect the overall size of the position vector in the space of $c_l$'s, which is given by $\sum_l c_l^2$, to be constrained by the quantization condition on the number of D3-branes, as mentioned above.

Eq.~\eqref{E:SDinstanton} is a solution to the full non-linear equations of motion, hence if we take the $c_l$ to be perturbatively small, eq.~\eqref{E:SDinstanton} also provides a solution to the linearized equations of motion, Maxwell's equations. In general, via AdS/CFT the solutions to these linearized equations determine the retarded two-point functions of the $\mathcal{O}_l^-$. In particular, a normal mode in the bulk is dual to a pole in the retarded two-point function. The self-dual solutions in eq.~\eqref{E:SDinstanton} have the behavior of normal modes: they are normalizable at large $\rho$ and regular near the Poincar\'e horizon $\rho=0$. More precisely, the $\left(\frac{l+1}{2},\frac{l-1}{2}\right)$ mode of eq.~\eqref{E:SDinstanton} corresponds to a gapless mode (in Fourier space, when $k=0$ the mode is gapless, $\omega=0$) in the retarded two-point function of $\mathcal{O}_l^-$. We thus identity these solutions with the ``R-spin diffusion'' modes  observed in ref.~\cite{Ammon:2011hz}. We thus have a nice way to think about the ``R-spin diffusion'' modes: they are perturbative excitations of the emergent moduli $c_l$.

The existence of a Higgs branch in these compressible states of the D3/D7 system raises a natural question: what is the metric on the Higgs branch? In general, a Higgs branch metric is the metric that the moduli ``see,'' which in our case means the metric in the space of $c_l$. To calculate that metric, we would need to write an ansatz for $A_{\alpha}(z)$ in which the $c_l$ depend on the field theory directions $(x^0,\dots,x^3)$, insert that ansatz into the D7-brane action, and expand to quadratic order in the $(x^0,\dots,x^3)$ derivatives of the $c_l$'s. As argued in refs.~\cite{Guralnik:2004ve,Guralnik:2004wq,Guralnik:2005jg}, the Higgs branch metric is the metric entering into these kinetic terms for the $c_l$'s. We will leave the calculation of the metric on the nonzero-density Higgs branch for future work.

Our regular instanton solutions suggest some degeneracy, in the following sense. If we imagine that the $N_c$ D3-branes generating the background metric and RR five-form are ``hidden'' behind the Poincar\'e horizon, then because of the D7-brane's worldvolume electric flux they appear to have the option of either dissolving into the D7-brane or not, with no change to the on-shell action. That leads to a natural speculation: perhaps the microstates producing the nonzero entropy in this system could be described holographically in terms of these D3-branes interacting with the D7-brane worldvolume electric flux. In other words, perhaps these gapless D3-brane modes are the degrees of freedom producing the nonzero entropy in this system. In a best-case scenario, the effective description of such degrees of freedom would be a (0+1)-dimensional CFT with an extensive degeneracy of zero-energy states. Notice that the same corrections in $N_c$ and/or $\lambda$ that may lift the moduli space may also lift the degeneracy of states producing the nonzero entropy.

In field theory terms, the nonzro-density Higgs branch we have uncovered is almost certainly an artifact of the large-$N_c$ and/or large-$\lambda$ limits. The moduli that parameterize the nonzero-density Higgs branch have no obvious symmetries to protect them from finite-$N_c$ and/or finite-$\lambda$ corrections. For example, upon including the back-reaction of the D7-branes, which is a correction in $N_f/N_c=1/N_c$, the so-called ``Fermi seasickness'' instability may appear \cite{Chen:2009kx,Hartnoll:2009ns}: the D7-brane's worldvolume electric flux may exert a sufficiently large force on the D3-branes sitting at $r=0$ to pull some of those D3-branes to nonzero $r$. In other words, upon including back-reaction, the D3-branes may not have the option to sit at the bottom of $AdS_5$ anymore.

We will postpone further discussion of our solutions to section~\ref{S:discuss}. For now let us apply the techniques of this section to the other D3/D$p$ systems.

\section{D3/D5 and Vector/Scalar Duals}
\label{S:d3d5}

We now turn to the D3/D5 system~\cite{Karch:2001cw},
\begin{center}
\begin{tabular}{c|cccccccccc}
& $x^0$ & $x^1$ & $x^2$ & $x^3$ & $x^4$ & $x^5$ & $x^6$ & $x^7$ & $x^8$ & $x^9$ \\
\hline
 D3 & X & X & X & X & & & & & & \\
 D5 & X & X & X &  & X & X & X & & & \\
\end{tabular}
\end{center}
The flavor fields break the $SO(6)$ R-symmetry down to $SO(3)_1\times SO(3)_2\simeq SO(4)$~\cite{DeWolfe:2001pq}, which is the R-symmetry of the remaining (2+1)-dimensional $\mathcal{N}=4$ SUSY. The $SO(3)_1$ corresponds to rotations in the $(x^4,x^5,x^6)$ directions while the $SO(3)_2$ corresponds to rotations in the $(x^7,x^8,x^9)$ directions. We will now relabel the directions $(x^4,x^5,x^6)$, along the D5-brane but transverse to the D3-branes, as $z^i$ with $i=1,2,3$.

In the Maldacena and probe limits, the D5-brane wraps an asymptotically $AdS_4\times\mathbb{S}^2$ submanifold inside of $AdS_5\times\mathbb{S}^5$. The action of the probe D5-brane is
\beq
\label{E:d3d5S}
S_5 = -T_5 \int d^6\xi\sqrt{\text{det}(-P[G]_{ab}+F_{ab})}+T_5 \int P[C_4]\wedge F\,.
\eeq
As in our study of D7-branes, we will consider an ansatz for the worldvolume fields that is more general than that of section~\ref{S:4ND}. We will demand the same symmetries as in that section, except for $SO(3)_1$ invariance and reflection symmetry about $x^3=0$. The most general ansatz consistent with the symmetries is then
\beq
\label{E:d3d5ansatz}
x^3(\xi) = x^3(z), \qquad A(\xi) = A_0(z) dx^0 + A_i(z) dz^i\,,
\eeq
with all other worldvolume fields vanishing. The scalar field $x^3(\xi)$ specifies the position of the D5-brane inside the $AdS_5$ part of the geometry. Substituting the ansatz in eq.~\eqref{E:d3d5ansatz} into eq.~\eqref{E:d3d5S}, we obtain the D5-brane action density, which involves an integral over the $\mathbb{R}^3$ spanned by the $z^i$,
\beq
\label{E:d3d5S2}
s_5 = -T_5 \int d^3z \left[ \sqrt{\text{det}(g_{ij} + Z^{-1/2}f_{ij}+Z^{-1}\partial_i x^3\partial_j x^3)}-\frac{1}{2}Z^{-1}\tilde{\epsilon}^{ijk}\partial_i x^3 f_{jk} \right]\,,
\eeq
where $\tilde{\epsilon}^{ijk}$ is the Levi-Civita symbol on $\mathbb{R}^3$ with orientation $\tilde{\epsilon}^{123}\equiv+1$, the factor $Z=1/\rho^4$, and we have defined an effective metric and field strength on $\mathbb{R}^3$,
\beq
\label{E:3dfields}
g_{ij} \equiv \delta_{ij} - \partial_i A_0 \partial_j A_0, \qquad f_{ij} \equiv \partial_i A_j - \partial_j A_i.
\eeq

Our goal is to find solutions for the fields $x^3(z), A_0(z),$ and $A_i(z)$. We can simplify our task by adapting the methods presented in section~\ref{S:d3d7} to this system. To that end, we note that the D5-brane action in eq.~\eqref{E:d3d5S2} may actually be brought into the same form as the D7-brane action in eq.~\eqref{D7action1}, by uplifing eq.~\eqref{E:d3d5S2} to an action defined on $\mathbb{R}^4$. We do this by introducing an extra direction, $z^4$, and uplifting the fields on $\mathbb{R}^3$ to fields on $\mathbb{R}^4$, in two steps. First, we define an effective metric $\hat{g}_{ij}$, a gauge field $\hat{a}$, and field strength $\hat{f}$, all on $\mathbb{R}^4$, as
\beq
\hat{g}_{ij} = g_{ij} + \delta_{~i}^4 \, \delta_{~j}^4\,, \qquad \hat{a} = A_i(z) dz^i + x^3(z) dz^4\,, \qquad \hat{f} = d\hat{a},
\eeq
where $g_{ij}$ was defined in eq.~\eqref{E:3dfields}. Notice the formal similarity between our uplift to $\mathbb{R}^4$ and T-duality in the $x^3$ direction, in particular, the worldvolume scalar $x^3(z)$ becomes a component of the gauge field $\hat{a}$, as in genuine T-duality. Second, we demand that none of the fields $A_0(z)$, $A_i(z)$, or $x^3(z)$ depend on the extra direction $z^4$. The action in eq.~\eqref{E:d3d5S2} may then be written formally as
\beq
\label{S:d3d5S3}
s_5 = -T_5 \int d^4z\left[ \sqrt{\text{det}(\hat{g}_{ij} + Z^{-1/2} \hat{f}_{ij})}-\frac{1}{8}Z^{-1}\tilde{\epsilon}^{ijkl}\hat{f}_{ij}\hat{f}_{kl} \right],
\eeq
where $\tilde{\epsilon}^{ijkl}$ is the Levi-Civita symbol on $\mathbb{R}^4$ with orientation $\tilde{\epsilon}^{1234}=+1$. To maintain the original normalization of the D5-brane action, we must demand that the integration over $z^4$ simply produces a factor of one, which we can achieve by choosing for example $z^4 \in [0,1]$. Having put the D5-brane action into the form of eq.~\eqref{S:d3d5S3}, we can immediately invoke the arguments of section~\ref{S:d3d7}. The topological bound on the DBI action in four dimensions, eq.~\eqref{mink2}, implies a bound on the action in eq.~\eqref{S:d3d5S3},
\beq
\label{E:d3d5bounds1}
s_5\leq -T_5 \int d^4z \left [\sqrt{\text{det}\,\hat{g}_{ij}}+\frac{1}{8}Z^{-1}\left( |\tilde{\epsilon}^{ijkl}\hat{f}_{ij}\hat{f}_{kl}| - \tilde{\epsilon}^{ijkl}\hat{f}_{ij}\hat{f}_{kl} \right)\right].
\eeq
As in our study of D7-branes, this bound is saturated only for $\hat{f}_{ij}$ self-dual with respect to the metric $\hat{g}_{ij}$. For such self-dual $\hat{f}_{ij}$, the D5-brane action density reduces to
\beq
\label{E:d3d5S4}
s_5 = -T_5\int d^3z \sqrt{\text{det}\,g_{ij}},
\eeq
which is independent of both the $A_i(z)$ and of $x^3(z)$. In fact, eq.~\eqref{E:d3d5S4} is just the action for DBI electrostatics in $\mathbb{R}^3$. Self-dual $\hat{f}_{ij}$ extremize the action, and hence also solve the equations of motion of $x^3(z)$ and $A_i(z)$. Moreover, for self-dual $\hat{f}_{ij}$ the equation of motion for $A_0(z)$ follows from variation of eq.~\eqref{E:d3d5S4}, which is the equation of motion for the electric potential in DBI electrostatics. Self-dual $\hat{f}_{ij}$ will contribute nothing to the D5-brane stress-energy tensor. We can obtain (a subset of all) solutions for $A_0(z)$ and $\hat{f}_{ij}$ by a simple recipe: we first solve for $A_0(z)$, which determines the effective metric, which in turn defines the self-duality condition for $\hat{f}_{ij}$, which we then need to solve.

The key difference between the D5-brane and the D7-brane is that $A_i(z)$ and $x^3(z)$ do not depend on $z^4$, so that the self-duality condition for $\hat{f}_{ij}$ becomes
\beq
\label{E:vsDuality}
\partial_i x^3 = \frac{1}{2}\epsilon_{ijk}f^{jk},
\eeq
where we have defined the Levi-Civita tensor $\epsilon^{ijk} \equiv \tilde{\epsilon}^{ijk}/\sqrt{\text{det}\,g_{ij}}$, and indices are raised and lowered with the metric $g_{ij}$. Eq.~\eqref{E:vsDuality} is that of \emph{vector/scalar duality}, wherein the vector $A_i(z)$ is Hodge dual to the scalar $x^3(z)$.

Unlike the self-duality condition for $\hat{f}_{ij}$, the vector/scalar duality condition in eq.~\eqref{E:vsDuality} is not conformally invariant, so to find solutions we will have to work harder than we did in section~\ref{S:d3d7}, even when $g_{ij}$ is conformally equivalent to the flat metric. Vector/scalar duality implies that $x^3(z)$ and the $A_i(z)$ are harmonic, so in particular $x^3(z)$ obeys the Laplace equation $\Box x^3(z)=0$ with $\Box$ the Laplacian built from the metric $g_{ij}$. We may thus solve first for the most general harmonic $x^3(z)$, then determine $f_{ij}$ algebraically via eq.~\eqref{E:vsDuality}, and then find the $A_i(z)$ by appropriately integrating the $f_{ij}$. (The converse is of course equivalent: we can solve for the most general harmonic $A_i(z)$, insert the resulting $f_{ij}$ into eq.~\eqref{E:vsDuality}, and then integrate to find $x^3(z)$.)

We will actually consider solutions more general than harmonic $x^3(z)$, \textit{i.e.} we will promote $x^3(z)$'s Laplace equation to a Poisson equation by introducing a source, $\mathcal{S}(z)$,
\beq
\label{E:x3Lap}
\Box \, x^3(z) = \mathcal{S}(z).
\eeq
Eq.~\eqref{E:x3Lap} is straightforward to solve, for example by the Green's function technique. If we input that solution into eq.~\eqref{E:vsDuality}, then we find a violation of $f_{ij}$'s Bianchi identity, or equivalently, $dF\neq 0$. For certain sources $\mathcal{S}(z)$, this violation of $F$'s Bianchi identity has a simple interpretation in string theory. Consider for example sources localized in $\mathbb{R}^3$. The most general form for a single source localized at a point $z'$ in $\mathbb{R}^3$ is
\beq
\label{E:Slocal}
\mathcal{S}(z) = \sum_{L=0}^{\infty} s^{i_1..i_L}\partial_{i_1}..\partial_{i_L} \delta(z-z'),
\eeq
where the $s^{i_1..i_L}$ are real constants. The integer $L$ corresponds to a multipole moment for the source $\mathcal{S}(z)$. A monopole source, $L=0$, will produce a $dF\neq0$ that implies $\int F \neq 0$, where the integration is over the two-cycle $\mathcal{C}_2$ dual (via de Rham's theorem) to $F$. We thus learn, unsurprisingly, that a monopole source indeed represents a magnetic monopole on the D5-brane worldvolume. In general, any nonzero $F$ on the D5-brane worldvolume sources $C_4$ through the WZ term in the D5-brane action eq.~\eqref{E:d3d5S}, thus a nonzero $F$ endows the D5-brane with a smeared distribution of D3-brane charge density. The monopole source with $\int F \neq 0$ thus represents some net D3-brane charge, which in string theory we interpret as some D3-branes blown up on $\mathcal{C}_2$. By charge conservation, these D3-branes must come from the $N_c$ D3-branes producing the background geometry and RR five-form. Moreover, charge conservation also requires that if $\mathcal{C}_2$ collapses to zero size then the net D3-brane charge must be carried off by D3-branes attached to the D5-brane at the point where $\mathcal{C}_2$ collapses. A higher multipole source, with integer $L>0$, will produce $dF\neq0$ but $\int F =0$, and so represents a distribution of D3-brane charge density on the D5-brane with zero \textit{net} D3-brane charge.

We have glossed over a crucial point: the source term in eq.~\eqref{E:x3Lap} ultimately comes from a source term in the D5-brane action eq.~\eqref{E:d3d5S}. Generically, such a source term will modify the topological bound on the D5-brane action, eq.~\eqref{E:d3d5bounds1}, in which case $x^3(z)$ and $A_i(z)$ obeying the vector/scalar duality condition in eq.~\eqref{E:vsDuality} may no longer extremize the action, and hence may no longer solve the equations of motion. Na\"ively, we might think that for a localized source, a vector/scalar dual pair $x^3(z)$ and $A_i(z)$ may solve the equations of motion at points away from the source, because $\mathcal{S}(z)$ has support at only one point and so $\Box x^3=0$ at all other points. That is not the case, however. Let us define a vector $V^i(z)$ as
\beq
V^i(z) \equiv \partial^i x^3 - \frac{1}{2} \epsilon^{ijk} f_{jk},
\eeq
so that the vector/scalar duality condition is $V^i(z)=0$. In the equations of motion for the $A_i(z)$, if we introduce localized sources and then integrate the equations of motion over a region including the sources, then we find that $V^i(z)$ must be nonzero even at points \textit{away} from the source. In other words, even with a localized source vector/scalar duality is violated \textit{away} from the source.\footnote{To illustrate this point more simply, let us consider a real function $\mathcal{F}$ of a real variable $\zeta \in (-\infty,+\infty)$, and suppose that $\mathcal{F}(\zeta)$ obeys a first-order equation $\mathcal{F}'(\zeta) = \mathcal{J}(\zeta)$, where here prime denotes $\partial_{\zeta}$ and $\mathcal{J}(\zeta)$ is some source function. This first-order equation is the analogue of eq.~\eqref{E:vsDuality}, and $\mathcal{F}'(\zeta)$ is the analogue of our $V^i(z)$, so here ``vector/scalar duality'' means $\mathcal{F}'(\zeta)=0$. Taking a derivative, we find $\mathcal{F}''(\zeta) = \mathcal{J}'(\zeta)$, which is the analogue of our equations of motion. Now suppose this second-order equation has a source localized at $\zeta =0$, so say $\mathcal{J}'(\zeta) \propto \delta(\zeta)$. That means $\mathcal{J}(\zeta)$ must be a step function, for example we can choose $\mathcal{J}(\zeta)$ to be zero for $\zeta<0$ and a nonzero constant for $\zeta >0$. We thus find $\mathcal{F}'(\zeta) \neq 0 $ for all $\zeta >0$: the ``vector/scalar duality'' is violated at points \textit{away} from the source at $\zeta=0$.} The general lesson is: when seeking solutions using our methods, we must treat any source terms with caution. 

To illustrate our method, including the careful treatment of sources, let us reproduce known solutions representing points on the \textit{zero-density} Higgs branch~\cite{Arean:2006vg,Arean:2007nh}. Taking $A_0(z)=0$, the effective metric in eq.~\eqref{E:2dfields} reduces to the flat metric on $\mathbb{R}^3$. The most general normalizable solution to eq.~\eqref{E:x3Lap} supported by delta-function ($L=0$) sources only is
\beq
\label{E:d3d5originalHiggs}
x^3(z) = \sum_m \frac{C_m}{|z-z_m'|},
\eeq
where the $C_m$ are finite constants. The solution in eq.~\eqref{E:d3d5originalHiggs} is clearly singular at the points $z_m'$, the locations of the delta-function sources, with the values of the $C_m$ determined by the strengths of these delta-functions. These sources threaten to invalidate the vector/scalar duality condition used to derive $x^3(z)$'s Laplace equation, even at points away from the $z_m'$, \textit{i.e.} the solution threatens its own existence by threatening to invalidate the equation it solves. We will argue below that in fact, to reproduce all points on the Higgs branch, we must excise the points $z_m'$ from $\mathbb{R}^3$, in which case the $C_m$ are actually fixed by boundary conditions near the points $z_m'$. For now we will discuss the physics of the solution in eq.~\eqref{E:d3d5originalHiggs} as if genuine sources are present.

As mentioned above, the localized sources supporting the solution in eq.~\eqref{E:d3d5originalHiggs} have a physically sensible interpretation in string theory as D3-branes blown up into the D5-brane. Via the vector/scalar duality condition eq.~\eqref{E:vsDuality}, we can obtain the $f_{ij}$ dual to the $x^3(z)$ in eq.~\eqref{E:d3d5originalHiggs}. Integrating the resulting $F$ over an $\mathbb{S}^2$ surrounding one (and only one) of the $z_m'$, we find $\int_{\mathbb{S}^2} F=- 4 \pi C_m$, so each singularity looks like a magnetic monopole. In string theory terms, the magnetic monopole charge is $\int_{\mathbb{S}^2} F= (4 \pi^2 \alpha') q_m$, where $q_m$ is the number of D3-branes attached to the D5-brane at $z_m'$. In the full string theory (not just classical supergravity), $q_m$ will be an integer, hence $C_m$ will also be quantized. In short, the solution in eq.~\eqref{E:d3d5originalHiggs} represents D3-branes blown up into the D5-brane, where the $z_m'$ denote the positions of the D3-branes and the $C_m$ encode the number of D3-branes at each $z_m'$.

\begin{figure}[t]
  \centering
  $\begin{array}{ccc}
  \includegraphics[width=0.3\textwidth]{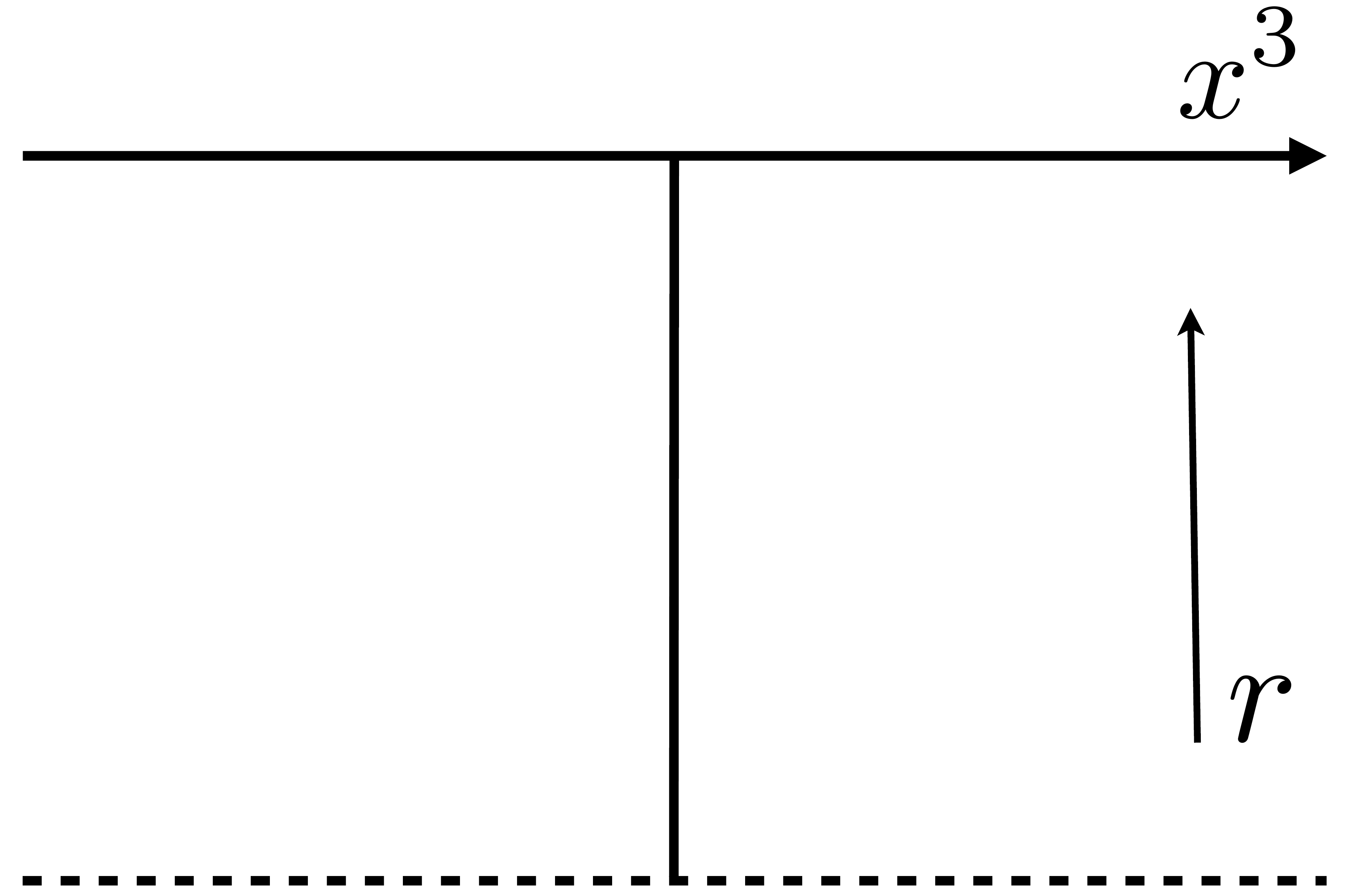} & \includegraphics[width=0.3\textwidth]{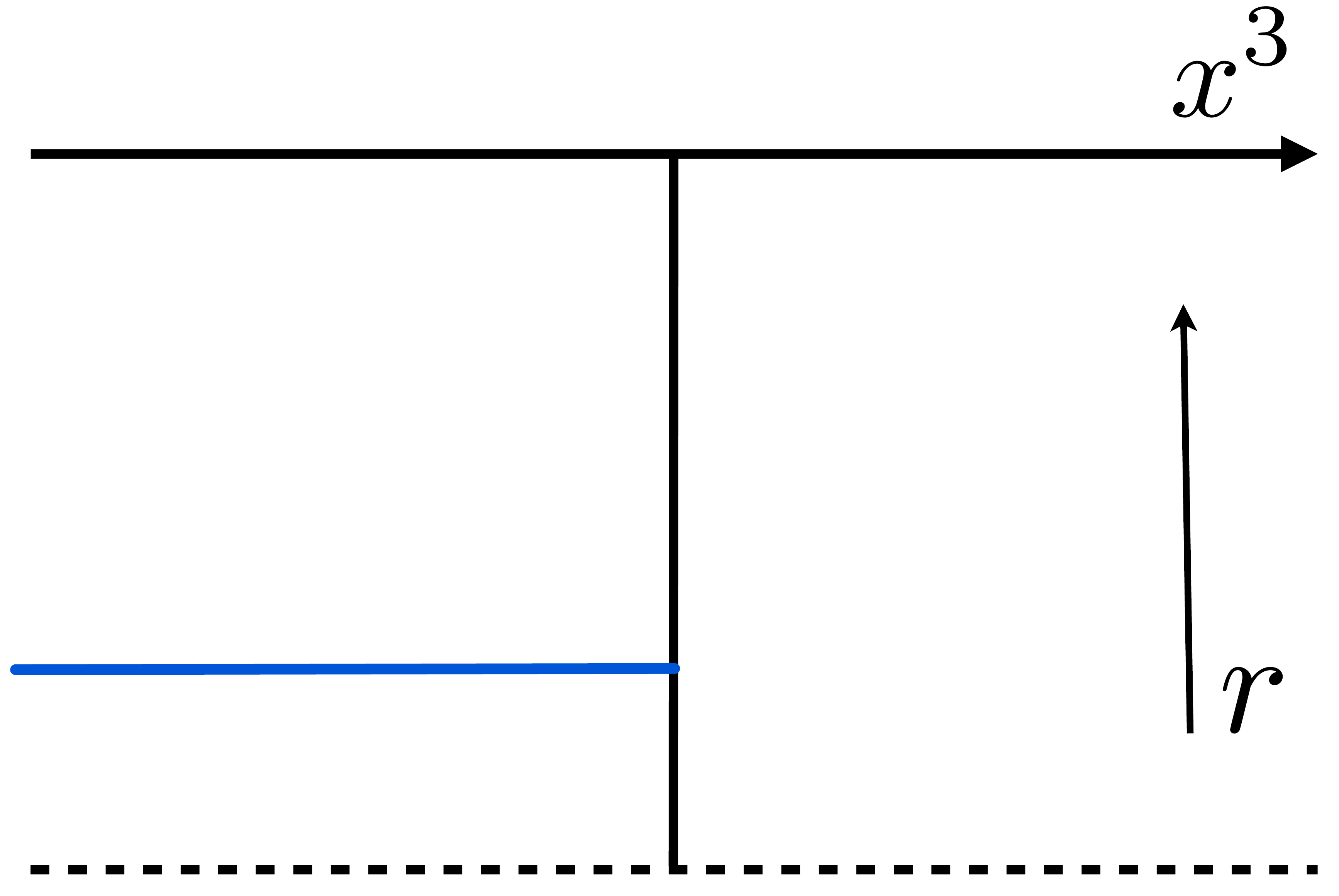} & \includegraphics[width=0.3\textwidth]{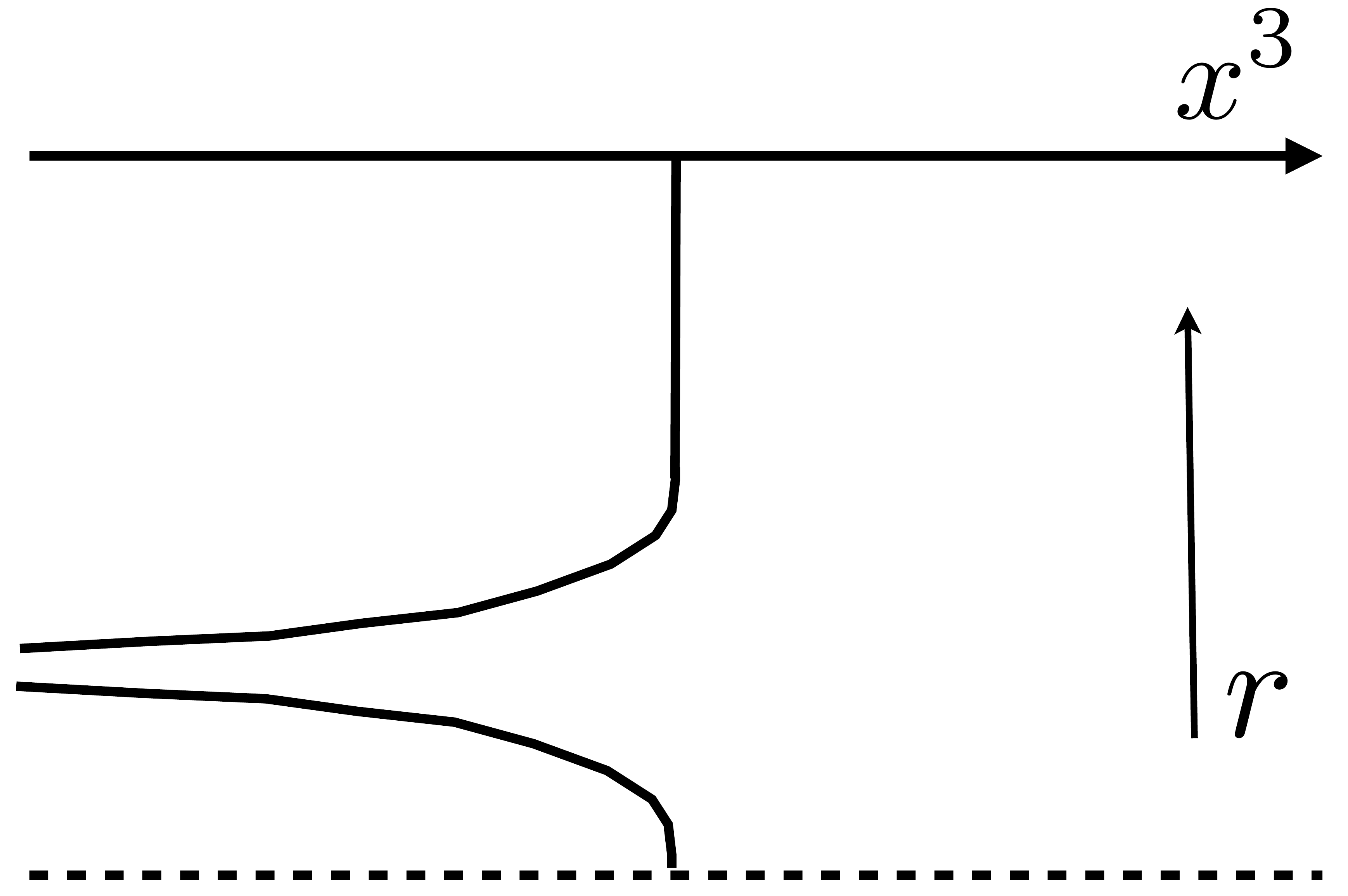} \\
  (a.) & (b.) & (c.)
  \end{array}$
    \caption{Cartoon pictures of some bulk D-brane configurations involved in describing the zero-density Higgs branch in the D3/D5 system. The horizontal axis is $x^3$ and the vertical axis is $r$, with all other directions of $AdS_5 \times \mathbb{S}^5$ suppressed. The horizontal solid black line is the $AdS_5$ boundary $r\to\infty$ while the horizontal dashed line is the Poincar\'e horizon $r=0$. (a.) The solid black vertical line represents the D5-brane localized at $x^3=0$. (b.) The solid blue horizontal half-line represents half D3-branes that have separated from the stack of $N_c$ D3-branes ``hiding'' at $r=0$, and that end on the D5-brane. These D3-branes exert a force on the D5-brane, which then bends and extends to $x^3 \to -\infty$, as depicted in (c.). The final configuration, in (c.), has only a D5-brane carrying D3-brane charge: the source of D3-brane charge is ``hidden'' at the end of the spike, the point at infinity $x^3 \to -\infty$.}
    \label{fig:d5figs}
\end{figure}

\begin{figure}[t]
  \centering
  $\begin{array}{ccc}
  \includegraphics[width=0.32\textwidth]{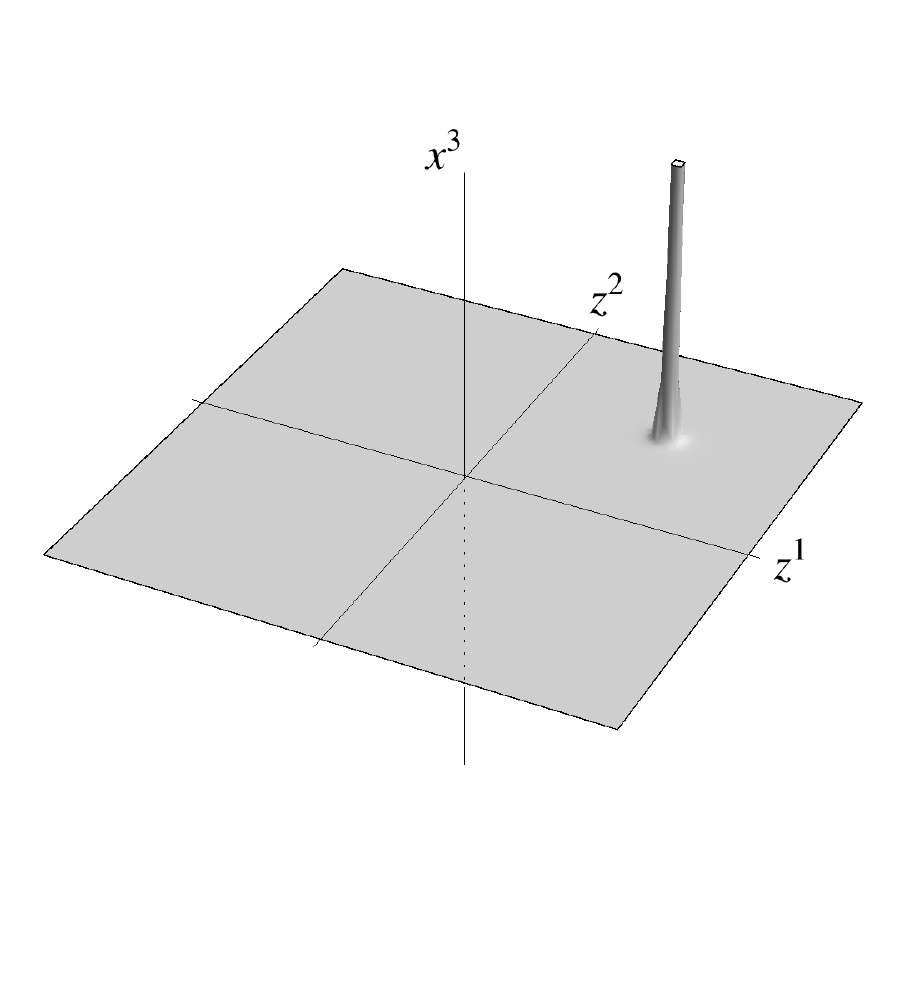} & \includegraphics[width=0.32\textwidth]{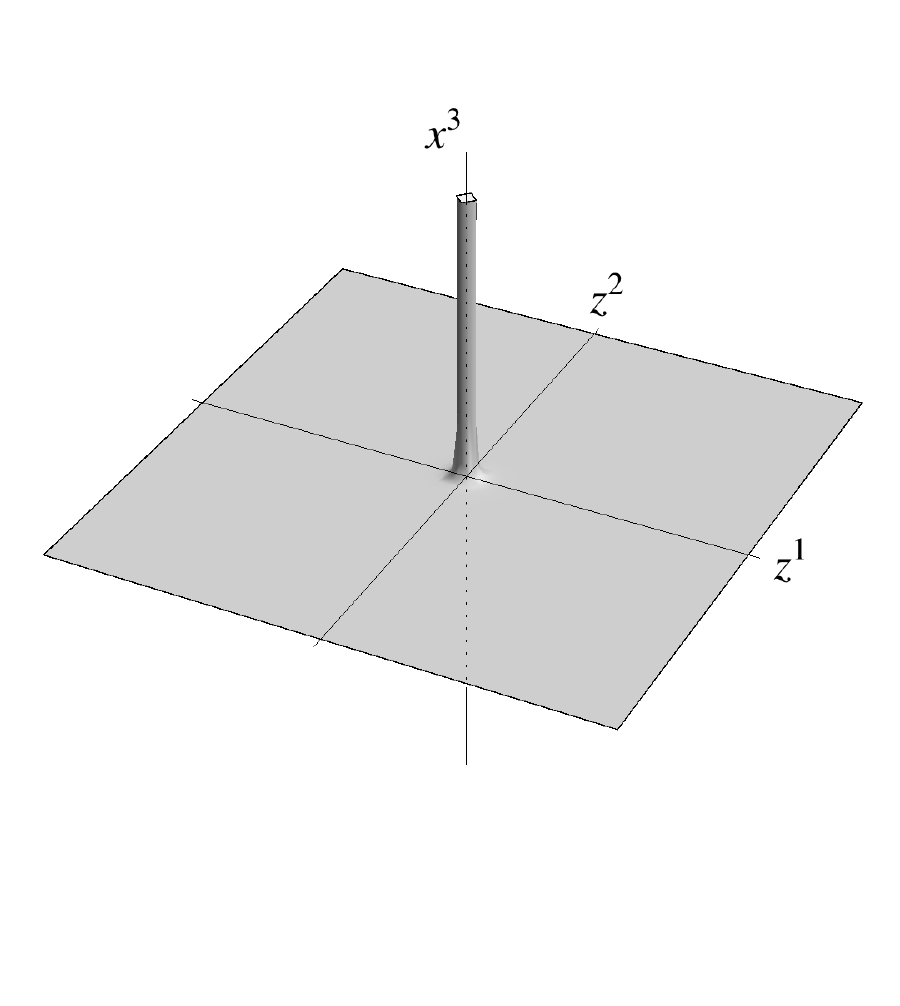} & \includegraphics[width=0.32\textwidth]{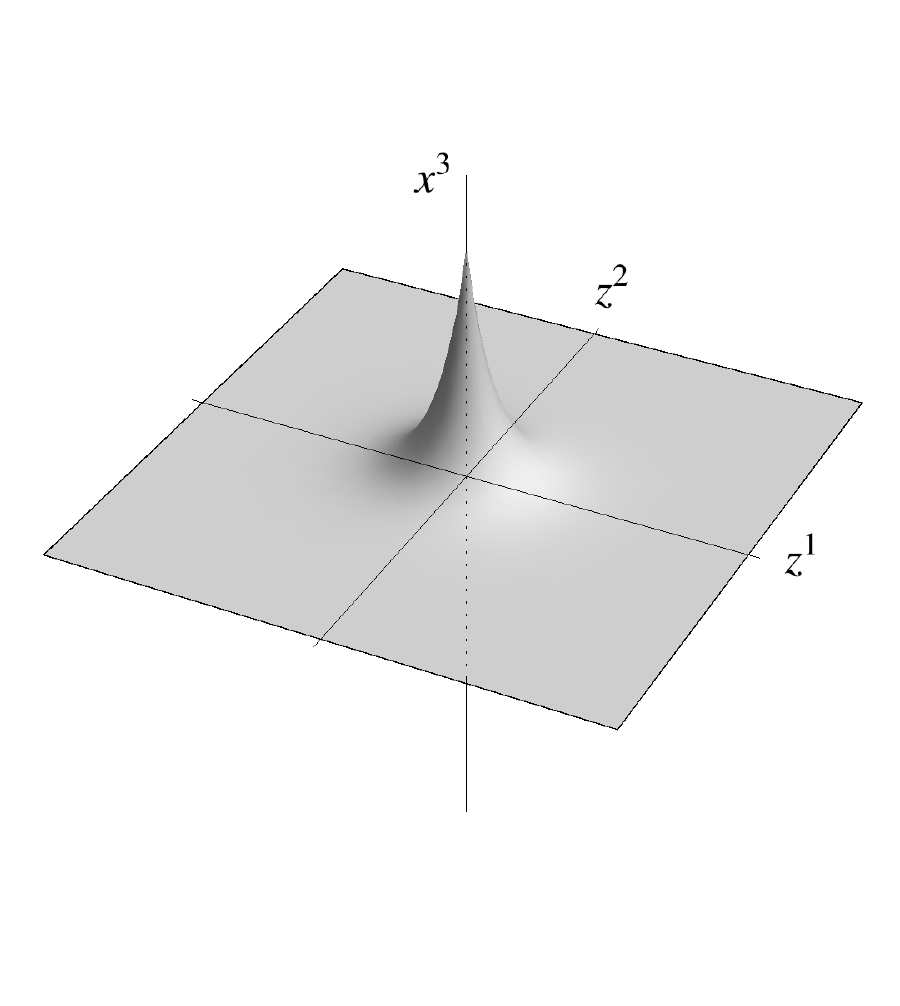} \\
  (a.) & (b.) & (c.)
  \end{array}$
\caption{Depictions of D5-brane solutions describing points on both the zero-density and nonzero-density Higgs branches. The vertical axis is $x^3$ and the horizontal axes are two directions, $z^1$ and $z^2$, of the $\mathbb{R}^3$ spanned by the $z^i$. The grey-shaded surface represents the D5-brane. (a.) A D5-brane that represents a generic point on the either the zero-density or nonzero-density Higgs branch: a D3-brane spike emerges from the D5-brane at some position in $\mathbb{R}^3$ and extends to $x^3(z) \to +\infty$. Such a spike solution is dual to a state in which an infinite number of scalar operators bilinear in the squarks have nonzero expectation values, as explained in the text. (b.) A solution describing a special point on the zero-density Higgs branch, in which the D3-brane spike sits exactly at the origin of $\mathbb{R}^3$. Solutions such as these can describe states in which only a single operator bilinear in the squarks has a nonzero expectation value. (c.) A solution describing a special point on the nonzero-density Higgs branch. Here the D5-brane has a kink, rather than a singularity, at the origin of $\mathbb{R}^3$. These solutions describe states in which two operators bilinear in the squarks have nonzero expectation values, as discussed below eq.~\protect\eqref{E:d3d5bdyData}. The solutions depicted in all of (a.), (b.) and (c.) carry a nonzero net D3-brane charge. Other spike and kink solutions also exist, carrying higher D3-brane multipole charge density distributions. }
\label{fig:d5figs2}
\end{figure}

Solutions of the form in eq.~\eqref{E:d3d5originalHiggs} describe the following physical picture in the bulk. We begin with a D5-brane extended along an $AdS_4 \times \mathbb{S}^2$ localized at $x^3=0$ inside $AdS_5 \times \mathbb{S}^5$, as shown in fig.~\ref{fig:d5figs} (a.). A stack of $N_c$ D3-branes extended along $(x^0,\dots,x^3)$ sits at $r=0$. Consider a solution like that in eq.~\eqref{E:d3d5originalHiggs}, with a \textit{single} singularity at a point $z'$ of strength $C<0$, which represents the endpoint of the following process. First, a number $q = -C/\pi\alpha'$ of D3-branes break on the D5-brane, and the resulting half D3-branes (extended along half the $x^3$ axis) separate from the stack and move up to nonzero $r$, as depicted in fig.~\ref{fig:d5figs} (b.). These D3-branes are localized in $\rho$ and on the $\mathbb{S}^2$, \textit{i.e.} the D3-branes sit at a point $z'$ in $\mathbb{R}^3$. Next, the D5-brane feels a net force due to the half D3-branes, and bends in the $x^3$ direction, or in other words, the D3-branes excite the worldvolume scalars on the D5-brane. The solution in eq.~\eqref{E:d3d5originalHiggs} then represents the final equilibrium configuration in which the D5-brane extends all the way to $x^3 \to -\infty$: at $z'$, the solution in  eq.~\eqref{E:d3d5originalHiggs} diverges to $-\infty$. We depict this final configuration in fig.~\ref{fig:d5figs} (c.), and in fig.~\ref{fig:d5figs2} (a.). The half D3-branes are absent in the final configuration, being replaced by the spike, which has magnetic monopole charge on the $\mathbb{S}^2$ and hence carries some D3-brane charge. We refer to such a spike on the D5-brane as a ``D3-brane spike.'' If $C>0$ then the D5-brane forms a spike going to $x^3 \to + \infty$, with magnetic \textit{anti}-monopole charge on the $\mathbb{S}^2$, representing half D3-branes extended along $x^3>0$ that are blown up into the D5-brane. A solution of the form in eq.~\eqref{E:d3d5originalHiggs} with multiple singularities represents multiple D3-brane spikes on the D5-brane.

Now let us return to the delta-function sources supporting the singularities in the solution of eq.~\eqref{E:d3d5originalHiggs}. We will argue that in fact we must \textit{excise} the points where these delta-functions are located, as is standard practice in various systems involving ``spike'' solutions on D-branes~\cite{Callan:1997kz,Constable:2001ag,Arean:2007nh}. Our argument is a proof by contradiction, using SUSY, as follows. In the D3/D5 theory, SUSY guarantees that the Higgs branch exists for all values of $N_c$, $N_f$, and $\lambda$. Taking the Maldacena and probe limits, and invoking holography, we thus know that some solutions for the D5-brane worldvolume fields must exist that describe all points on the Higgs branch. Obviously such solutions must preserve SUSY, which in particular means they must satisfy a $\kappa$-symmetry condition (for a nice review, see ref.~\cite{CaminoMartinez:2002tj}). We now come to the crucial point: in the appendix, we prove that when $A_0(z)=0$, the fields $x^3(z)$ and $A_i(z)$ satisfy the $\kappa$-symmetry condition \textit{if and only if} they satisfy the vector/scalar duality condition in eq.~\eqref{E:vsDuality}. In other words, when $A_0(z)=0$, the condition for preservation of SUSY is \textit{equivalent} to vector/scalar duality. We thus know that the solutions describing points on the Higgs branch must obey vector/scalar duality, and hence must have $\Box x^3(z)=0$. In the absence of sources, the only solution to $\Box x^3 = 0$ is the trivial solution, which cannot reproduce all points on the Higgs branch. We are thus (apparently) forced to introduce sources in $\mathbb{R}^3$. If we do, however, then those sources will violate vector/scalar duality, and hence also SUSY, \textit{even at points away from the sources}. We thus have a contradiction: SUSY seems to demand the presence of sources that break SUSY. To escape this contradiction, the only option is to \textit{remove} the points in $\mathbb{R}^3$ where the offending sources are located. If we do so, then vector/scalar duality, and hence SUSY, can be preserved at all remaining points in $\mathbb{R}^3$, and we can obtain non-trivial solutions by imposing appropriate boundary conditions near the excised points.

We can also make two less rigorous, but more intuitive, arguments for ignoring the delta-function sources. The first is to notice that the singularities in eq.~\eqref{E:d3d5originalHiggs} are in fact artifacts of our coordinate choice. To see that in a simple example, consider a single D3-brane blown up into the D5-brane at exactly $\rho=0$, described by a solution $x^3(z) \propto 1/\rho$. We can change coordinates on the worldvolume of the D5-brane, choosing $x^3$ to be a worldvolume coordinate rather than $\rho$, in which case the solution becomes $\rho(x^3) \propto 1/x^3$, which is completely regular, and hence requires no source, as $|x^3| \to \infty$. The delta-function sources are thus merely artifacts of our coordinate choice, and hence are unphysical. Crucially, the field strength two-form $F$ is gauge- and coordinate-invariant, hence a violation of its Bianchi identity at $x^3 \to -\infty$ cannot be removed by a gauge or coordinate transformation; the violation of the Bianchi identity is physical.

The second intuitive argument for ignoring the delta-function sources is to notice that the singularity in the $x^3(z)$ of eq.~\eqref{E:d3d5originalHiggs} represents a D3-brane spike, and at the end of each spike is a source of D3-brane charge. These sources of D3-brane charge are sitting at a fixed $\rho$ and at $|x^3| \to \infty$, \textit{i.e.} at the point at infinity, so excising them seems natural. In practice, that means excluding any explicit source terms representing them in $\mathbb{R}^3$ by excising the points in $\mathbb{R}^3$ where those sources are located. As explained in section~\ref{S:4ND}, we can also reach the point at infinity by sitting at a finite value of $(x^1,x^2,x^3)$ and moving towards $\rho = 0$, so we should excise any sources localized at $\rho=0$ as well.

Solutions for $x^3(z)$ of the form in eq.~\eqref{E:d3d5originalHiggs}, and the associated Hodge dual $A_i(z)$, are the most general solutions that are supported by only $L=0$ sources, that obey vector/scalar duality when $A_0(z)=0$, and that are normalizable at the $AdS_5$ boundary. Normalizable solutions to the $A_0(\rho)=0$ vector/scalar duality condition supported by sources with $L\geq1$ also exist. These require higher multipole D3-brane sources at the point at infinity, which we excise as we did in the $L=0$ case. These solutions will have free parameters, analogous to the $C_m$, fixed by boundary conditions near the excised points. Given our arguments above, we can conclude that all of these solutions preserve SUSY, which guarantees that the value of the on-shell action (suitably regulated) for all of these solutions is zero. By taking suitable linear combinations of these solutions, we can construct the most general normalizable SUSY solutions for $x^3(z)$ and $A_i(z)$.

These solutions must describe all points on the SUSY Higgs branch. To translate to the field theory precisely, we need to know what operators are dual to the fields $x^3(z)$ and $A_i(z)$. The field/operator correspondence for the D3/D5 system was worked out in ref.~\cite{DeWolfe:2001pq} for the case where the D5-brane carries zero net D3-brane charge. In what follows we review this correspondence, suitably generalized to the case where the D5-brane carries nonzero net D3-brane charge~\cite{Arean:2007nh}. First, we must decompose $x^3(z)$ and $A_i(z)$ into scalar and vector spherical harmonics on $\mathbb{S}^2$. In $A_{\rho}(z)=0$ gauge, vector/scalar duality implies that $x^3(z)$ and $A_i(z)$ may be decomposed as
\beq
\label{E:x3Decomp}
x^3(z) = \sum_{l=0}^{\infty} x^3_l(\rho)\mathcal{Y}^l(\mathbb{S}^2), \qquad A_{\alpha}(z) = A_{l=0}(\rho)\delta^{\phi}_{\phantom{\phi}\alpha}\sin\theta+ \sum_{l=1}^{\infty}A_l(\rho) \epsilon_{\alpha\beta}\partial^{\beta}\mathcal{Y}^l(\mathbb{S}^2),
\eeq
where Greek indices $\alpha,\beta=\theta,\phi$ are indices on the $\mathbb{S}^2$, which are raised and lowered by the two-sphere metric
\beq
ds^2_{\mathbb{S}^2}=d\theta^2+\cos^2\theta d\phi^2, \qquad \theta \in \left[-\frac{\pi}{2},\frac{\pi}{2}\right], \qquad \phi \in [0,2\pi),
\eeq
$\epsilon^{\alpha\beta}$ is the Levi-Civita tensor on $\mathbb{S}^2$ with orientation $\epsilon^{\theta\phi} = +1/\cos\theta$, and the $\mathcal{Y}^l(\mathbb{S}^2)$ are the scalar spherical harmonics in the spin-$l$ representation of $SO(3)_1$. The modes $x^3_l(\rho)$ and $A_l(\rho)$, with spin-$l$ under the $SO(3)_1$ that acts on the $\mathbb{S}^2$, are dual to operators with spin-$l$ under the $SO(3)_1$ subgroup of the R-symmetry in the field theory. The $l=0$ terms in $x^3(z)$ and $A_{\alpha}(z)$ are special: they encode the total D3-brane charge blown up on the D5-brane, and must be discussed separately. 

For $l\geq 1$, the theory contains two scalar spin-$l$ operators, $\mathcal{O}^{\pm}_l$, dual to two different linear combinations $\varphi^{\pm}_l(\rho)$ of $x^3_l(\rho)$ and $A_l(\rho)$ given by ref.~\cite{DeWolfe:2001pq} as
\begin{subequations}
\label{E:d3d5phidefs}
\begin{align}
\varphi^+_l(\rho) &\equiv l \, A_l(\rho)+ \rho \, x^3_l(\rho)\,,  \qquad \hspace{.9cm} l\geq 1\,, 
\\ 
\varphi^-_l(\rho) &\equiv  (l+1)A_l(\rho) - \rho \, x^3_l(\rho)\,, \qquad l \geq 1\,.
\end{align}
\end{subequations}
In a large-$\rho$ asymptotic expansion, $\varphi^+_l(\rho)$ has a leading, non-normalizable term $\propto \rho^{l+1}$ and a sub-leading, normalizable term $\propto \rho^{-(l+4)}$, while $\varphi^-_l(\rho)$ has a leading non-normalizable term $\propto \rho^{l-3}$ and a sub-leading, normalizable term $\propto \rho^{-l}$. The coefficients of the non-normalizable terms are dual to sources for the opertors $\mathcal{O}^{\pm}_l$ while the coefficients of the normalizable terms are dual to expectation values of the $\mathcal{O}^{\pm}_l$. We thus conclude that the operator $\mathcal{O}_l^+$ has dimension $\Delta_+=l+4$, and $\mathcal{O}_l^-$ has dimension $\Delta_-=l$. Crudely speaking, these operators consist of $l$ of the adjoint scalar fields in the $\mathcal{N}=4$ vector multiplet, with symmetrized, traceless $SO(3)_1$ indices, restricted to the defect at $x^3=0$ and sandwiched between a squark and anti-squark. The precise forms of these operators are discussed in ref.~\cite{DeWolfe:2001pq}. 

For $l=0$, the function $A_{l=0}(\rho)$ is fixed by flux quantization to be proportional to the total D3-brane charge $Q$ blown up on the D5-brane inside a ball of radius $\rho$ centered at the origin of $\mathbb{R}^3$. As $\rho\to\infty$, the function $A_{l=0}(\rho)$ goes to a constant, $\pi \alpha' Q$. When $A_{l=0}(\rho)$ is nonzero, the equation of motion for $x^3_{l=0}(\rho)$ is an inhomogeneous second order equation. This result has two immediate consequences. First, $x^3_{l=0}(\rho)$ is forced to assume a non-trivial profile in equilibrium. For the SUSY solutions we are presently studying, the non-trivial profile for $x^3_{l=0}(\rho)$ is fixed by vector/scalar duality, and at $\rho\to\infty$ approaches $x^3_{l=0}(\rho) = -\pi \alpha' Q/\rho$. Second, the fluctuation of $x^3_{l=0}(\rho)$ around the equilibrium state is dual to an operator with spin-$0$ under $SO(3)_1$. With suitable normalization, $x^3_{l=0}(\rho)$ behaves as a scalar field in $AdS_4$ with mass-squared equal to four, and so is dual to a dimension-four operator.

We can now say more precisely what the solutions for $x^3(z)$ of the form in eq.~\eqref{E:d3d5originalHiggs}, and the associated Hodge dual $A_i(z)$, and their higher-$L$ generalizations, represent in the field theory. Consider first the case of a single source with multipole moment $L$ sitting at the Poincar\'e horizon $\rho=0$. We depict a solution supported by such a source, with $L=0$, in figure.~\ref{fig:d5figs2} (b.). In that case only the mode $x_l^3(\rho)$ with $l=L$ will be nonzero, and indeed will diverge at $\rho=0$, that is, the D5-brane will have a single D3-brane spike exactly at $\rho=0$, with some D3-brane $L^{\textrm{th}}$-multipole source at the end of the spike. Upon applying Hodge duality, we will find for the gauge field that only the mode $A_l(\rho)$ with $l=L$ is nonzero. Via eq.~\eqref{E:d3d5phidefs}, we might then na\"ively conclude that both of the operators $\mathcal{O}^{\pm}_L$ will have nonzero expectation values. A closer examination reveals that only the $\mathcal{O}_L^-$ operators acquire expectation values; the solutions for the bulk fields $x^3_L(\rho)$ and $A_L(\rho)$ conspire in such a way that the $\phi^+_L(\rho)$ and so also the $\langle \mathcal{O}_L^+\rangle$ vanish. Furthermore, since the D5-brane carries some D3-brane charge density, we also expect some of the adjoint scalars of the $\N=4$ SYM vector multiplet to acquire nonzero expectation values. These expectation values will be $x^3$-dependent, with support only in one of the regions $x^3<0$ or $x^3>0$ (whichever side that the D5-brane has the spike), peaked around the defect $x^3=0$ and going to zero as $|x^3| \rightarrow \infty$~\cite{Arean:2006vg,Arean:2006pk}. The expectation values of the squark bilinear and adjoint scalar operators will generically break $SO(3)_1 \times SU(N_c)$ to a subgroup, although the pattern of symmetry breaking can be rather subtle. For example, with an $L=0$ source the amount of D3-brane charge will be different in the $x^3>0$ and $x^3<0$ regions, in which case in the field theory the rank of the gauge group will jump from $N_c$ on one side of the defect to $N_c-Q$ on the other~\cite{Arean:2006vg,Arean:2006pk}, when the D5-brane carries $Q$ total units of D3-brane charge.

The $L^{\textrm{th}}$ multipole source at $\rho=0$ preserves some subgroup of the $SO(3)_1$ isometry. Now imagine displacing that source from $\rho=0$ to some nonzero $\rho$. We depict a solution supported by such a source, with $L=0$, in figure.~\ref{fig:d5figs2} (a.). Clearly such a source will generically break more of the $SO(3)_1$ isometries than when the source was located at $\rho=0$. As a result, such a source will produce a solution in which infinitely many modes $x_l(\rho)$ will be nonzero, and similarly for the $A_l(\rho)$. Translating to the field theory, we expect an \emph{infinite number} of the $\mathcal{O}^{-}_l$ to have nonzero expectation values. The D5-brane will again generically carry some nonzero D3-brane charge density, so as before we expect some of the adjoint scalars to acquire nonzero, $x^3$-dependent expectation values with support only in one of the regions $x^3>0$ or $x^3<0$, and peaked around $x^3=0$. Here we expect these expectation values to approach \textit{nonzero} constants as $|x^3|\to\infty$: in the absence of the D5-brane, a state with D3-branes distributed at points in $\mathbb{R}^3$ would be dual to a point on the Coulomb branch. At such a point some of the adjoint scalars acquire nonzero constant expectation values, so for a D5-brane with a D3-brane spike at nonzero $\rho$, we expect that the dual field theory state should approach a point on the Coulomb branch far from the defect $|x^3|\to\infty$.

Generalizing the above cases to include any number of sources of all possible $L$ at all points in $\mathbb{R}^3$, we find bulk solutions that reproduce all points on the SUSY Higgs branch. We find that at a generic point on the Higgs branch, an infinite number of the $\mathcal{O}^{-}_l$ have nonzero expectation values and some of the adjoint scalars have $x^3$-dependent expectation values peaked around $x^3=0$ and going to nonzero constants as $|x^3|\to\infty$. Special points on the Higgs branch exist where only a finite number of $\mathcal{O}^{-}_l$ have nonzero expectation values. At these points some of the adjoint scalars have $x^3$-dependent expectation values peaked around $x^3=0$ and going to zero as $|x^3|\to\infty$.

The singularities in the $x^3(z)$ of eq.~\eqref{E:d3d5originalHiggs} are qualitatively different from our $A_0(z)=0$ self-dual instanton solutions in the D3/D7 system, eq.~\eqref{E:selfdualSoln}. In the D3/D5 system, the singularities in $x^3(z)$ are physically acceptable. In the D3/D7 system, the point-like instanton solutions in eq.~\eqref{E:selfdualSoln} are genuinely singular: the field strength derived from the solution in eq.~\eqref{E:selfdualSoln} is singular, as are its derivatives, at the core on the instanton. We thus cannot remove these singularities by any gauge transformation. Moreover, in the $\mathbb{R}^4$ spanned by the D7-brane, the singularities are not at the point at infinity, rather they are at a finite distance from the origin, so we cannot argue for their excision. In short, in the D3/D7 system, the singularities are authentic, and render the solutions physically unacceptable, in contrast to the singularities in the solutions in the D3/D5 system. This statement is of course consistent with the fact that the D3/D7 theory has no SUSY Higgs branch while the D3/D5 theory does.

The singularities in the $x^3(z)$ of eq.~\eqref{E:d3d5originalHiggs} have taught us important and general lessons about how to treat potentially dangerous source terms. Foremost among those lessons is that we should excise from the D$p$-brane worldvolume any source sitting at the point at infinity, whether that point is approached by fixing values of $(x^1,x^2,x^3)$ and taking $\rho \to 0$ to reach the Poincar\'e horizon, or by fixing $\rho$ and taking any of $(x^1,x^2,x^3)$ to infinity. In the D3/D5 system with $A_0(z)=0$, SUSY demands the excision of such sources. To be consistent with the SUSY solutions in the D3/D5 system, we have throughout this paper excised any sources at the point at infinity, even in cases without SUSY, including the string sources producing the worldvolume electric flux in eq.~\eqref{eq:solA0} and the D3-brane sources supporting the various self-dual solutions in sections~\ref{S:d3d7}, \ref{S:d3d5}, and \ref{S:d3d3}.

Let us now consider solutions for the worldvolume fields that describe compressible states. Following our recipe for constructing solutions, we first require a solution for $A_0(z)$ that describes a compressible state. Fortunately, we have such a solution: the known ground state represented by the solution in eq.~\eqref{eq:solA0}, with $n=2$,
\beq
\label{E:d3d5solA0}
A_0'(\rho) = \frac{1}{\sqrt{1+\rho^4/\rho_0^4}}\,, \qquad \rho_0^4 = \frac{d^2}{T_5^2\text{vol}(\mathbb{S}^2)^2}.
\eeq
The effective metric corresponding to this solution for $A_0(z)$ is
\beq
\label{E:d3d5effG}
g_{ij}dz^i dz^j = \frac{\rho^4}{\rho^4+\rho_0^4}d\rho^2+\rho^2 ds^2_{\mathbb{S}^2}.
\eeq
As for the effective metric in section~\ref{S:d3d7}, $g_{ij}$ is conformally equivalent to the flat metric: upon redefining the radial coordinate as in eq.~\eqref{rhobar},
\beq
\label{E:d3d5rhobar}
\bar{\rho} \equiv  \rho \left(\frac{1+\sqrt{1+\rho_0^4/\rho^4}}{2}\right)^{1/2},
\eeq
we find
\beq
\label{E:d3d5flatg}
g_{ij}dz^i dz^j = \Omega(\bar{\rho})^2(d\bar{\rho}^2 + \bar{\rho}^2ds^2_{\mathbb{S}^2}), \qquad \Omega(\bar{\rho}) = \left( 1-\frac{\rho_0^4}{4\bar{\rho}^4}\right)^{1/2}\,.
\eeq
The radial coordinate $\rho\in \mathbb{R}^+$, while from eq.~\eqref{E:d3d5rhobar} we see that $\bar{\rho}\in [2^{-1/2}\rho_0,\infty)$. The conformal factor $\Omega(\bar{\rho})$ vanishes at the lower endpoint $\bar{\rho} = 2^{-1/2}\rho_0$, so the effective metric is actually conformally equivalent to $\mathbb{R}^3\backslash\mathbb{B}^3$, that is, $\mathbb{R}^3$ with a three-ball $\mathbb{B}^3$ of radius $2^{-1/2}\rho_0$ excised. The effective metric in eq.~\eqref{E:d3d5flatg} is singular at $\rho=0$, having Ricci scalar $+6\rho_0^4/\rho^6$, but the singularity will not produce any singularities in any of the physical quantities that we will study.

The final step of our recipe is to solve the vector/scalar duality condition, eq.~\eqref{E:vsDuality}, defined by the effective metric in eq.~\eqref{E:d3d5effG}. To do so, we first solve for a harmonic scalar $x^3(z)$ and then impose eq.~\eqref{E:vsDuality}. We can easily construct explicit solutions as follows. Due to the $SO(3)_1$ isometry of the effective metric in eq.~\eqref{E:d3d5effG}, the equation of motion eq.~\eqref{E:x3Lap} is separable into radial and angular pieces. Decomposing $x^3(z)$ into scalar spherical harmonics on $\mathbb{S}^2$ as in eq.~\eqref{E:x3Decomp}, the equation $\Box x^3(z)=0$ implies a second-order linear ordinary differential equation for each of the $x_l^3(\rho)$,
\beq
\frac{d^2 x^3_l}{d\rho^2} + \frac{2\rho^3}{\rho^4+\rho_0^4}\frac{dx^3_l}{d\rho}-l(l+1)\frac{\rho^2}{\rho^4+\rho_0^4}x^3_l=0,
\eeq
with solutions
\beq
\label{E:d3d5x3Sol}
x^3_l(\rho) = b_l\,_2F_1\left(\frac{l+1}{4},-\frac{l}{4};\frac{3}{4};-\frac{\rho^4}{\rho_0^4}\right) +c_l\frac{\rho}{ \rho_0 }\,_2F_1\left(\frac{1-l}{4},\frac{l+2}{4};\frac{5}{4};-\frac{\rho^4}{\rho_0^4}  \right),
\eeq
where $b_l$ and $c_l$ are real constants. We then solve for the $A_i(z)$ by choosing the gauge $A_{\rho}=0$ and substituting eqs.~\eqref{E:x3Decomp} and \eqref{E:d3d5x3Sol} into the vector/scalar duality condition, eq.~\eqref{E:vsDuality},
\beq
\label{E:d3d5Asol}
A_{\alpha}(z) = \rho_0 \, c_0 \,\delta_{~\alpha}^{\phi}  \sin\theta + \sum_{l=1}^{\infty}\frac{\sqrt{\rho^4+\rho_0^4}}{l(l+1)}\frac{dx^3_l}{d\rho}\epsilon_{\alpha \beta} \partial^\beta \mathcal{Y}^l(\mathbb{S}^2).
\eeq

With our solutions for $x^3(z)$ and $A_{\alpha}(z)$, we can construct the $\varphi^{\pm}_l(\rho)$, dual to the operators $\mathcal{O}^{\pm}_l$, via eq.~\eqref{E:d3d5phidefs}. We do not want any external sources in the field theory besides the chemical potential $\mu$, so we will demand that the coefficients of any non-normalizable terms in the $\varphi^{\pm}_l(\rho)$ vanish. We can actually accomplish that as follows. In a large-$\rho$ asymptotic expansion, the $x^3_l(\rho)$ in eq.~\eqref{E:d3d5x3Sol} has a leading, non-normalizable term $\propto \rho^{l}$. The coefficient of that term will vanish if we demand
\beq
\label{E:d3d5cl}
b_l = -\sqrt{2} \, \frac{\Gamma\left(\frac{5}{4}\right)\Gamma\left(\frac{l+1}{2}\right)}{\Gamma\left(\frac{3}{4}\right)\Gamma\left(\frac{l+2}{2}\right)} \, c_l,
\eeq
so normalizability fixes $b_l$ in terms of $c_l$. A straightforward exercise then shows that, if we impose eq.~\eqref{E:d3d5cl}, then the coefficients of the leading, non-normalizable terms in $\varphi^{\pm}_l(\rho)$ also vanish. Once we impose eq.~\eqref{E:d3d5cl}, the leading large-$\rho$ behaviors of our solutions for $\varphi^{\pm}_l(\rho)$ are thus
\begin{subequations}
\label{E:d3d5bdyData}
\begin{align}
\varphi_l^+(\rho) &= c_l\rho_0 \left(\frac{\rho_0}{\rho} \right)^{l+4} \frac{\Gamma\left(\frac{-5-2l}{4}\right) \, \Gamma\left(\frac{1+l}{2}\right) \, \sin\left(\frac{\pi}{4} + \frac{l\pi}{2}\right)}{\sqrt{\pi} \, 2^{\frac{9}{2}+\frac{l}{2}} \, \Gamma\left(\frac{3}{4}\right)} \, \left(1+O(\rho^{-4})\right),& l\geq1, \\
\varphi_l^-(\rho) & = c_l\rho_0 \left(\frac{\rho_0}{\rho} \right)^{l} \frac{\Gamma\left(\frac{3}{4} - \frac{l}{2}\right) \, \Gamma\left(l\right) \, \sin\left(\frac{\pi}{4} +\frac{l\pi}{2}\right)}{2^{-\frac{1}{2}+\frac{3l}{2}} \, \Gamma\left(\frac{2+l}{2}\right) \, \Gamma\left(\frac{3}{4}\right)}\,\left(1+O(\rho^{-4})\right), & l\geq1.
\end{align}
\end{subequations}
We then expect the dual operators to have expectation values proportional to $c_l$: $\langle \mathcal{O}_l^{+}\rangle \propto c_l \rho_0^{l+5}$ and $\langle \mathcal{O}_l^-\rangle \propto c_l \rho_0^{l+1}$. In each case the precise proportionality constant may be computed by holographic renormalization. Notice that nonzero $c_l$ produces nonzero expectation values for \textit{both} of $\mathcal{O}^{\pm}_l$: these operators mix on the nonzero-density Higgs branch.

Once we impose normalizability, the functions of $\rho$ in our solutions eqs.~\eqref{E:d3d5x3Sol} and~\eqref{E:d3d5Asol} are non-singular for all $\rho$. The solutions depend on $\mathbb{S}^2$ directions via the spherical harmonics $\mathcal{Y}^l(\mathbb{S}^2)$, however, so if we take a derivative in an $\mathbb{S}^2$ direction and then take $\rho \to 0$, the solution (and its derivatives) always approaches a finite constant, but the value of that constant depends on the direction in $\mathbb{R}^3$ along which we approach $\rho = 0$. These solutions therefore have kink singularities at $\rho=0$, as depicted in fig.~\ref{fig:d5figs2} (c.), and must be supported by some multipole sources sitting at $\rho=0$. We must promptly excise these sources, as explained above. Intuitively, we can imagine that our solutions represent some multipole distributions of D3-branes blown up into the D5-brane at $\rho=0$. However, recall that any $l\geq1$ solution has zero \textit{net} D3-brane charge.

The solution for $x^3(z)$ given by eqs.~\eqref{E:x3Decomp} and~\eqref{E:d3d5x3Sol}, with arbitrary $b_l$ and $c_l$, is the most general solution to the Laplace's equation $\Box x^3(z)=0$. We can thus construct any solution with arbitrary sources (which we then excise) by stitching together solutions of the form in eqs.~\eqref{E:x3Decomp} and~\eqref{E:d3d5x3Sol} piecewise among various regions of $\mathbb{R}^3\backslash\mathbb{B}^3$, matching the coefficients $b_l$ and $c_l$ in different regions via boundary conditions.

In broad terms, similar to the D3/D7 system, normalizable solutions for vector/scalar dual pairs fall into two classes, those with singularities outside of $\mathbb{B}^3$, meaning at nonzero $\rho$, and those with singularities ``inside the ball,'' although the latter is really just a mnemonic device. Unlike the D3/D7 system, however, solutions with singularities outside the ball are admissable. We present an illustration of solutions with singularities inside of and outside of $\mathbb{B}^3$ in fig.~\ref{F:d3d7_higgs}

For solutions with singularities outside the ball, eq.~\eqref{E:x3Lap} is approximately the equation of motion for the electric potential in three-dimensional electrostatics with a point charge at $z'$, hence the solution for $x^3(z)$ near $z'$ is the Coulomb potential of an electric charge, which diverges as $|z-z'|^{-1}$. These solutions describe D3-brane spikes developing at nonzero $\rho$, without changing the value of the D5-brane's on-shell action. We depict such a solution in fig.~\ref{fig:d5figs2} (a.). These solutions are dual to states in which an infinite number of the $\mathcal{O}^{\pm}_l$ acquire nonzero expectation values and some of the adjoint scalars have $x^3$-dependent expectation values peaked around $x^3=0$ and going to \textit{nonzero} constants as $|x^3|\to\infty$.

Solutions with singularities inside the ball, given by our eqs.~\eqref{E:x3Decomp}, \eqref{E:d3d5x3Sol}, \eqref{E:d3d5Asol}, and \eqref{E:d3d5cl}, are more subtle. To explain why, let us use the language of electrostatics. Suppose we put some charges inside the ball and then compute the electric potential that they produce. In the region outside of the ball, we can produce the same electric potential using some distribution of charge on the \textit{surface} of the ball. A point charge at the center of the ball preserves spherical symmetry and hence is equivalent to only a monopole moment on the surface of the ball. A point charge displaced from the center produces an infinite number of higher multipole moments. In our system, the charges are D3-branes and the potential is our solution for $x^3(z)$. These solutions are dual to states in which we can give the $\mathcal{O}^{\pm}_l$ nonzero expectation values independently for every $l$. As in the D3/D7 system, varying the strengths, values of $L$, and positions of the D3-brane sources corresponds to moving around in the space of these expectation values. We also expect that in these states the adjoint scalars acquire $x^3$-dependent expectation values peaked around $x^3=0$ and going to zero as $|x^3|\to\infty$. These solutions are non-singular only in the presence of the $A_0(\rho)$ in eq.~\eqref{E:d3d5solA0}: in the limit $A_0(\rho)\to0$, these solutions will describe a D5-brane with D3-brane spikes exactly at $\rho=0$. In effect, the worldvolume electric flux de-singularizes these solutions by ``hiding'' the singularity ``inside the ball,'' turning the D3-brane spike, such as the one depicted in fig.~\ref{fig:d5figs2} (b.), into a D3-brane kink, such as the one depicted in fig.~\ref{fig:d5figs2} (c.).

We have thus reproduced the result of ref.~\cite{UW:WIP}, that somehow, miraculously, the entire zero-density Higgs branch survives the introduction of nonzero density. Moreover, many of the same statements we made about the nonzero-density Higgs branch in the D3/D7 system apply also for the nonzero-density Higgs branch in the D3/D5 system. For example, our solutions with singularities inside the ball are normal modes that should correspond to poles in the retarded two-point functions of the $\mathcal{O}_l^{\pm}$, indicating the existence of ``R-spin modes'' in these compressible states of the D3/D5 theory. The fact that D3-branes at $\rho=0$ have the option to dissolve into the D5-branes or not, with no change to the on-shell action, suggests a degeneracy, so perhaps these gapless modes are responsible for the ground-state entropy in these compressible states of the D3/D5 theory. We suspect that the nonzero-density Higgs branch is an artifact of the large-$N_c$ and/or large-$\lambda$ limits: the moduli have no obvious symmetry to protect them, and will probably be lifted by finite-$N_c$ and/or finite-$\lambda$ corrections.

\section{D3/D3 and Holomorphic Scalars}
\label{S:d3d3}

Let us now consider the D3/D3 system~\cite{Constable:2002xt},
\begin{center}
\begin{tabular}{c|cccccccccc}
& $x^0$ & $x^1$ & $x^2$ & $x^3$ & $x^4$ & $x^5$ & $x^6$ & $x^7$ & $x^8$ & $x^9$ \\
\hline
 D$3$ & X & X & X & X & & & & & & \\
 D$3'$& X & X &  & & X & X &  & & & \\
\end{tabular}
\end{center}
The flavor fields break the $SO(6)$ R-symmetry down to $SO(2)\times SO(4)$, where the $SO(2)$ and $SO(4)$ act as rotations in the $(x^4,x^5)$ and $(x^6,..,x^9)$ directions, respectively, and the $SO(2)$ factor is part of the residual R-symmetry. For convenience we will relabel the directions $(x^4,x^5)$, along the D$3'$-brane but transverse to the D3-branes, as $z^i$ with $i=1,2$.

In the Maldacena and probe limits, the D$3'$-brane wraps an asymptotically $AdS_3\times\mathbb{S}^1$ submanifold of $AdS_5\times\mathbb{S}^5$. The action of the D$3'$-brane is
\beq
\label{E:d3d3S}
S_3 = -T_3\int d^4\xi \sqrt{\text{det}(-P[G]_{ab}+F_{ab})}+T_3\int P[C_4].
\eeq
As in the previous two sections, we proceed by considering configurations more general than those reviewed in section~\ref{S:4ND}: we impose the same symmetries as we did in that section, except for $SO(2)$ invariance and the reflection symmetries about $x^2=0$ and $x^3=0$. The most general ansatz for the worldvolume fields is then
\beq
\label{E:d3d3ansatz}
x^2(\xi)=x^2(z), \qquad x^3(\xi)=x^3(z), \qquad A(\xi)=A_0(z)dx^0+A_i(z)dz^i,
\eeq
with all other worldvolume fields vanishing. Substituting the ansatz eq.~\eqref{E:d3d3ansatz} into eq.~\eqref{E:d3d3S}, we find the D$3'$-brane action density
\beq
\label{E:d3d3S2}
s_3 = -T_3\int d^2z\left[ \sqrt{\text{det}(g_{ij}+Z^{-1/2}f_{ij} + Z^{-1}(\partial_i x^2\partial_jx^2+\partial_ix^3\partial_j x^3))}-Z^{-1}\tilde{\epsilon}^{ij}\partial_i x^2\partial_j x^3 \right],
\eeq
where $\tilde{\epsilon}^{ij}$  is the Levi-Civita symbol on $\mathbb{R}^2$ with orientation $\tilde{\epsilon}^{12}\equiv+1$, the factor $Z(\rho)=1/\rho^{4}$, and we have defined an effective metric and field strength on $\mathbb{R}^2$,
\beq
\label{E:2dfields}
g_{ij}=\delta_{ij} - \partial_iA_0\partial_jA_0, \qquad f_{ij}=\partial_iA_j-\partial_jA_i.
\eeq

Our goal is to find solutions for the fields $x^2(z), x^3(z), A_0(z)$, and $A_i(z)$. Once again we can simplify our task by adapting the methods of section~\ref{S:d3d7} to this system. As in our study of the D5-brane, we may put the D$3'$-brane action density eq.~\eqref{E:d3d3S2} into the same form as the D7-brane action density eq.~\eqref{D7action1} by uplifting eq.~\eqref{E:d3d3S2} to an action defined on $\mathbb{R}^4$. To do so, we first introduce two extra directions $z^3,z^4\in [0,1]$, and then uplift the fields on $\mathbb{R}^2$ to fields on $\mathbb{R}^4$ in two steps. First, we define the effective metric $\hat{g}_{ij}$, gauge field $\hat{a}$, and field strength $\hat{f}$, all on $\mathbb{R}^4$, as
\beq
\hat{g}_{ij}=g_{ij} +\delta_{~i}^3\,\delta_{~j}^3+\delta_{~i}^4\,\delta_{~j}^4\, \qquad \hat{a} = A_i(z) dz^i+x^2(z)dz^3+x^3(z)dz^4\,,\qquad \hat{f}=d\hat{a}.
\eeq
Second, we demand that $x^2(z)$, $x^3(z)$, $A_0(z)$, and $A_i(z)$ do not depend on the extra directions $z^3$ and $z^4$. As for the D5-brane we studied in section~\ref{S:d3d5}, our uplift is formally similar to T-duality, in this case in the $x^2$ and $x^3$ directions. The action density eq.~\eqref{E:d3d3S2} may then be formally written as
\beq
\label{E:d3d3S3}
s_3 = -T_3\int d^4z\left[\sqrt{\text{det}(\hat{g}_{ij}+Z^{-1/2}\hat{f}_{ij})}-\frac{1}{8}Z^{-1}\tilde{\epsilon}^{ijkl}\hat{f}_{ij}\hat{f}_{kl}\right],
\eeq
where $\tilde{\epsilon}^{ijkl}$ is the Levi-Civita symbol on $\mathbb{R}^4$ with orientation $\tilde{\epsilon}^{1234}=+1$. We may then straightforwardly apply the topological bound eq.~\eqref{mink2} to find
\beq
\label{E:d3d3bounds}
s_3\leq -T_3\int d^4z\left[ \sqrt{\text{det}\hat{g}_{ij}} + \frac{1}{8}Z^{-1}\left( |\tilde{\epsilon}^{ijkl}\hat{f}_{ij}\hat{f}_{kl}|-\tilde{\epsilon}^{ijkl}\hat{f}_{ij}\hat{f}_{kl}\right)\right].
\eeq
This bound is saturated for $\hat{f}_{ij}$ self-dual with respect to the metric $\hat{g}_{ij}$. For such self-dual $\hat{f}_{ij}$, the D$3'$-brane action density becomes
\beq
\label{E:d3d3S4}
s_3 = - T_3\int d^2z\sqrt{\text{det}\,g_{ij}}.
\eeq
Upon substituting eq.~\eqref{E:2dfields} into eq.~\eqref{E:d3d3S4}, we see that this action is independent of $x^2(z)$, $x^3(z)$, and $A_i(z)$, and is in fact the action of DBI electrostatics in two dimensions. Self-dual $\hat{f}_{ij}$ extremize the action, and hence also solve the $A_i(z)$ equations of motion. Moreover, for self-dual $\hat{f}_{ij}$ the equation of motion for $A_0(z)$ follows from variation of eq.~\eqref{E:d3d3S4}, which is the equation of motion for the electric potential in DBI electrostatics. Self-dual $\hat{f}_{ij}$ will contribute nothing to the D$3'$-brane stress-energy tensor. We can obtain (a subset of all) solutions for $A_0(z)$ and $\hat{f}_{ij}$ by a simple recipe: we first solve for $A_0(z)$, which determines the effective metric, which in turn defines the self-duality condition for $\hat{f}_{ij}$, which we then need to solve.

The key difference between the D$3'$-brane and the D7-brane is that $x^2(z)$, $x^3(z)$, and the $A_i(z)$ do not depend on $z^3$ or $z^4$, so the self-duality condition for $\hat{f}_{ij}$ becomes
\beq
\label{E:holo}
F_{12}=0, \qquad \partial_i x^2 = -\epsilon_{ij}\partial^j x^3,
\eeq
where we have defined the Levi-Civita tensor $\epsilon^{ij}\equiv\tilde{\epsilon}^{ij}/\sqrt{\text{det}\,g_{ij}}$, and indices are raised and lowered with the metric $g_{ij}$. The condition $F_{12}=0$ implies that the $A_i(z)$ are locally gauge-equivalent to zero. To understand the condition on $x^2(z)$ and $x^3(z)$ in eq.~\eqref{E:holo}, let us re-write the effective metric. In two dimensions every metric is locally conformally equivalent to the flat metric, so upon introducing complex coordinates $u$, $\bar{u}$, we can write the effective metric as
\beq
\label{E:holoCoord}
g_{ij}\,dz^idz^j = \Omega^2(u,\bar{u}) \, du \, d\bar{u}.
\eeq
Next we combine the two real fields $x^2(z)$ and $x^3(z)$ into a single complex field,
\beq
w(u,\bar{u}) \equiv x^2(z) + i x^3(z).
\eeq
Given that both $x^2(z)$ and $x^3(z)$ are real functions of the $z^i$, we have $\bar{w}(u,\bar{u})=x^2(z)-ix^3(z)=w(\bar{u},u)$. The condition on $x^2(z)$ and $x^3(z)$ in eq.~\eqref{E:holo} then becomes
\beq
\label{E:wholo}
\partial_{\bar{u}}w(u,\bar{u})=0,
\eeq
which is the condition for holomorphicity of $w(u,\bar{u})$.

The only entire holomorphic functions of the complex plane are constants, so if we want non-trivial solutions for $w(u,\bar{u})$ then we must consider functions with singularities. The simplest such singularities are poles of $L^{\textrm{th}}$ degree at various points in $\mathbb{R}^2$. A holomorphic function $w(u)$ with such isolated singularities will in fact be harmonic except at the locations of the singularities, \textit{i.e.} will solve a Poisson equation with localized sources. The most general form of that Poisson equation, for a single source localized at a point $z'$, is
\beq
\label{E:wLap}
\Box w(z) = \sum_{L=1}^{\infty}s_L \, (\partial_u)^L \delta (z-z'),
\eeq
where $\Box$ is the Laplacian with respect to the metric $g_{ij}$ and the $s_L$ are arbitrary constants. Note that we exclude $L=0$ sources on the right hand side of eq.~\eqref{E:wLap}; the Green's function of the scalar Laplacian in two dimensions goes as $\ln |z-z'|^2$ and so is not holomorphic and is therefore inconsistent with eq.~\eqref{E:wholo}. In string theory terms, on the right-hand side of eq.~\eqref{E:wLap} the terms with $L$ derivatives correspond to an $(L-1)^{th}$ multipole moment of D3-branes that merge with the D$3'$-brane at $z'$. By charge conservation, any net D3-brane charge on the D$3'$-brane must come from the stack of $N_c$ D3-branes at the bottom of $AdS_5$.

To illustrate our method, let us reproduce the known solutions that describe points on the zero-density SUSY Higgs branch~\cite{Constable:2002xt,Arean:2007nh}. Taking $A_0(z)=0$, the effective metric eq.~\eqref{E:2dfields} reduces to the flat metric on $\mathbb{R}^2$, so that the holomorphic coordinate built out of $z^1$ and $z^2$ is just $u=z^1+iz^2$. Per our discussion above, the most general normalizable solution to eq.~\eqref{E:wLap} with only D3-monopole ($L=1$) sources is
\beq
\label{E:wSol}
w(u) = \sum_m \frac{C_m}{u-u_m}, \qquad \bar{w}(u)=w(\bar{u}),
\eeq
for finite constants $C_m$. The solution in eq.~\eqref{E:wSol} clearly diverges at the points $u_m$, the locations of the sources, whose strengths fix the $C_m$. The singularities in $w(u)$ represent D3-branes that have merged with the D$3'$-brane, producing a spike of D$3'$-brane carrying nonzero D3-brane charge. We thus expect the $C_m$ to obey a quantization condition, since the number of D3-branes is quantized. As explained in detail in section~\ref{S:d3d5}, in fact we must excise the points where the sources are located, and fix the $C_m$ by imposing boundary conditions at points near the $u_m$. Ultimately, then, the singularities in $w(u)$ are physically admissable. As shown in ref.~\cite{Arean:2007nh}, the holomorphicity condition eq.~\eqref{E:wholo} is equivalent to the $\kappa$-symmetry condition, hence the solution in eq.~\eqref{E:wSol} preserves SUSY. We can thus also conclude that for the solution in eq.~\eqref{E:wSol} the suitably regulated on-shell action is zero. Solutions also exist that are supported by sources with any multipole moment $L$. Such solutions represent mutlipole moments of D3-branes that merge with the D$3'$-brane. Most of our statements, suitably modified for arbitrary $L$, apply for these solutions as well. By linear superposition we can obtain the most general normalizable SUSY solutions, supported by any number of sources with any $L$ at any points in $\mathbb{R}^2$.

These most general solutions must describe all points on the SUSY Higgs branch. To translate to the field theory, we need to know what operators are dual to the fields $x^2(z)$ and $x^3(z)$. The field/operator correspondence in the D3/D3 system was worked out in ref.~\cite{Constable:2002xt}. If we decompose $w(z)$ into Fourier modes on the $\mathbb{S}^1$,
\beq
\label{E:wDecomp}
w(z) = \sum_{l=-\infty}^{\infty} w_l(\rho) \, e^{i l \theta},
\eeq
then each of the $\varphi_l(\rho)\equiv \rho\,w_l(\rho)$ is dual to a scalar operator of charge $l$ under the $SO(2)$ symmetry. In a large-$\rho$ asymptotic expansion, the $\varphi_l(\rho)$ with $l\leq-1$ have a leading, non-normalizable term $\propto \rho^{-(l+3)}$ and a sub-leading, normalizable term $\propto \rho^{l+1}$, while $\varphi_l(\rho)$ with $l\geq-1$ have a leading, non-normalizable term $\propto \rho^{l+1}$ and a sub-leading, normalizable term $\propto \rho^{-(l+3)}$. We thus conclude that the $l\leq-1$ modes are dual to operators $\mathcal{O}_{l\leq-1}$ of dimension $-(l+1)$ while the $l\geq-1$ modes are dual to operators $\mathcal{O}_{l\geq-1}$ of dimension $l+3$. Crudely speaking, the $\mathcal{O}_{l\leq -1}$ consist of $|l+1|$ adjoint scalar fields of the $\N=4$ vector multiplet restricted to the defect at $x^2=x^3=0$ and sandwiched between a squark and anti-squark. The precise forms of these operators are discussed in ref.~\cite{Constable:2002xt}.

We can now translate the data of the bulk solutions we have found to that of the SUSY Higgs branch. We will be brief, as the results are qualitatively similar to those of the SUSY Higgs branch of the D3/D5 theory we reviewed in section~\ref{S:d3d5}. The first class of solutions are those with $(L-1)^{\textrm{th}}$ D3-multipole moments localized at $\rho=0$, at which the embedding $w(u)$ diverges. The $SO(2)$ symmetry that rotates $\mathbb{S}^1$ is broken to a discrete subgroup, and the only $\mathcal{O}_l$ with a non-vanishing expectation value is the single operator with $l=-L$. The second class of solutions are those supported by $(L-1)^{\textrm{th}}$ D3-multipole moments localized away from $\rho=0$, so that $w(u)$ diverges away from the origin of $\mathbb{R}^2$. For these embeddings the $SO(2)$ symmetry is broken to a smaller subgroup than that preserved by a multipole located at the origin. As a result, an \emph{infinite number} of the $\mathcal{O}_l$ (with $l\leq -1$) acquire nonzero expectation values. In both classes of solutions, the D$3'$-brane is endowed with nonzero D3-brane charge, and generically represent multipole moments of D3-branes merging with the D$3'$-brane at various positions in $\mathbb{R}^2$. As a result, some of the adjoint scalars of $\N=4$ SYM acquire nonzero expectation values~\cite{Constable:2002xt} which depend on the $x^2$ and $x^3$ directions. Together, the squark and adjoint scalar expectation values will generically break $SO(2) \times SU(N_c)$ to a subgroup. 

As we mentioned above, the most general normalizable SUSY-preserving bulk solution may be obtained by linear superposition of the $w(u)$ corresponding to any number of sources of all possible $L$ located at arbitrary points in $\mathbb{R}^2$. The set of these bulk solutions reproduces all points on the SUSY Higgs branch.

Our ($1+1$)-dimensional squarks are massless. Massless scalars in ($1+1$)-dimensions generically have strong infra-red fluctuations, producing logarithmic large-distance divergences in their correlation functions. We thus expect such logarithmic large-distance divergences in the correlators of the scalar operators that parameterize the Higgs branch, which are bilinear in the squarks. As a result, we expect that the quantum mechanical vacuum wave function will not remain localized at a single point on the Higgs branch, but rather will spread out over the entire Higgs branch~\cite{Constable:2002xt}. This is not apparent in our calculation because of the large-$N_c$ limit, which suppresses the strong infra-red fluctuations of the squark fields.

Let us now consider solutions for the worldvolume fields that describe compressible states. Following our recipe for constructing solutions for the fields $x^2(z), x^3(z), A_0(z)$, and $A_i(z)$, we begin by finding a profile for $A_0(z)$ that extremizes eq.~\eqref{E:d3d3S4}. Fortunately, we already have such a solution, eq.~\eqref{eq:solA0} with $n=1$,
\beq
A_0'(\rho) = \frac{1}{\sqrt{1+\rho^2/\rho_0^2}}, \qquad \rho_0^2 = \frac{d^2}{T_3^3\text{vol}(\mathbb{S}^1)^2},
\eeq
where $\rho^2=(z^1)^2+(z^2)^2$. The effective metric eq.~\eqref{E:2dfields} associated with this solution is
\beq
\label{E:d3d3effG}
g_{ij}dz^idz^j = \frac{\rho^2}{\rho^2+\rho_0^2}d\rho^2+\rho^2d\theta^2,
\eeq
where $\theta \in [0,2\pi)$. This effective metric is conformally equivalent to the flat metric: upon redefining the radial coordinate as in eq.~\eqref{rhobar},
\beq
\label{E:d3d3rhobar}
\bar{\rho} \equiv \rho \left(\frac{1+\sqrt{1+\rho_0^2/\rho^2}}{2} \right),
\eeq
and introducing the complex coordinate $u\equiv \bar{\rho} \, e^{i\theta}$, we can write the effective metric in eq.~\eqref{E:d3d3effG} as
\beq
\label{E:d3d3effGflat}
g_{ij} \, dz^i dz^j = \Omega(\bar{\rho})^2dud\bar{u}, \qquad \Omega(\bar{\rho}) \equiv 1-\frac{\rho_0^2}{4\bar{\rho}^2}.
\eeq
The radial coordinate $\rho \in \mathbb{R}^+$, however from eq.~\eqref{E:d3d3rhobar} we see that $\bar{\rho} \in [\rho_0/2,\infty)$. At the lower endpoint, $\bar{\rho} = \rho_0/2$, the conformal factor $\Omega(\bar{\rho})$ vanishes, so the effective metric is actually conformally equivalent to $\mathbb{R}^2$ with a two-ball $\mathbb{B}^2$ of radius $\rho_0/2$ excised. The effective metric in eq.~\eqref{E:d3d3effG} is singular at $\rho=0$, having Ricci scalar $+2\rho_0^2/\rho^4$, but this singularity will not produce any singularities in any of the physical quantities that we will study.

The second step of our recipe is to solve eq.~\eqref{E:holo} for $x^2(z)$ and $x^3(z)$. That is trivial to do, however: according to eq.~\eqref{E:wholo}, any function $w(z) = x^2(z) + i x^3(z)$ that is holomorphic in the coordinate $u = \bar{\rho} \, e^{i \theta}$ provides a solution for $x^2(z)$ and $x^3(z)$. The most general normalizable solution supported by sources of arbitrarily high $L$ localized at $\rho=0$ is
\beq
\label{E:wNonSingular}
w(z) = \sum_{L=1}^{\infty} \frac{c_L}{u^L} = \sum_{L=1}^{\infty} \frac{c_L \, 2^L e^{-iL\theta}}{\rho^L(1+\sqrt{1+\rho_0^2/\rho^2})^L}\,,
\eeq
for finite constants $c_L$ that are fixed by the strengths of the sources. In the second equality of eq.~\eqref{E:wNonSingular} we used $u=\bar{\rho}\,e^{i\theta}$ and then we used eq.~\eqref{E:d3d3rhobar} to convert $\bar{\rho}$ into $\rho$. The sources supporting the solution in eq.~\eqref{E:wNonSingular} represent $(L-1)^{\textrm{th}}$ multipole moments of D3-brane charge; the solution in eq.~\eqref{E:wNonSingular} represents some multipole distribution of D3-branes that have merged with the D$3'$-brane at $\rho=0$. As we argued in section.~\ref{S:d3d5}, we must excise the point at infinity, $\rho=0$, where these sources are located, in which case the $c_L$ are fixed by boundary conditions near $\rho=0$.

The normalizable solution in eq.~\eqref{E:wNonSingular} has only modes $w_l(\rho)$ with $l\leq-1$. The mode $w_l(\rho)$ corresponds to term in eq.~\eqref{E:wNonSingular} with $l=-L$, so we expect that, after careful holographic renormalization, each $c_L$ will be proportional to the expectation value of $\mathcal{O}_{-L}$. Solutions of the form in eq.~\eqref{E:wNonSingular} thus describe points on a moduli space in the compressible states of our theory. We can also construct more general solutions supported by sources with any multipole moment $L$ at any point in $\mathbb{R}^2$. All such solutions represent some multipole distribution of D3-branes that merge with the D$3'$-brane at points in $\mathbb{R}^2$.

In broad terms, as in the D3/D7 and D3/D5 systems, solutions fall into two classes. The first possess singularities outside of $\mathbb{B}^2$, meaning at nonzero $\rho$, and the second, given by eq.~\eqref{E:wNonSingular}, have singularities inside of $\mathbb{B}^2$, although since the points in $\mathbb{B}^2$ are excised the latter is really just a mnemonic device. We present an illustration of solutions with singularities outside and inside the ball in fig.~\ref{F:d3d7_higgs}. Solutions with singularities outside the ball correspond to embeddings where D3-multipoles merge with the D$3'$-brane away from $\rho=0$. Like the corresponding states on the SUSY Higgs branch, these states are characterized by an infinite number of operators $\mathcal{O}_{l\leq -1}$ with nonzero expectation values. Furthermore, the D$3'$-brane stretches to spatial infinity where the D3-multipoles merge. The solutions with singularities ``inside the ball'' are completely non-singular and correspond to embeddings where multipole moments of D3-brane merge at the Poincar\'e horizon $\rho=0$. They are similar to the corresponding SUSY embeddings, in that the expectation values of the $\mathcal{O}_{l\leq -1}$ may be independently dialed, but differ in that the D$3'$-brane does not diverge at the D3-multipoles. Instead the D$3'$-brane intersects the Poincar\'e horizon at finite values of $x^2$ and $x^3$. Just as in the D3/D7 and D3/D5 systems, the worldvolume electric flux has ``de-singularized'' the singularity at $\rho=0$, so that the infinite spike of D$3$-brane becomes a D$3$-brane kink. The existence of both classes of solutions together indicates that somehow, miraculously, the SUSY Higgs branch survives the breaking of SUSY and persists at nonzero density.

Finally, many statements we made about the nonzero-density Higgs branch in the D3/D7 and D3/D5 systems apply also for the nonzero density Higgs branch in the D3/D3 system. Our solutions with singularities inside the ball, eq.~\eqref{E:wNonSingular}, are normal modes that should correspond to poles in the retarded two-point functions of the $\mathcal{O}_{l\leq-1}$, indicating the existence of ``R-spin modes'' in these compressible states of the D3/D3 theory. The fact that D3-branes at $\rho=0$ have the option to merge with the D$3'$-brane or not, with no change to the on-shell action, suggests a degeneracy, so perhaps these gapless modes are responsible for the ground state entropy in these compressible states of the D3/D3 theory. We suspect that the nonzero-density Higgs branch is an artifact of the large-$N_c$ and/or large-$\lambda$ limits, and will probably be lifted by finite-$N_c$ and/or finite-$\lambda$ corrections. Indeed, as explained above, we expect that even the zero-density SUSY Higgs branch will not survive finite-$N_c$ corrections, due to the spreading of the wave function over the entire SUSY Higgs branch.

\section{Classifying Holographic Matter?}
\label{S:class}

In all three of the D3/D$p$ systems, we have discovered a moduli space of compressible states. How generic are such nonzero-density moduli spaces? As a step towards answering that question, we have studied a variety of compressible states described holographically by probe D-branes with worldvolume electric flux in various supergravity backgrounds. We will call these states collectively ``holographic matter,'' extending our definition of that term beyond the D3/D$p$ systems alone.

We have considered two supergravity backgrounds. The first are the solutions representing the near-horizon geometries of D$q$-branes, with $q=1,2,3,4,6$ \cite{Itzhaki:1998dd}. Each of these backgrounds includes a non-trivial metric with a holographic radial coordinate, $N_c$ units of RR flux, $\star F_{q+2}=F_{10-q}$, on an internal space, and for $q \neq 3$ a non-trivial dilaton. Type II supergravity in these backgrounds is dual to $(q+1)$-dimensional, maximally-SUSY $SU(N_c)$ Yang-Mills theory, in the 't Hooft limit. We did \textit{not} consider solutions with any field theory directions compactified, and in particular for $q=4$ we studied systems more like the ``stringy NJL'' model~\cite{Antonyan:2006vw} rather than systems like the Sakai-Sugimoto model~\cite{Sakai:2004cn}. In the $q\neq3$ cases, the duality is only reliable over a limited range of energy scales in the field theory, where the coupling is large and where deviation from conformal invariance arises solely from the dimensionful Yang-Mills coupling. Holography has revealed that the $q\neq3$ theories exhibit hyperscaling violation of a ``hidden'' conformal symmetry: for details, see refs.~\cite{Huijse:2011ef,Dong:2012se}. The second supergravity background we have considered is the $AdS_4\times\mathbb{CP}^3$ solution of type IIA supergravity, which has $N_c$ units of $F_4=dC_3$ flux in the $AdS_4$ and nonzero $F_2=dC_1$ flux on the $\mathbb{CP}^1 \in \mathbb{CP}^3$. In this case the dual field theory is (2+1)-dimensional $\N=6$ superconformal $U(N_c) \times U(N_c)$ Chern-Simons-matter theory, the ABJM theory~\cite{Aharony:2008ug}, in the 't Hooft limit and with large (but not too large) Chern-Simons levels.

A variety of different probe D$p$-branes can be embedded in these backgrounds. In the near-horizon D$q$-brane geometries, a probe D$p$-brane is characterized completely by the number of ND directions counted with respect to the D$q$-branes. A 4ND D$p$-brane \cite{Myers:2006qr} preserves half the SUSY of the background, and is dual to a fundamental-representation hypermultiplet. The cases $p=q+4,q+2,q$ correspond to hypermultiplets of codimension zero, one, and two (which obviously requires $q\geq2$), respectively. A 6ND D$p$-brane breaks all SUSY, and is dual to a fundamental-representation fermion alone, with no scalar superpartner. The cases $p=q+6,q+4,q+2,q$ correspond to fermions of codimension zero, one, two (when $q\geq2$), and three (when $q\geq3$), respectively. The 6ND D$p$-branes are generally unstable, in the sense that usually one of the worldvolume scalars is tachyonic. For example, in the simplest 6ND D3/D7 system, the D7-brane is extended along $AdS_4 \times \mathbb{S}^4$ inside $AdS_5 \times \mathbb{S}^5$, and the D7-brane worldvolume scalar representing fluctuations in the $\mathbb{S}^5$ direction transverse to the $\mathbb{S}^4$ has a mass violating the Breitenlohner-Freedman bound of $AdS_4$~\cite{Rey_Talk,Davis:2008nv}. Often such a tachyon can be avoided by wrapping the 6ND D$p$-brane on a non-trivial cycle and/or by introducing nonzero worldvolume gauge field flux~\cite{Myers:2008me,Bergman:2010gm,Jokela:2011eb}. In some cases such a tachyon is absent for trivial reasons, for example in the 6ND D4/D8 system the D8-branes wrap the entire compact space of the background and so obviously no such tachyon exists. The 8ND D$p$-branes are a little more exotic. These preserve half the SUSY of the background, but are dual to fermions alone, with no scalar superpartners. SUSY is preserved because the fermions are neutral under SUSY. The 8ND D$p$-branes fall into two classes, based on the dimensionality of the fermions. The first class are D$(8-q)$ branes~\cite{Camino:2001at}, dual to (0+1)-dimensional fermions that may be interpreted as Kondo-like impurities in the $(q+1)$-dimensional SYM theory~\cite{Kachru:2009xf,Jensen:2011su,Harrison:2011fs,Benincasa:2012wu}. The second class of 8ND D$p$-branes are D$(10-q)$-branes, which are dual to $(1+1)$-dimensional fundamental-representation chiral fermions.\footnote{A glaring omission from our list are 2ND probe D$p$-branes. They break all SUSY, and are dual to fundamental-representation scalars alone. To date, many basic questions about them (in the context of holography) have not been answered. For example, being non-SUSY, an obvious question is whether these D$p$-branes are stable, and, if not, then what is the endpoint of the instability? If they are unstable, can they be stabilized, for example by introducing some appropriate worldvolume flux? Answering these questions is clearly beyond the scope of this paper, so we have omitted the 2ND probe D$p$-branes from the analysis of this section.}

The $AdS_4 \times \mathbb{CP}^3$ solution of type IIA supergravity does not arise simply as the near-horizon geometry of a stack of D-branes alone, hence probe D$p$-branes in $AdS_4 \times \mathbb{CP}^3$ cannot be characterized as easily as in the near-horizon D$q$-brane backgrounds. Nevertheless, $N_c$ D2-branes are involved in constructing the $AdS_4 \times \mathbb{CP}^3$ solution, so we can partially characterize probe D$p$-branes by counting the number of ND directions with respect to those D2-branes. Many kinds of probe D$p$-branes have been studied in $AdS_4 \times \mathbb{CP}^3$: we refer the intrepid reader to refs.~\cite{Fujita:2009kw,Hohenegger:2009as,Gaiotto:2009tk,Hikida:2009tp,Ammon:2009wc,Jensen:2010vx} for details. For simplicity, we will focus on only three probe D$p$-branes: (i.) a D6-brane wrapping $AdS_4\times\mathbb{RP}^3$, (ii.) a D4-brane wrapping $AdS_3\times\mathbb{CP}^1$, and (iii.) a D8-brane wrapping $AdS_3\times\mathbb{CP}^3$. With respect to the D2-branes involved in producing the background, these are 4ND, 4ND, and 8ND, respectively. Starting now, we include these cases when we refer to 4ND or 8ND D$p$-branes.

For all of the above probe D$p$-branes, we have performed an analysis similar to that of sections~\ref{S:4ND},~\ref{S:d3d7},~\ref{S:d3d5}, and~\ref{S:d3d3}, that is, we introduced worldvolume electric flux and studied the implications of topological bounds on the D$p$-brane action~\cite{Gibbons:2000mx}. We will omit the details of our analysis, which are unilluminating, and instead simply summarize the patterns that we find. As in the 4ND D3/D$p$ systems, the probe D$p$-brane action for the field strength $\hat{f}$ (including dummy indices where necessary) reduces to a sum of two terms, a DBI term and an axionic term, on a space with an effective metric. Importantly, the dimension of this space is the number of ND directions, so for 4ND, 6ND, and 8ND D$p$-branes, the axionic term is of the form $\hat{f} \wedge \hat{f}$, $\hat{f}\wedge \hat{f} \wedge \hat{f}$, and $\hat{f}\wedge \hat{f} \wedge \hat{f} \wedge \hat{f}$, respectively. In the 4ND systems, we find that self-dual field strengths non-trivially saturate the topological bound on the D$p$-brane action. In the 6ND and 8ND systems, the topological bound is saturated if and only if the field strength $\hat{f}$ vanishes. Our principal result is that all of the 4ND D$p$-branes exhibit nonzero-density moduli spaces,\footnote{An exception is the 4ND D4-brane along $AdS_3 \times \mathbb{CP}^1 \subset AdS_4 \times\mathbb{CP}^3$, as we discuss below.} while the 6ND and 8ND D$p$-branes do not. As an ancillary result, notice that those 6ND systems that require fluxes for stabilization will never saturate the topological bound.

Of course, that the 6ND and 8ND D$p$-branes lack nonzero-density moduli spaces is no surprise: we already know that the physics of these D$p$-branes, in the presence of nonzero worldvolume electric flux, is very different from that of the 4ND D$p$-branes. For instance, for the 6ND D$p$-branes,  even after any obvious tachyon has been stabilized, a new instability appears in the presence of nonzero worldvolume electric flux: in field theory terms, the na\"ive, translationally-invariant compressible ground state has a known spatially-modulated instability, \textit{i.e.} is unstable against ``striping.'' Crucially, in bulk terms this instability is due to the WZ term in the D$p$-brane action. In the 6ND D4/D8 system, for example, this instability is just the Nakamura-Ooguri-Park (NOP) spatially-modulated instability~\cite{Nakamura:2009tf,Ooguri:2010xs}, realized here via the D8-brane action's $\text{P}[C_3]\wedge F\wedge F \wedge F$ term, which reduces to a Chern-Simons term $A\wedge F \wedge F$ in the non-compact directions of the worldvolume theory. The resulting Chern-Simons level is large enough to trigger the NOP instability. In the 6ND D3/D7 system, the non-compact part of the worldvolume is four-dimensional, and a four-dimensional analogue of the NOP instability occurs: the D7-brane action's $\text{P}[C_4] \wedge F \wedge F$ term leads to a four-dimensional axionic term that is ultimately responsible for a spatially-modulated instability~\cite{Bergman:2011rf}. We have reason to believe that the 6ND D2/D8 system also exhibits a spatially-modulated instability of similar origin \cite{striped_WIP}.

As for the 8ND D$p$-branes, when the D$p$-brane is dual to a $(0+1)$-dimensional fermion, the known compressible ground state preserves SUSY~\cite{Camino:2001at}, whereas when the D$p$-brane is dual to $(1+1)$-dimensional chiral fermions, the baryon number $U(1)$ is anomalous. In the latter case, the dynamics of the current are dominated by the anomaly and, up to non-perturbative effects\footnote{Strictly speaking, the calculations in ref.~\cite{Jensen:2010em} apply either when the $U(1)$ is non-compact or when the chemical potential $\mu$ is small. When the $U(1)$ is compact, the partition function may receive additional non-perturbative contributions from the twisting of winding modes in the presence of nonzero $\mu$.}, the $U(1)$ charge is topological and decouples from the rest of the physics~\cite{Kraus:2006wn,Jensen:2010em}.

We summarize our observations in table~\ref{T:468nd}.
\begin{table*}[tc]
\begin{center}
\begin{tabular}{c||c|c|c}
  & 4ND & 6ND & 8ND \\
\hline Higgs branch & X & & \\
\hline Striped instability & & X & \\
\hline SUSY & & & X: $(0+1)$d fermion \\
\hline Chiral anomaly & X: D4's in ABJM & & X: $(1+1)$d fermion
\end{tabular}
\caption{A summary of the known compressible ground states described by probe D$p$-branes with worldvolume electric flux in either the near-horizon geometry of very many D$q$-branes or the $AdS_4 \times \mathbb{CP}^3$ background of type IIA supergravity. We have characterized these D$p$-branes by the number of ND directions with respect to the D-branes producing the background. These ND numbers appear in the top row. An `X' indicates that the known stable solution for the D$p$-brane worldvolume fields describes a compressible ground state with the salient feature listed in the leftmost column.}
\label{T:468nd}
\end{center}
\end{table*}
The astute reader will notice a special case in table~\ref{T:468nd}: the D4-brane along $AdS_3 \times \mathbb{CP}^1$ inside $AdS_4\times\mathbb{CP}^3$~\cite{Fujita:2009kw}. This case is special because the D4-brane action includes two WZ terms that together help determine the physics of the ground state. The first WZ term, $P[C_3]\wedge F$, conspires with the DBI term in the D4-brane action to admit a Higgs branch at zero baryon density, just as in the D3/D5 system. The crucial difference between the D4-brane and the D3/D5 system is the nature of the compressible matter. The D4-brane wraps the $\mathbb{CP}^1$ threaded by RR two-form flux, so that the $P[C_1]\wedge F\wedge F$ WZ term becomes a Chern-Simons term on the non-compact part of the worldvolume, which is $AdS_3$. As a result the baryon number current is chiral, so that the $U(1)$ charge is essentially topological. States with nonzero baryon number correspond to embeddings with a Wilson line, so that the effective metric does not depend on the charge density. Consequently the zero-density Higgs branch trivially survives at nonzero density.

Table~\ref{T:468nd} displays a clear correlation between the number of ND directions of the probe D$p$-brane and the character of the known compressible ground state. Moreover, that correlation appears to be intimately related to the form of the WZ terms in the D$p$-brane actions. This ``experimental'' evidence suggests that a classification of holographic matter, as we have defined it, may be possible, via a classification of WZ terms that can appear on probe D$p$-branes, or in other words via a K-theory classification of D-branes~\cite{Witten:1998cd,Horava:1998jy,Sen:2004nf} with worldvolume electric flux. WZ terms encode anomalies in the D$p$-brane worldvolume theory, and roughly speaking are dual to anomalies in the field theory. In field theory terms, then, table~\ref{T:468nd} raises the possibility that holographic matter could be classified by anomalies.

Remarkably, topological insulators, defined as states \textit{incompressible} in their bulk but containing gapless, topologically-protected edge modes, can be classified either by K-theory~\cite{Kitaev:2009mg} or equivalently by anomalies~\cite{2012PhRvB..85d5104R}. Either classification leads to the so-called ``periodic table'' of topological insulators. Furthermore, the classification has been reproduced via a K-theory classification of intersecting D-branes and O-planes (not in the Maldacena limit\footnote{The classification can probably be extended to theories in the Maldacena limits, accounting for \emph{e.g.} fractional topological insulators~\cite{Maciejko:2010tx,Swingle:2010rf}. For holographic examples of topological insulators, see also refs.~\cite{HoyosBadajoz:2010ac,Karch:2010mn,Ammon:2012dd}.})~\cite{Ryu:2010hc,Ryu:2010fe}. Our table~\ref{T:468nd} is a hint that a similar classification may also be possible for D-brane systems describing \textit{compressible} states. In other words, table~\ref{T:468nd} may be the beginning of a periodic table of holographic matter.

A number of questions arise about any such putative classification of holographic matter, however. For topological insulators, anomalies provide a definition equivalent to that in terms of gapless edge modes, so of course classifying anomalies necessarily means classifying topological insulators. In other words, anomalies alone are sufficient to characterize topological insulators completely. Are anomalies really sufficient to characterize holographic matter? Given the form of the WZ terms in the D$p$-brane action, can we immediately deduce what the compressible ground state must be, or do we need more information? Are the cases in table~\ref{T:468nd} merely special cases where WZ terms happen to play an important role, or do they reveal some fundamental (and heretofore obscure) principle relating the form of anomalies to the character of compressible ground states? Regrettably, we will leave these important questions for future research.

\section{Summary and Open Questions}
\label{S:discuss}

We have studied $(3+1)$-dimensional $\mathcal{N}=4$ SYM theory with gauge group $SU(N_c)$, in the large-$N_c$ and large-coupling limits, coupled to a single massless fundamental-representation hypermultiplet propagating along an $(n+1)$-dimensional defect, with $n=3,2,1$, in the probe limit. We studied states in these theories with a finite baryon number charge density. These states were compressible and had a high degree of symmetry, including translational and rotational symmetry, parity invariance, and other global symmetries. Using holography, we demonstrated that in fact a moduli space of such states exists, parameterized by the expectation values of certain scalar operators bilinear in the squarks. More precisely, we demonstrated that neither the free energy nor the energy depends on the values of these scalar expectation values. At a generic point on the moduli space, the R-symmetry and $SU(N_c)$ gauge invariance are broken to subgroups, hence we are justified in calling these moduli spaces Higgs branches.

The holographic descriptions of our compressible states were D$p$-branes, with $p=2n+1$, extended along an asymptotically $AdS_{n+2} \times \mathbb{S}^n$ submanifold of $AdS_5 \times \mathbb{S}^5$, with nonzero worldvolume electric flux at the $AdS_{n+2}$ boundary. We constructed explicit normalizable solutions for the worldvolume fields dual to the scalars whose expectation values parameterize the Higgs branch. The crux of our calculation followed from an observation made by Gibbons and Hashimoto~\cite{Gibbons:2000mx}: a four-dimensional DBI action for a $U(1)$ field strength $\hat{f}$ satisfies a topological bound, which is saturated for either self-dual or anti-self-dual $\hat{f}$. In all three of our cases, we were able to represent each of our solutions as a four-dimensional field strength $\hat{f}$ ``living'' on the manifold  $\mathbb{R}^{n+1}\backslash\mathbb{B}^{n+1} \times \mathbb{R}^{3-n}$, that is, the $\mathbb{R}^{n+1}$ spanned by the $AdS_{n+2}$ radial coordinate and the $\mathbb{S}^n$ with all points inside the ball $\mathbb{B}^{n+1}$ excised, and where the $\mathbb{R}^{3-n}$ factor indicates dummy variables. The metric on the $(n+1)$-dimensional space, which we called the effective metric, was determined by the worldvolume electric flux, and was in fact conformally equivalent to the flat metric. We uplifted this metric to one on the four-dimensional space by introducing dummy directions and putting the flat metric on them. Our solutions for $\hat{f}$ were self-dual with respect to this four-dimensional metric, and thus extremized the action. The D$p$-brane's on-shell action and stress-energy tensor were independent of our self-dual $\hat{f}$'s. For the D7-brane ($n=3$), self-dual $\hat{f}$ correspond to dissolved D3-branes. Solutions for $\hat{f}_{ij}$ with singularities outside the $\mathbb{B}^4$ are singular, and hence unphysical. Only solutions with singularities ``inside the ball,'' which are non-singular in the physical region, correspond to points on the nonzero-density Higgs branch. For the D5-brane ($n=2$), if we eliminate the dummy variables then self-dual $\hat{f}$ reduce to (Hodge) dual vector/scalars, and in the simplest cases correspond to half-D3 branes attached to the D5-brane. For the D$3'$-brane ($n=1$), if we eliminate the dummy variables then self-dual $\hat{f}$ reduce to holomorphic scalars, and correspond to the merger of the D$3'$-branes with the D3-branes producing the background geometry and RR five-form flux. For $n=2,1$, solutions with singularities outside the ball plus those with singularities inside the ball together describe all points on the nonzero-density Higgs branch.

Our results raise many questions. Here we will mention only a few, approximately arranged in order of (what we perceive as) increasing difficulty:

\begin{itemize}

\item What is the metric on the nonzero-density Higgs branch in each of our systems?

\item What is the spectrum of fluctuations at a generic point of the nonzero-density Higgs branch? More specifically, is holographic zero sound present at any points on the nonzero-density Higgs branch besides the origin?

\item What is the precise quantization condition on the free parameters of our self-dual field strength solutions, such as the $c_l$ in eq.~\eqref{E:SDinstanton}?

\item What happens if we heat our systems up to some nonzero temperature $T$? On general grounds, we expect any nonzero $T$ to lift a moduli space, \textit{i.e.} to push the moduli towards specific values, as occurs in SUSY gauge theories with $T=0$ moduli spaces, like the $\mathcal{N}=2$ $SU(2)$ SYM theory studed in ref.~\cite{Paik:2009iz}.

\item What happens to the nonzero-density Higgs branch in the D3/D$p$ systems in the presence of a nonzero magnetic field $B$? Does nonzero $B$ lift the moduli space?

\item What if the hypermultiplets are massive? First of all, notice that to maintain a nonzero density we must demand that the mass be less than the chemical potential $\mu$~\cite{Karch:2007br}. To write an ansatz for the D$p$-brane worldvolume fields describing nonzero mass, we must break $SO(5-n)$ symmetry, and allow for the scalars $y^M$ on the D$p$-brane to become non-trivial. The scalar $\sqrt{\sum_M^{5-n} (y^M)^2}$ is holographically dual to the mass operator, so we want a non-normalizable solution for $\sqrt{\sum_M^{5-n} (y^M)^2}$. Allowing for $SO(n+1)$ symmetry to be broken, the most general ansatz we can write for the scalars is $y^M(z)$. The effective metric then becomes
\beq
g_{ij} = \delta_{ij} - \partial_i A_0 \partial_j A_0 + \delta_{MN} \, \partial_i y^M \partial_j y^N. \nonumber
\eeq
Most of our analysis then proceeds unchanged, for example we find that self-dual $\hat{f}$'s saturate the topological bound on the D$p$-brane action, and that the D$p$-brane action evaluated on self-dual $\hat{f}$'s depends only on $A_0(z)$ and the $y^M(z)$, but is independent of $\hat{f}$. For $A_0(z)$ and $y^m(z)$ describing compressible states, the self-duality (or vector/scalar or holomorphicity) condition is straightforward to solve. We still find a nonzero-density Higgs branch. Notice that at zero density with massive flavors these systems have no Higgs branches, rather, they have mixed Coulomb-Higgs branches. The reason why is easy to see in terms of D-brane physics. Consider for example the D3/D7 system with $N_f>1$ so that a zero-density SUSY Higgs branch exists. A nonzero mass corresponds to separating the D3- and D7-branes in an overall transverse direction. In that case, D3-branes can only dissolve into the D7-brane if they move on top of the D7-brane, indicating the some adjoint scalar(s) of $\N=4$ SYM acquired a nonzero expectation value, and hence only a mixed Coulomb-Higgs branch exists in this case. A similar argument applies for the D3/D5 and D3/D3 systems as well. At nonzero density, our systems have an ``extra'' ingredient, namely a density of strings stretched between the D-branes. These pull the D$p$-brane towards the D3-branes, and indeed the D$p$-brane must always intersect the D3-branes~\cite{Kobayashi:2006sb}. In this case a nonzero mass appears as a nonzero asymptotic separation. In the D3/D7 example, the D3-branes can then dissolve into the D7-branes, even when the asymptotic separation is nonzero, so in the field theory at nonzero density we can find a Higgs branch alone, rather than a mixed Coulomb-Higgs branch.

\item What about solutions of the general form written for example in eq.~\eqref{D7fields}, which could describe many field theory states with different external sources for the dual fields, beyond just $\mu$? What do solutions of the self-duality (vector/scalar duality, or holomorphicity) conditions imply for such states?

\item What is the physical meaning of the effective metric, on both sides of the correspondence? Is the effective metric related to the metric on the nonzero-density Higgs branch, and if so, how?

\item Are the states we have studied \textit{global} minima of the free energy, or do states with lower free energy exist? At the origin of the Higgs branch, as we discussed in section~\ref{S:4ND} the D3/D$p$ theories are known to be thermodynamically stable~\cite{Benincasa:2009be}. For the D3/D7 system, a holographic calculation revealed that at the origin of the moduli space the spectrum is tachyon-free, hence this state is stable with respect to dynamical (\textit{i.e.} finite-frequency and finite-momentum) fluctuations~\cite{Ammon:2011hz}. Is that true for the D3/D5 and D3/D3 systems? Away from the origin of the moduli space, are any of the D3/D$p$ systems stable against either thermodynamic or dynamical fluctuations? A straightforward way to determine the absolute minimum of the free energy is, of course, to scan through the entire space of all compressible states. Our work amounts to a small step toward this goal, in that we have scanned through a large subspace of compressible states with a high degree of symmetry, including rotational, translational, and parity invariance in $\mathbb{R}^{n,1}$.

As we mentioned in sections~\ref{S:d3d7}-\ref{S:d3d3}, the D$p$-brane action for our solutions is just that of DBI electrostatics in $\mathbb{R}^{n+1}$. The states that we have studied are those which have a charge at the origin of $\mathbb{R}^{n+1}$, the string endpoints sitting at the D3-branes, which is dual to the charge density $\langle J^0\rangle$ in the field theory. The string endpoints source the analogue of the Coulomb potential in ordinary electrostatics, and the resulting solution is sometimes referred to as a BIon in the literature~\cite{Gibbons:1997xz}. Solutions describing BIons distributed throughout $\mathbb{R}^{n+1}$ represent various different states with the same net charge density as the known ground states. To answer the question of thermodynamic stability, we must scan through the space of such solutions and determine which corresponding field theory state has the lowest free energy.

For the D3/D5 system, the authors of ref.~\cite{UW:WIP} used methods similar to those in section~\ref{S:d3d5} to demonstrate the existence of the nonzero-density Higgs branch of the D3/D5 theory. They further argued, based on arguments given by Gibbons in ref.~\cite{Gibbons:1997xz}, that states might exist in the theory with the same charge as the known ground state, but with lower free energy. The essence of the argument is the observation that, na\"ively, a single BIon can lower its energy by fissioning into multiple BIons, which suggests that indeed states with lower free energy may exist. In the bulk these states would be described by a D5-brane with D3-brane spikes centered at disparate locations in $\mathbb{R}^3$, with BIons centered at the points where the D5-brane stretches to infinity. We refer the reader to ref.~\cite{UW:WIP} for more information, however we point out that the arguments of ref.~\cite{UW:WIP}, suitably modified, also apply to the D3/D7 and D3/D3 systems. It would be extremely interesting to construct such multi-BIon solutions explicitly and determine whether the dual field theory states have free energy lower than the known ground state.

\item What is the low-energy effective description of holographic matter? Holographic calculations indicate that such a theory is some (0+1)-dimensional CFT~\cite{Jensen:2010ga,Nickel:2010pr,Ammon:2011hz}. Any putative effective theory must reproduce the bizarre physics of holographic matter, such as the extensive ground state degeneracy, holographic zero sound, and now also a moduli space. Indeed, the existence of moduli suggests that the low-energy theory may be the theory on the D3-branes moving around inside the D$p$-brane, in a similar fashion to the description in the D1/D5 system in terms of the sigma model describing D1-branes dissolved in the D5-brane worldvolume~\cite{Strominger:1996sh}. As we discussed in section~\ref{S:d3d7} for the D3/D7 case, the low-energy theory may arise from the gapless modes of D3-branes dissolved into the D7-brane. More specifically, we could imagine writing a theory for the D3-brane worldvolume fields interacting with the strings stretched between the D3-branes and the D7-brane. Presumably we would obtain some kind of non-linear sigma model. In a best-case scenario, that sigma model would exhibit (0+1)-dimensional conformal invariance and enough zero-energy states to reproduce the extensive ground state degeneracy. A natural question is whether the target-space metric in such a sigma model is related to the effective metric and/or the Higgs branch metric. A concrete way to approach the low-energy effective description of holographic matter may be to leave holography temporarily, and consider the D3/D7 intersection at small $N_c$ in the presence of a BIon. In that system, we have performed a calculation similar to that in section~\ref{S:d3d7}, which shows that a Higgs branch exists in the classical limits. We should be able to access the low-energy spectrum and interactions directly in this system, upon quantizing open strings in this background. Hopefully such an exercise will indeed reveal a sigma model with target space $\mathbb{R}^4\backslash \mathbb{B}^4$, (0+1)-dimensional conformal invariance, and extensive ground state degeneracy. If so, then we may have a good chance of identifying the right low-energy effective theory in the Maldacena and probe limits.

\end{itemize}

\acknowledgments
\noindent We would like to thank Han-Chih Chang, Aleksey Cherman, Johanna Erdmenger, Nick Evans, Sean Hartnoll, Clifford Johnson, Andreas Karch, Shu Lin, Dam Son, David Tong and Hyun Seok Yang for valuable discussions and correspondence. We would also like to thank Shamit Kachru and Matthias Kaminski for reading and commenting on a preliminary draft of this paper. A.O'B. would especially like to thank Andreas Karch for collaboration on closely related topics that led to this paper. K.J. is grateful to the organizers of the workshop \textit{Applications of Gauge-Gravity Duality} at the Technion for their hospitality while this work was in progress. A.O'B. is grateful to the Crete Center for Theoretical Physics for their hospitality and support while this work was in progress. The work of M.A. was supported by National Science Foundation grant PHY-07-57702. K.J. was supported in part by NSERC, Canada. K.K. acknowledges support via an NWO Vici grant of K. Skenderis. His work is part of the research program of the Stichting voor Fundamenteel Onderzoek der Materie (FOM), which is financially supported by the Nederlandse Organisatie voor Wetenschappelijk Onderzoek (NWO). J.L. is supported by the ERC STG grant 279943, “Strongly Coupled Systems”. The work of A.O'B. was supported in part by the European Research Council grant ``Properties and Applications of the Gauge/Gravity Correspondence'' and in part by the European Union grant FP7-REGPOT-2008-1-CreteHEPCosmo-228644.

\appendix
\section*{Appendix: $\kappa$-symmetry and Vector/Scalar Duality}
\addcontentsline{toc}{section}{Appendix: $\kappa$-symmetry and Vector/Scalar Duality}

In this appendix we present the technical details of the $\kappa$-symmetry condition for the probe D5-brane discussed in section~\ref{S:d3d5}. Following ref.~\cite{Skenderis:2002vf}, we will prove that for our ansatz in eq.~\eqref{E:d3d5ansatz}, with $A_0(z)=0$, the D5-brane satisfies the $\kappa$-symmetry condition, and hence preserves maximal SUSY, if and only if the worldvolume fields obey the vector/scalar duality condition in eq.~\eqref{E:vsDuality}. We will follow the conventions used in ref.~\cite{Ammon:2012dd}.

To ensure spacetime SUSY the probe D5-brane's worldvolume fields must satisfy the $\kappa$-symmetry condition
\beq
( 1 - \Gamma ) \epsilon = 0 \, ,
\eeq
where $\epsilon$ are the Killing spinors generating the 32 supersymmetries preserved by the $AdS_5 \times \mathbb{S}^5$ background, and $\Gamma$ is the $\kappa$-symmetry projector for the probe D5-brane~\cite{Bergshoeff:1997kr}
\beq
\label{E:appgammadef}
\Gamma = \frac{1}{\sqrt{- \det\left( P[G]_{ab} + F_{ab} \right)}} \sum\limits_{n=0}^\infty \frac{1}{n! 2^n} \gamma^{i_1 j_1 \dots i_n j_n} F_{i_1 j_1} \dots F_{i_n j_n} J_{5}^{(n)}\,,
\eeq
where the $\gamma_a$ are the pullbacks to the D5-brane worldvolume of the curved-space gamma matrices. The $\gamma_a$ are related to the flat-space gamma matrices $\Gamma_A$ as
\beq
\gamma_a = (\partial_a x^m) E_{\phantom{A}m}^A \Gamma_A\,,
\eeq
where $m,A \in \{0, \dots, 9\}$ and $E_{\phantom{A}m}^A$ is the vielbein associated with the metric in eq.~\eqref{eq:metric}. For a D5-brane along $(x^0,x^1,x^2)$ and $(x^4,x^5, x^6)$ the factor $J_5^{(n)}$ in eq.~\eqref{E:appgammadef} is
\beq
J_5^{(n)} = (-1)^n (\sigma_3)^{n+1} i\sigma_2 \otimes \gamma_{012456}\,,
\eeq
where $\sigma_2$ and $\sigma_3$ are Pauli matrices.

As shown in refs.~\cite{Kehagias:1998gn,Grana:2000jj}, for the 16 Poincar\'e supersymmetries of $AdS_5 \times \mathbb{S}^5$ the corresponding spinor $\epsilon$ can be written in terms of a constant spinor $\epsilon_0$,
\beq
\epsilon = r^{-1/2} \epsilon_0\, ,
\eeq
where both $\epsilon$ and $\epsilon_0$ are doublets of ten-dimensional Majorana-Weyl spinors. The spinor $\epsilon_0$ satisfies
\beq
\label{eq:spinorcond1}
\left( i \sigma_2 \otimes \Gamma_{0123}\right) \epsilon_0 = \epsilon_0,
\eeq
as does the spinor $\epsilon$. Notice that the Pauli matrices act on the doublet index.

For the D5-brane studied in section~\ref{S:d3d5}, when $A_0(z)=0$ the $\kappa$-symmetry projector is
\beq
\Gamma = \frac{1}{\sqrt{- \det\left( P[G]_{ab} + F_{ab} \right)}} \left[ \sigma_1 \otimes \gamma_{012456} - \frac{i}{2} F_{ij}  \sigma_2 \otimes \gamma^{ij}\gamma_{012456} \right] \, .
\eeq
The relevant gamma matrices $\gamma_a$ are
\begin{subequations}
\begin{align}
\gamma_\mu &= \rho \Gamma_\mu\, , \qquad\qquad\qquad\ \ \mbox{for} \quad \mu \in \{ 0,1,2 \}\,, \\
\gamma_i &= \frac{1}{\rho} \Gamma_i + \rho \Gamma_3 \partial_i x^3\, , \qquad \mbox{for} \quad i \in \{ 4,5,6 \}\,,
\end{align}
\end{subequations}
and so we find
\begin{subequations}
\begin{align}
\label{eq:Gammamat1}
\gamma_{012456} &= \Gamma_{012456} \left( 1 + \rho^2 \partial_i x^3 \Gamma_j \Gamma_3 \delta^{ij}\right)\,, \\  \label{eq:Gammamat2}
\gamma_{012i} &= \rho^2 \Gamma_{012i} + \rho^4 \Gamma_{0123} \partial_i x^3\,.
\end{align}
\end{subequations}
Henceforth we raise and lower the indices $i$ of the $\Gamma_i$ with a Kronecker delta. Using eqs.~\eqref{eq:Gammamat1} and~\eqref{eq:Gammamat2}, and using $\gamma^{ij} \gamma_{012456} = - \tilde{\epsilon}^{ijk} \gamma_{012k}$ with $\tilde{\epsilon}^{456}=1$, we obtain for the $\kappa$-symmetry projector
\beq
\Gamma = \frac{1}{\sqrt{- \det\left( P[G]_{ab} + F_{ab} \right)}} \left[ \sigma_1 \otimes \Gamma_{012456} \left( 1 +\rho^2 \partial^i x^3 \Gamma_{i3} \right) + \frac{i}{2} \rho^2 \tilde{\epsilon}^{ijk} F_{ij} \sigma_2 \otimes \Gamma_{0123} \left( \Gamma_{3k} + \rho^2 \partial_k x^3 \right)\right]\,.
\eeq
For our probe D5-brane in $AdS_5 \times \mathbb{S}^5$, the constant spinor $\epsilon_0$ must satisfy eq.~\eqref{eq:spinorcond1} as well as
\beq
\label{eq:spinorcond2}
\left( \sigma_1 \otimes \Gamma_{012456} \right) \epsilon_0 = \epsilon_0\,,
\eeq
which is equivalent to
\beq
\label{eq:spinorcond3}
\left( -\sigma_3 \otimes \Gamma_{456} \right) \epsilon_0 = (1 \otimes \Gamma_3) \epsilon_0\,.
\eeq
We wish to study D5-brane embeddings which preserve the maximum possible amount of SUSY. This may be accomplished by only imposing eqs.~\eqref{eq:spinorcond1} and~\eqref{eq:spinorcond3} with no other constraints on $\epsilon_0$, in which case the D5-brane preserves eight of the sixteen Poincar\'e supercharges of the background. Using eqs.~\eqref{eq:spinorcond1} and \eqref{eq:spinorcond3}, we find
\beq
\Gamma \epsilon_0 = \frac{1}{\sqrt{- \det\left( P[G]_{ab} + F_{ab} \right)}}  \left[ \epsilon_0 \left( 1+ \frac{1}{2} \tilde{\epsilon}^{ijk} F_{ij} \rho^4 \partial_k x^3 \right) + (1\otimes \Gamma_{3i}) \epsilon_0 \rho^2 \left( \partial^i x^3 - \frac{1}{2} \tilde{\epsilon}^{ijk} F_{jk} \right) \right]\,.
\eeq
For an arbitrary spinor $\epsilon_0$ satisfying eqs.~\eqref{eq:spinorcond1} and~\eqref{eq:spinorcond3}, with no additional constraints, $\epsilon_0$ and $(1\otimes \Gamma_{3i})\epsilon_0$ must be linearly independent, hence we must impose that the coefficient of $(1\otimes \Gamma_{3i})\epsilon_0$ vanish, giving
\beq
\partial^i x^3 - \frac{1}{2} \tilde{\epsilon}^{ijk} F_{jk} = 0\,,
\eeq
which is precisely the vector/scalar duality condition in eq.~\eqref{E:vsDuality}. Notice that with $A_0(z)=0$ the metric in eq.~\eqref{E:3dfields} is flat, in which case $\epsilon^{ijk}$ as defined below eq.~\eqref{E:vsDuality} is the same as $\tilde{\epsilon}^{ijk}$. Imposing the vector/duality condition, we find
\beq
\sqrt{- \det\left( P[G]_{ab} + F_{ab} \right)} = 1+ \frac{\rho^4 }{2} \tilde{\epsilon}^{ijk} F_{ij}\partial_k x^3\,,
\eeq
and therefore
\beq
\Gamma \epsilon_0 = \epsilon_0\,,
\eeq
which shows that \textit{if} we impose the vector/scalar duality condition \textit{then} the D5-brane preserves maximal SUSY. Notice also that \textit{if} we do not impose the vector/scalar duality condition \textit{then} we cannot satisfy $\Gamma \epsilon_0 = \epsilon_0$ for a generic constant spinor satisfying eqs.~\eqref{eq:spinorcond1} and~\eqref{eq:spinorcond3}. We thus conclude that the probe D5-brane preserves the maximal amount of SUSY \textit{if and only if} the vector/scalar duality condition in eq.~\eqref{E:vsDuality} is satisfied.

\bibliographystyle{JHEP}
\bibliography{d3dpinstantons}
\end{document}